\documentclass[dvipdfmx]{pasj01}

\usepackage{lscape}
\usepackage[position=top,margin=20pt,singlelinecheck=off]{subfig}
\usepackage{color}
\usepackage{dcolumn}
\usepackage{url}
\usepackage{xspace} 

\newcommand{\Tab}[1]{Table \ref{#1}}
\newcommand{\Fig}[1]{Figure \ref{#1}}
\newcommand{\Sec}[1]{Section \ref{#1}}

\newcommand{\chandra}{{Chandra }}

\newcommand{\telec}{$T_\mathrm{e}$\xspace}
\newcommand{\tioni}{$T_\mathrm{Z}$\xspace}

\newenvironment{contribution}{\section*{Author contributions}\fontsize{8}{11}\selectfont}{\par}

\begin{document} 
\Received{}
\Accepted{}

\title{Temperature Structure in the Perseus Cluster Core Observed with Hitomi
\thanks{Corresponding authors are
Shinya \textsc{Nakashima},
Kyoko \textsc{Matsushita}, 
Aurora \textsc{Simionescu}, 
Mark \textsc{Bautz}, 
Kazuhiro \textsc{Nakazawa}, 
Takashi \textsc{Okajima},
and 
Noriko \textsc{Yamasaki}
}
}

\author{Hitomi Collaboration,
Felix \textsc{Aharonian}\altaffilmark{1,2,3},
Hiroki \textsc{Akamatsu}\altaffilmark{4},
Fumie \textsc{Akimoto}\altaffilmark{5},
Steven W. \textsc{Allen}\altaffilmark{6,7,8},
Lorella \textsc{Angelini}\altaffilmark{9},
Marc \textsc{Audard}\altaffilmark{10},
Hisamitsu \textsc{Awaki}\altaffilmark{11},
Magnus \textsc{Axelsson}\altaffilmark{12},
Aya \textsc{Bamba}\altaffilmark{13,14},
Marshall W. \textsc{Bautz}\altaffilmark{15}
% ...
%
%
}

\altaffiltext{1}{Dublin Institute for Advanced Studies, 31 Fitzwilliam Place, Dublin 2, Ireland}
\altaffiltext{2}{Max-Planck-Institut f{\"u}r Kernphysik, P.O. Box 103980, 69029 Heidelberg, Germany}
\altaffiltext{3}{Gran Sasso Science Institute, viale Francesco Crispi, 7 67100 L'Aquila (AQ), Italy}
\altaffiltext{4}{SRON Netherlands Institute for Space Research, Sorbonnelaan 2, 3584 CA Utrecht, The Netherlands}
\altaffiltext{5}{Institute for Space-Earth Environmental Research, Nagoya University, Furo-cho, Chikusa-ku, Nagoya, Aichi 464-8601}
\altaffiltext{6}{Kavli Institute for Particle Astrophysics and Cosmology, Stanford University, 452 Lomita Mall, Stanford, CA 94305, USA}
\altaffiltext{7}{Department of Physics, Stanford University, 382 Via Pueblo Mall, Stanford, CA 94305, USA}
\altaffiltext{8}{SLAC National Accelerator Laboratory, 2575 Sand Hill Road, Menlo Park, CA 94025, USA}
\altaffiltext{9}{NASA, Goddard Space Flight Center, 8800 Greenbelt Road, Greenbelt, MD 20771, USA}
\altaffiltext{10}{Department of Astronomy, University of Geneva, ch. d'\'Ecogia 16, CH-1290 Versoix, Switzerland}
\altaffiltext{11}{Department of Physics, Ehime University, Bunkyo-cho, Matsuyama, Ehime 790-8577}
\altaffiltext{12}{Department of Physics and Oskar Klein Center, Stockholm University, 106 91 Stockholm, Sweden}
\altaffiltext{13}{Department of Physics, The University of Tokyo, 7-3-1 Hongo, Bunkyo-ku, Tokyo 113-0033}
\altaffiltext{14}{Research Center for the Early Universe, School of Science, The University of Tokyo, 7-3-1 Hongo, Bunkyo-ku, Tokyo 113-0033}
\altaffiltext{15}{Kavli Institute for Astrophysics and Space Research, Massachusetts Institute of Technology, 77 Massachusetts Avenue, Cambridge, MA 02139, USA}
% ...

\email{shinya.nakashima@riken.jp}

\KeyWords{galaxies: clusters: individual (Perseus)  --- X-rays: galaxies: clusters --- methods: observational } 

\maketitle

\begin{abstract}
The present paper investigates the temperature structure of the X-ray emitting plasma in the core of the Perseus cluster using the 1.8--20.0 keV data obtained with the Soft X-ray Spectrometer (SXS) onboard the Hitomi Observatory.
A series of four observations were carried out, with a total effective exposure time of 338~ks and covering a central region $\sim7\arcmin$ in diameter.
The SXS was operated with an energy resolution of $\sim$5~eV (full width at half maximum) at 5.9~keV.
Not only fine structures of K-shell lines in He-like ions but also transitions from higher principal quantum numbers are clearly resolved from Si through Fe. This enables us to perform temperature diagnostics using the line ratios of Si, S, Ar, Ca, and Fe,
and to provide the first direct measurement of the excitation temperature and ionization temperature in the Perseus cluster.
The observed spectrum is roughly reproduced by a single temperature thermal plasma model in collisional ionization equilibrium, but detailed line ratio diagnostics reveal slight deviations from this approximation.
In particular, the data exhibit an apparent trend of increasing ionization temperature with increasing atomic mass, as well as small differences between the ionization and excitation temperatures for Fe, the only element for which both temperatures can be measured.
The best-fit two-temperature models suggest a combination of  3 and 5 keV gas, which is consistent with the idea that the observed small deviations from a single temperature approximation are due to the effects of projection of the known radial temperature gradient in the cluster core along the line of sight.
Comparison with the Chandra/ACIS and the XMM-Newton/RGS results on the other hand suggests that additional lower-temperature components are present in the ICM but not detectable by Hitomi SXS given its 1.8--20 keV energy band.
\end{abstract}

\section{Introduction}

The X-ray emitting hot intracluster medium (ICM) dominates the baryonic mass in galaxy clusters, and its thermodynamical properties are crucial for studying the evolution of large-scale structure in the Universe.
Discontinuities in the ICM temperature and density profiles reveal ongoing cluster mergers \citep{2000ApJ...541..542M,2001ApJ...551..160V,2007PhR...443....1M,2013PASJ...65...16A}, while the pressure profiles in the cluster outskirts are also key to understanding their growth \citep{2010A&amp;A...517A..92A,2011Sci...331.1576S,2013A&amp;A...550A.131P,2017MNRAS.469.1476S}.
The thermodynamical properties of the dense ICM at the centers of so-called ``cool-core'' clusters are even more complex; despite the fact that radiative cooling in these regions should be very efficient, 
stars are being formed at a rate smaller than that expected from the amount of hot ICM (e.g., \cite{2003ApJ...590..207P}).
The heating mechanism responsible for compensating the radiative cooling is under debate, and various ideas have been proposed, such as feedback from the active galactic nuclei (AGN) in the brightest cluster galaxies (e.g., \cite{2007ARA&amp;A..45..117M}), energy transfer from moving member galaxies (e.g, \cite{2001PASJ...53..401M,2013ApJ...767..157G}), and cosmic-ray streaming with Alfv\'en waves (e.g., \cite{2013MNRAS.432.1434F}). 
While less effective than expected, some radiative cooling likely does occur, and the presence of multi-phase ICM in cool-core clusters is also reported \citep{1994PASJ...46L..55F,2007MNRAS.381.1381S,2009ApJ...701..377T,2012ApJ...749..186G,2016MNRAS.457...82S,2016MNRAS.461.2077P}.

To date, temperature measurements of the ICM have been mainly performed by fitting broad-band spectra (typically 0.5--10.0~keV band) obtained from X-ray CCDs. 
Because of the moderate energy resolution of this type of spectrometers, temperatures are mainly determined by shapes of the continuum and the Fe L-shell lines complex. 
However, the continuum shape is subject to uncertainties due to background modeling and/or effective area calibration (e.g., \cite{2007A&amp;A...465..345D,2008A&amp;A...486..359L,2010A&amp;A...523A..22N,2015A&amp;A...575A..30S}).

An independent estimate of the gas temperature can be obtained from the flux ratios of various emission lines, the so-called line ratio diagnostic;
a ratio between different transitions in the same ion such as Ly$\alpha$-to-Ly$\beta$ indicates the excitation temperature, and a ratio of lines from different ionization stages such as He$\alpha$-to-Ly$\alpha$ represents the ion fraction (also referred to as the ionization temperature).
These temperatures should match the temperature from the continuum shape when the observed plasma is truly single temperature in collisional ionization equilibrium (CIE).
If there is a disagreement between those temperatures, deviation from a single CIE plasma is suggested: multi-temperature and/or non-equilibrium ionization (NEI).
For instance, \citet{2002A&amp;A...386...77M} utilized the Si and S K-shell lines to measure the temperature profile in M~87.
Ratios of K-shell lines from Fe were used for the Ophiuchus Cluster \citep{2008PASJ...60.1133F}, the Coma Cluster \citep{2011PASJ...63S.991S} and A754 \citep{2016PASJ...68S..23I}. 
In practice, this method has been applied to a relatively small number of lines 
because of line blending and because only the fluxes of the strongest lines are free from uncertainties in the exact continuum calibration and background subtraction.

The XMM-Newton Reflection Grating Spectrometers (RGS) offer higher spectral resolution and enable us to perform diagnostics with O K-shell and Fe L-shell lines, which are sensitive to the temperature range of $kT<1$~keV (e.g., \cite{2016MNRAS.461.2077P}).
However, the energy band of the RGS is limited to energies below 2~keV, and the energy resolution is degraded for diffuse sources due to the dispersive and slit-less nature of these spectrometers.
Therefore, observations with a non-dispersive high-resolution spectrometer covering a broad energy band are desired for a precise characterization of the multi-temperature structure in the ICM.

The Hitomi satellite launched on February 2016 performed the first cluster observation of this kind, using its Soft X-ray Spectrometer (SXS).
This non-dispersive microcalorimeter achieved spectral resolution of $\sim$5~eV in orbit (Porter et al. 2017), and observed the core of the Perseus cluster as its first light target.
In the observed region, fine ICM substructures such as bubbles, ripples, and weak shock fronts were previously revealed by deep \chandra imaging \citep[and references therein]{2011MNRAS.418.2154F}. These features are thought to be due to the activity of the AGN in the cD galaxy NGC~1275, which is pumping out relativistic 
electrons that disturb and heat the surrounding X-ray gas. The presence of multiple phases structure in the ICM spanning a range of temperatures between $kT=0.5-8$~keV is also reported \citep{2007MNRAS.381.1381S,2016MNRAS.461.2077P}.

The first measurement of Doppler shifts and broadening of the Fe-K emission lines from the Hitomi first-light data, reported in \citet{2016Natur.535..117H} (hereafter the First paper), revealed that the line-of-sight velocity dispersion of the ICM in the core regions is unexpectedly low and subsonic.  
Constraints on an unidentified feature at 3.5 keV suggested to originate from dark matter (e.g., \cite{2014ApJ...789...13B}) are described by \citet{2017ApJ...837L..15A}.
Using the full set of the Perseus data and the latest calibration,
we have performed X-ray spectroscopy over the full Hitomi SXS band and report a series of follow-up papers.
In this paper, we concentrate on measurements of the temperature structure in the cluster core.
The high spectral resolution of the SXS allowed us to estimate the gas temperature based on seventeen independent line ratios from various chemical elements (Si through Fe).
Companion papers report results on the metal abundances (Hitomi Collaboration 2017a, henceforth  the Z~paper), velocity fields (Hitomi Collaboration 2017b, the V~paper), properties of the AGN in NGC1275 (Hitomi Collaboration 2017c, the AGN~paper), the atomic code comparison (Hitomi Collaboration 2017d, the Atomic~paper), and the detection of resonance scattering (Hitomi Collaboration 2017e, the RS~paper).

Throughout this paper, we assume a cluster redshift of 0.017284 (see Appendix~1 of the V paper) and a Hubble constant of 70~km~s$^{-1}$~Mpc$^{-1}$.
Therefore, 1$\arcmin$ corresponds to the physical scale of 21~kpc.
We use the 68\% ($1\sigma$) confidence level for errors, but upper and lower limits are shown at the 99.7\% ($3\sigma$) confidence level. 
X-ray energies in spectra are denoted at the observed (hence redshifted) frame rather than the object's rest-frame.

\section{Observation and Data Reduction}
\label{obs}
\subsection{Hitomi Observation}
\label{obs:hitomi-obs}
\begin{figure*}
 \begin{center}
\subfloat{\includegraphics[width=8cm]{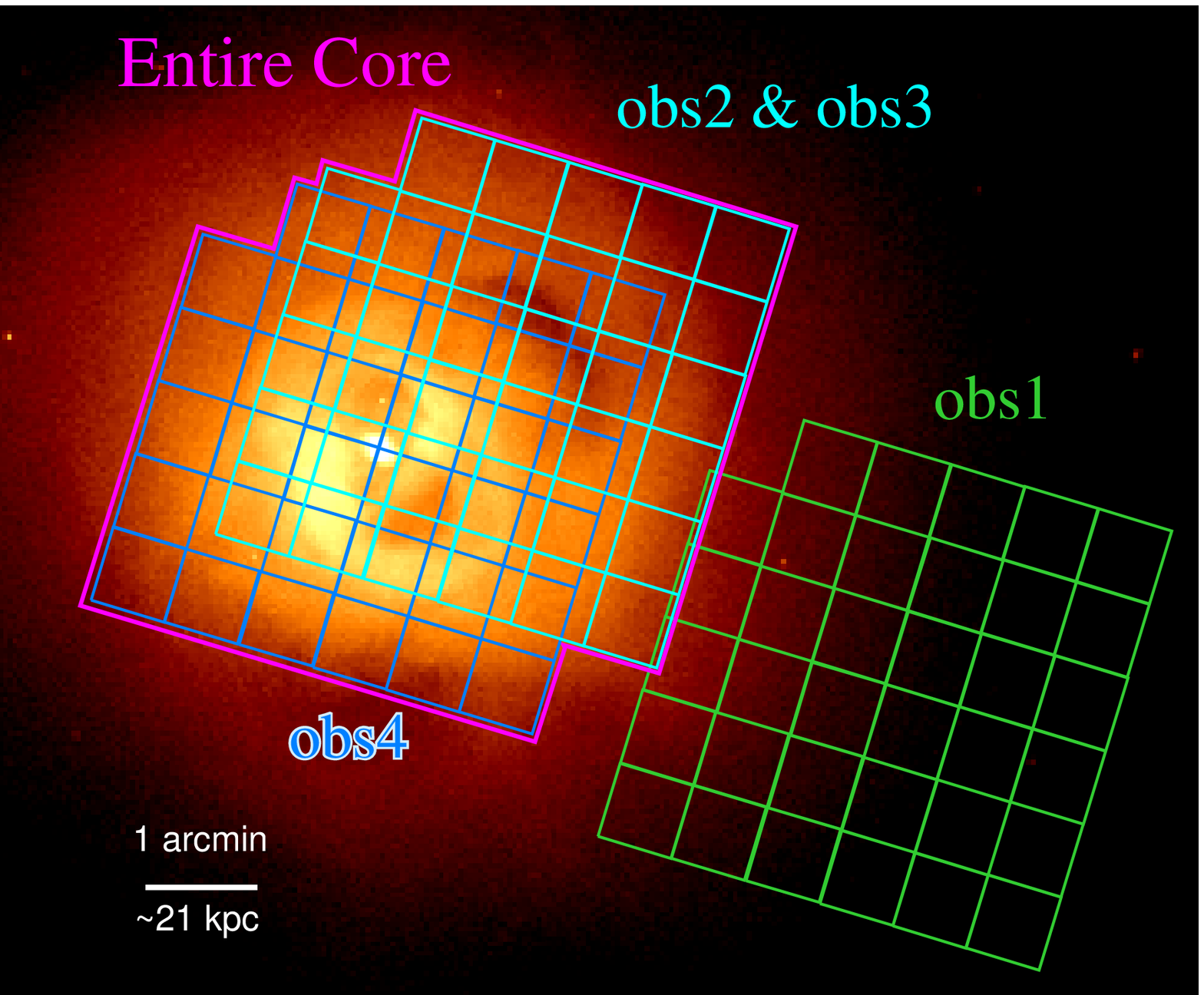}}
\hspace{0.1cm}
\subfloat{\includegraphics[width=8cm]{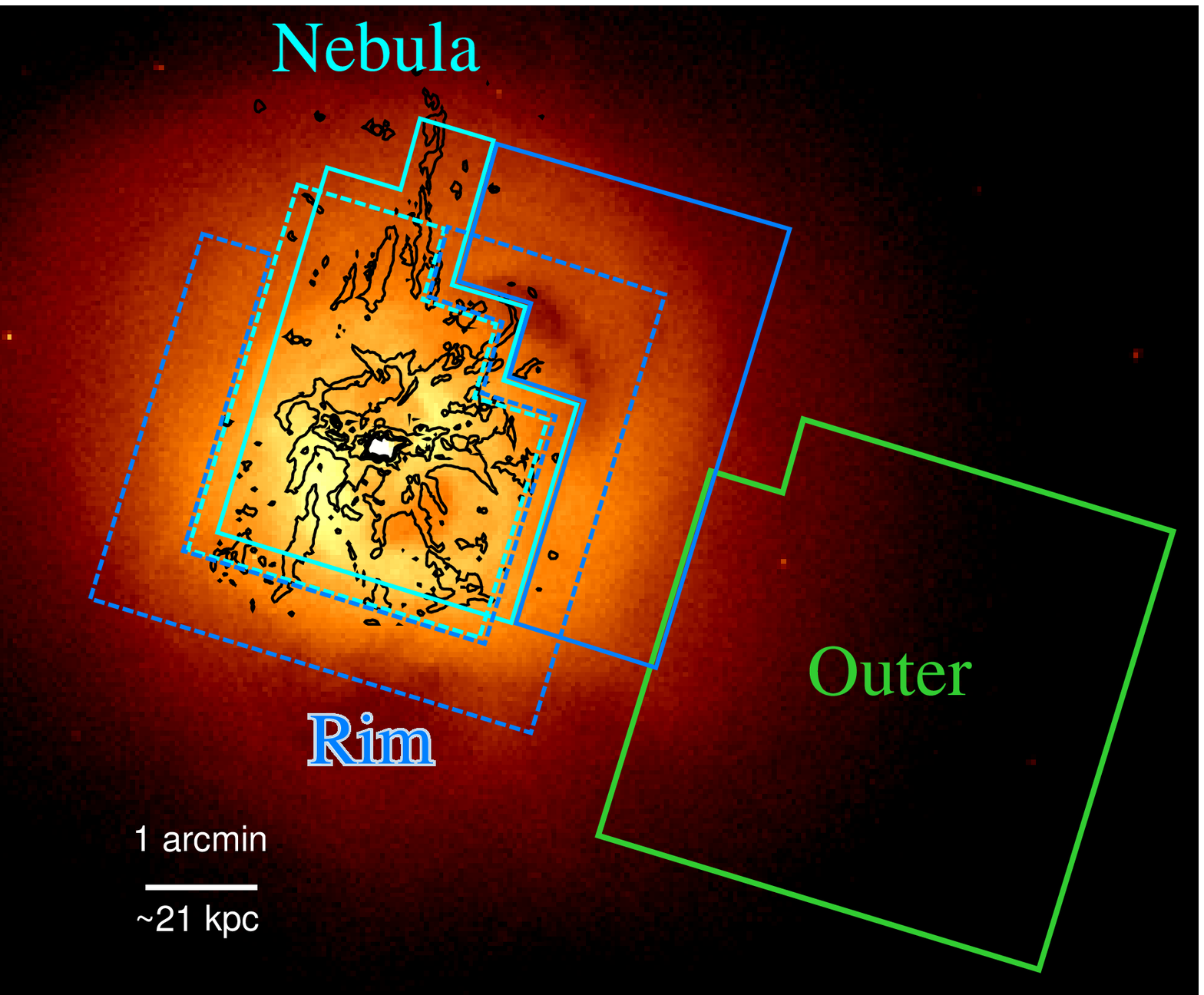}}
 \end{center}
\caption{(left) SXS FoVs of the Hitomi observations overlaid on the Chandra X-ray color image in the 1.8--9.0 keV band.
The green, cyan, and blue polygons indicate obs1, obs2 and obs3, and obs4, respectively.
The 35 square boxes in each FoV correspond to the SXS pixels.
The Entire core region covering the whole obs2/obs3 and obs4 is also shown in magenta.
(right) Analysis regions used in \Sec{ana:devided-1t} overlaid on the same Chandra image.
The H$\alpha$ emission obtained with the WIYN 3.5~m telescope \citep{2001AJ....122.2281C} is also shown in the black contours.
The cyan, blue, and green polygons corresponds to the Nebula, Rim, and Outer regions, respectively.
For Nebula and Rim regions, we used slightly different sky regions between obs2/obs3 and obs4; 
the regions with solid line are for obs2/obs3 and those with dashed line are for obs4 (see text for details).
}
\label{fig:pointing}
\end{figure*}

We observed the Perseus cluster four times with Hitomi/SXS during the commissioning phase in 2016 February and March (\Tab{tab:obslog}).
The aim points of each observation are shown in \Fig{fig:pointing}. 
The first light observation of Hitomi (obs1), is offset by $\sim3\arcmin$ from the center of the Perseus cluster because the attitude control system was not commissioned at that time.
In the next observation (obs2), the pointing direction was adjusted so that the Perseus core was in the SXS field-of-view (FoV).
The same region was observed again after extension of the Hitomi Hard X-ray Detector's optical bench (obs3).
The obs3 is divided into the three sequential data sets (100040030, 100040040, and 100040050) solely for convenience in pipeline processing.
In the final observation (obs4), the aim point was fine-tuned again to place the Perseus core at the center of the SXS FoV.

The SXS sensor is a $6\times6$ pixel array (Kelley et al. 2017).
Combined with the X-ray focusing mirror \citep{okajima16}, the SXS has a $3\arcmin\times3\arcmin$ FoV with an angular resolution of $1.2\arcmin$ (half power diameter).	
One corner pixel is always illuminated by a dedicated $^{55}$Fe source to track the gain variation with detector temperature, and is not used for astrophysical spectra.
The SXS achieved the unprecedented energy resolution of $5$~eV (full width at half maximum) at 5.9 keV in orbit (Porter et al. 2017).
The required energy bandpass of the SXS was 0.3--12 keV. 
During the early-mission observations discussed here, a gate valve remained closed to minimize the risk of contamination from outgassing in the spacecraft. The valve includes a Be window that absorbs most X-rays below ~ 2 keV \citep{eckart17b}. 

The other instruments on Hitomi (Takahashi et al. 2017) were not yet operational during most or all of the Perseus observations described here.
  
\begin{table*}
  \tbl{List of observations.}{%
  \begin{tabular}{clcccc}
      \hline
      Name & \multicolumn{1}{c}{Observation ID} & $\alpha_{2000.0}$ & $\delta_{2000.0}$  & Observation Date & Effective Exposure \\ 
      &  & (deg) & (deg) &  & (ks) \\ 
      \hline
      \multicolumn{6}{c}{Hitomi/SXS} \\
      obs1 & 100040010 & 49.878 & 41.484 & 2016-02-24 -- 2016-02-25 & 49 \\
      obs2 & 100040020 & 49.935 & 41.519 & 2016-02-25 -- 2016-02-27 & 97 \\
      obs3 & 100040030, 100040040, 100040050& 49.936 & 41.520 & 2016-03-04 -- 2016-03-06 & 146 \\
      obs4 & 100040060 & 49.955 & 41.512 & 2016-03-06 -- 2016-03-07 & 46 \\
      \hline
      \multicolumn{6}{c}{Chandra/ACIS-I} \\
      $\cdots$ & 11714 & 49.928 & 41.569 & 2009-12-07 -- 2009-12-08 &  92 \\
      \hline
      \multicolumn{6}{c}{XMM-Newton/RGS} \\
      $\cdots$ & 0085110101, 0085110201 & 49.951 & 41.512 & 2001-01-30 -- 2001-01-31 &  72 \\
      $\cdots$ & 0305780101 & 49.950 & 41.513 & 2006-01-29 -- 2006-01-31 &  125  \\
      \hline
    \end{tabular}}\label{tab:obslog}
\end{table*}

\subsection{Hitomi Data Reduction}
\label{obs:hitomi-reduction}

\begin{figure}
 \begin{center}
  \includegraphics[width=8cm]{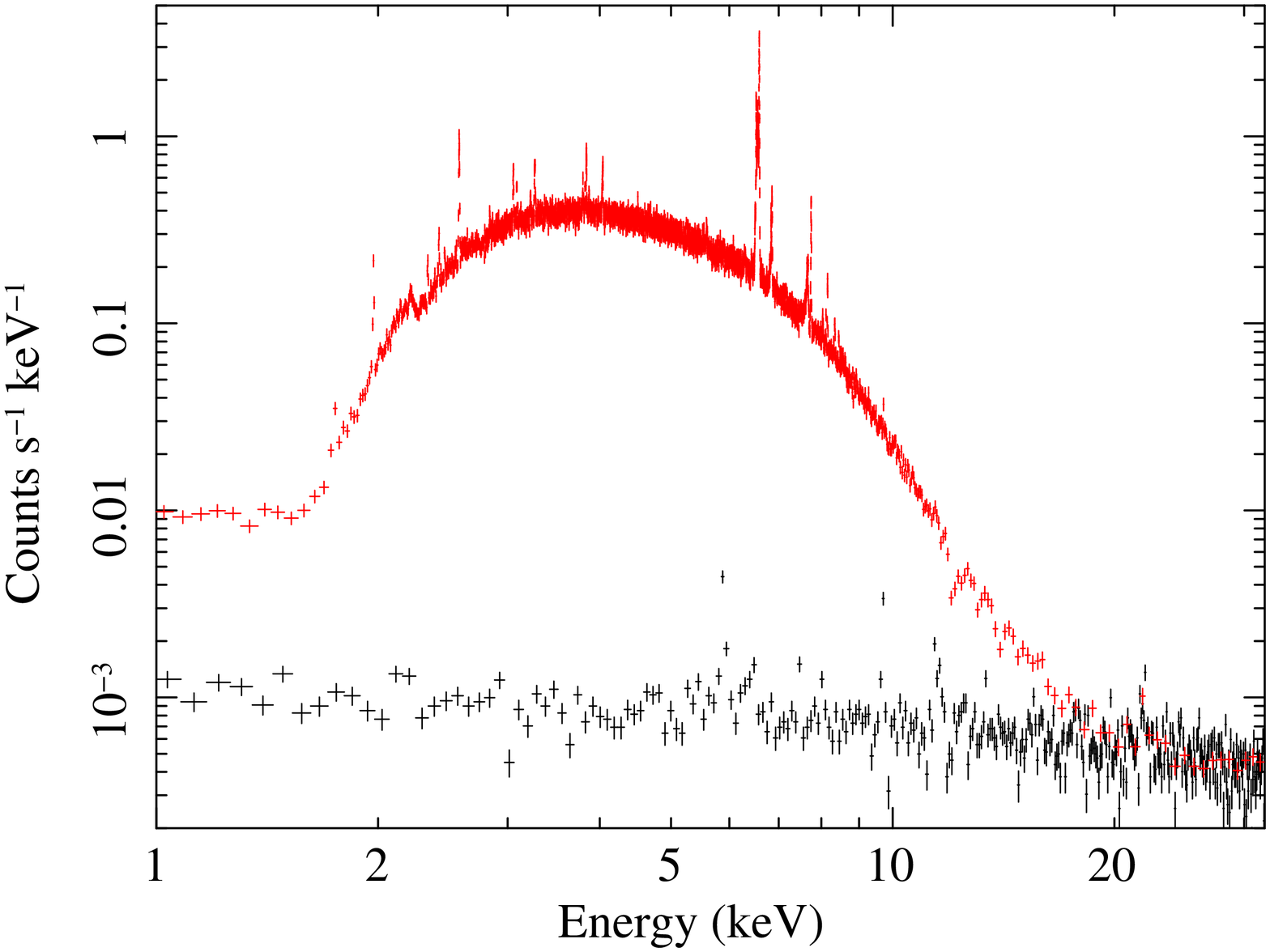} 
 \end{center}
\caption{
SXS 1--32 keV spectrum in the Entire core region (red).
The corresponding non X-ray background estimated by \texttt{sxsnxbgen} is also plotted in black. 
}
\label{fig:spectrum-nxb}
\end{figure}

We used the cleaned event list provided by the pipeline processing version 03.01.006.007, and applied the additional screening described below using the HEAsoft version 6.21, Hitomi software version 6, and Hitomi calibration database version 7\footnote{See https://heasarc.gsfc.nasa.gov/docs/hitomi/analysis for the Hitomi software and calibration database} (Angelini et al. 2017). 

The SXS recorded signals up to 32~keV, but the standard pipeline processing reduces the energy coverage to the 0--16~keV band in order to achieve a sufficiently fine energy bin with the realistic number of channels in the nominal energy band (32768 bins with 0.5~eV~bin$^{-1}$).
However, the SXS was sensitive to bright sources above 16 keV because of its very low non-X-ray background \citep{kilbourne17}. 
We thus used a coarser bin size of 1.0~eV~bin$^{-1}$ to extend the energy coverage up to 32~keV instead.  
This was technically achieved by the \texttt{sxsextend} ftools task.
We confirmed that choosing the coarser bin size has no impact on our analysis due to intrinsic thermal and velocity broadening of lines. 
  
We then applied event screening based on a pulse rise time versus energy relationship tuned for the wider energy coverage\footnote{See the Hitomi data reduction guide for details (https://heasarc.gsfc.nasa.gov/docs/hitomi/analysis). }.
We also selected only high primary grade events, for which arrival time between signal pulses was sufficiently large and hence the best spectroscopic performance was achieved. 
The branching ratio to other grades was less than 2\% for the Perseus observations, so this grade selection hardly reduced the effective exposure.

Since the in-flight calibration of the SXS is limited, there is uncertainty of the gain scale especially at energies far from 5.9~keV. 
In addition, the SXS was not in thermal equilibrium during obs1 and obs2, and thus a $\sim$2~eV gain shift was seen even at 5.9~keV (Fujimoto et al. 2017). 
In order to correct for the gain scale, we applied the pixel-by-pixel redshift correction and the gain correction using a parabolic function as described in Appendix \ref{appendix:gain-correction}.
 
We defined the four spectral analysis regions shown as the colour polygons in \Fig{fig:pointing}.
The Entire core region is the sum of the FoVs of obs2, obs3, and obs4 to maximize the photon statistics.
In order to investigate the spatial variation of the temperature, we divided the Entire core region into two sub-regions:
the Nebula region associated with the H$\alpha$ nebula \citep{2001AJ....122.2281C}, and the Rim region located just outside the core, including the bubble seen north-west of the cluster center.
The aim point of  obs4 is different from that of obs2/3 by $\sim60\arcsec$; thus, for the Nebula and Rim regions, spectra of obs2/3 and obs4 were extracted using slightly different spatial regions, and later co-added.
Lastly the fourth region, which we refer to as the Outer region, is the entire FoV of obs1.

Non X-ray backgrounds (NXB) corresponding to each region were produced from the Earth eclipsed durations using \texttt{sxsnxbgen}.
The redistribution matrix file (RMF) and the auxiliary response file (ARF) for spectral analysis were generated by \texttt{sxsmkrmf} and \texttt{aharfgen}, respectively.
As an input to the ARF generator, we used the 1.8--9.0 keV Chandra image in which the AGN region ($r=10\arcsec$) is replaced with average adjacent brightness. 
The spectrum of the Entire core region with the corresponding non X-ray background is shown in \Fig{fig:spectrum-nxb}. 
The cluster is clearly detected above the NXB up to 20 keV.
The attenuation below $\sim2$~keV due to the closed gate valve can also be seen.
For our analysis, we thus focus on the energy band spanning 1.8--20.0~keV.

\subsection{Chandra and XMM-Newton Archive Data}
\label{obs:archive}
For comparison with the Hitomi results, we also analyzed archival data from Chandra and XMM-Newton.
Details of the observations are summarized in \Tab{tab:obslog}.

We reprocessed the Chandra data with CIAO version 4.9 software package and calibration database version 4.7.4.
Spectra were extracted from the Nebula and Rim regions shown in \Fig{fig:pointing}.
A $9\arcsec$ radius circle around the central AGN region was excluded from the analysis taking advantage of Chandra's spatial resolution.
The spectra were binned so that each bin includes at least 100 counts.
Background spectra were generated from the blank-sky observations provided in the calibration database, and were scaled so that their count rates in the 10--12~keV band match the source spectra.

We followed the data analysis methods of the CHEERS collaboration \citep{2017arXiv170705076D} for the reduction of the XMM-Newton/RGS data with the SAS version 14.0.0 software package.
We extracted RGS source spectra in a region centered on the peak of the source emission, with a width of $0.8\arcmin$ in the cross-dispersion direction. While this is much smaller than the region probed by the SXS, a narrower extraction region in the cross-dispersion direction provides spectra that are least broadened by the spatial extent of the source, and thus have the best resolution. To further correct for this broadening, we used the \texttt{lpro} model component in SPEX to convolve the spectral models with the surface brightness profile extracted from the XMM-Newton MOS1 detector.
We used background spectra generated by the SAS \texttt{rgsbkgmodel} task. The template background files were scaled using the count rates measured in the off-axis region of CCD9, in which the soft protons dominate the light curve.

\section{Analysis and Results}
\label{ana}
The procedures described below were used for the spectral analysis presented in this section, unless stated otherwise.
Spectral fits were performed using the Xspec 12.9.1h package \citep{1996ASPC..101...17A} employing the modified C-statistic \citep{1979ApJ...228..939C} in which a Poisson background spectrum is taken into account (also referred to as the W-statistic). 
We used the atomic databases of the \texttt{AtomDB} version 3.0.9 \citep{2012ApJ...756..128F} and \texttt{SPEXACT} version 3.03.00 \citep{1996uxsa.conf..411K} for calculations of plasma models. 
We take differences between the model predictions as an estimate of model uncertainties. 
A python program was used to generate APEC format table models from SPEX \footnote{http://www.mpe.mpg.de/~jsanders/code/}, allowing us to perform a direct comparison of the results using a consistent treatment of all other assumptions and fit procedures. 

Photoelectric absorption by cold matter in our Galaxy was modeled using the \texttt{TBabs} code version 2.3 \citep{2000ApJ...542..914W}, in which fine-structures of absorption edges and cross-sections of dust grains and molecules are included.
Its hydrogen column density was fixed at $1.38\times10^{21}$~cm$^{-2}$ in accordance with the all-sky H~\emissiontype{I} survey \citep{2005A&amp;A...440..775K}.
We also considered the contaminating emission from the AGN in NGC\,1275.
Its spectrum was modeled using a power-law continuum and a neutral Fe~K$\alpha$ line with parameters fixed at the values described in the AGN~paper.
Its  flux was estimated by ray-tracing simulations (\texttt{aharfgen}).

\subsection{Line Ratio Diagnostics}
\label{ana:line}

\begin{figure*}
\begin{center}
\subfloat{\includegraphics[width=7.5cm]{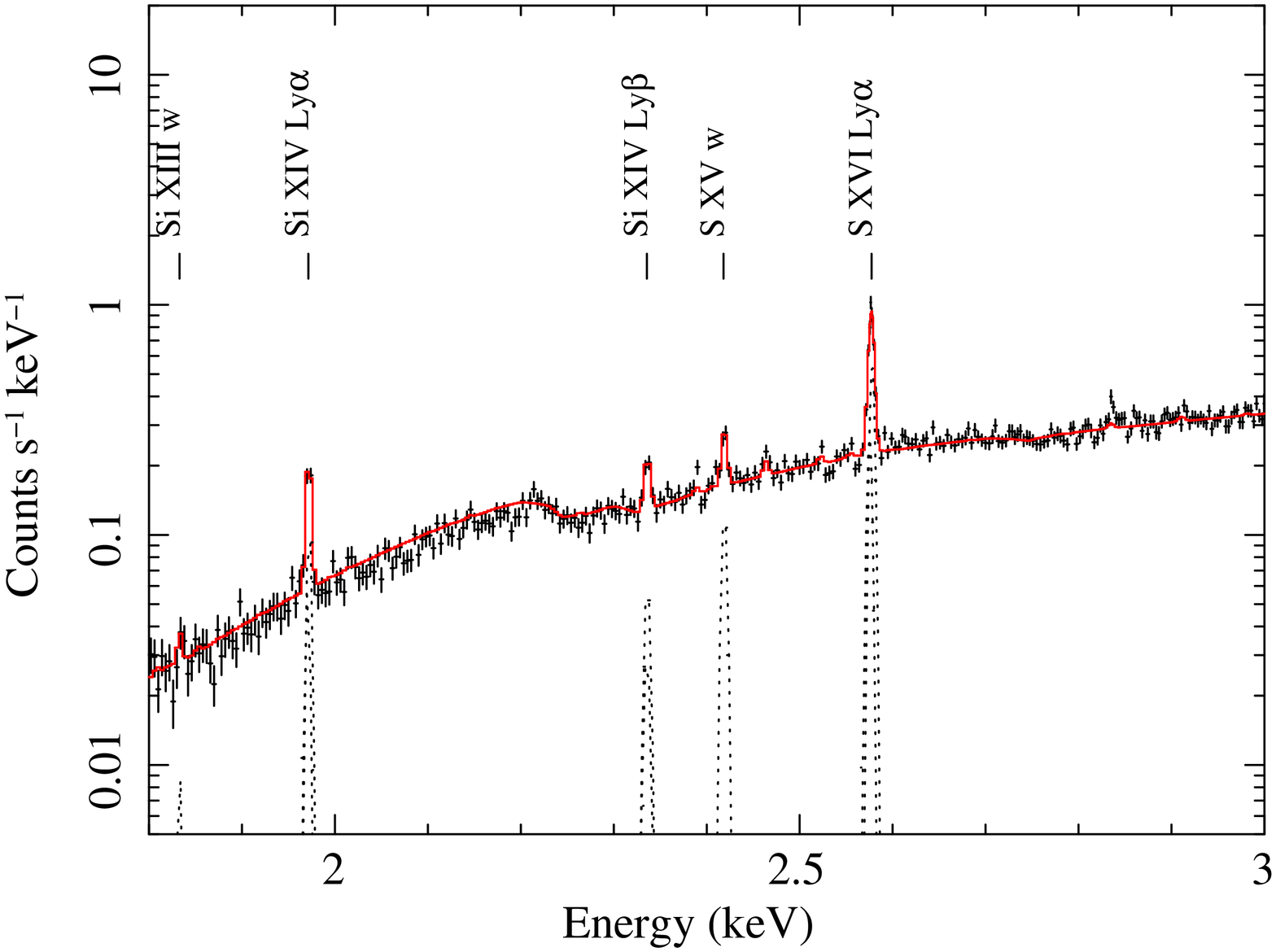}}
\hspace{0.1cm}
\subfloat{\includegraphics[width=7.5cm]{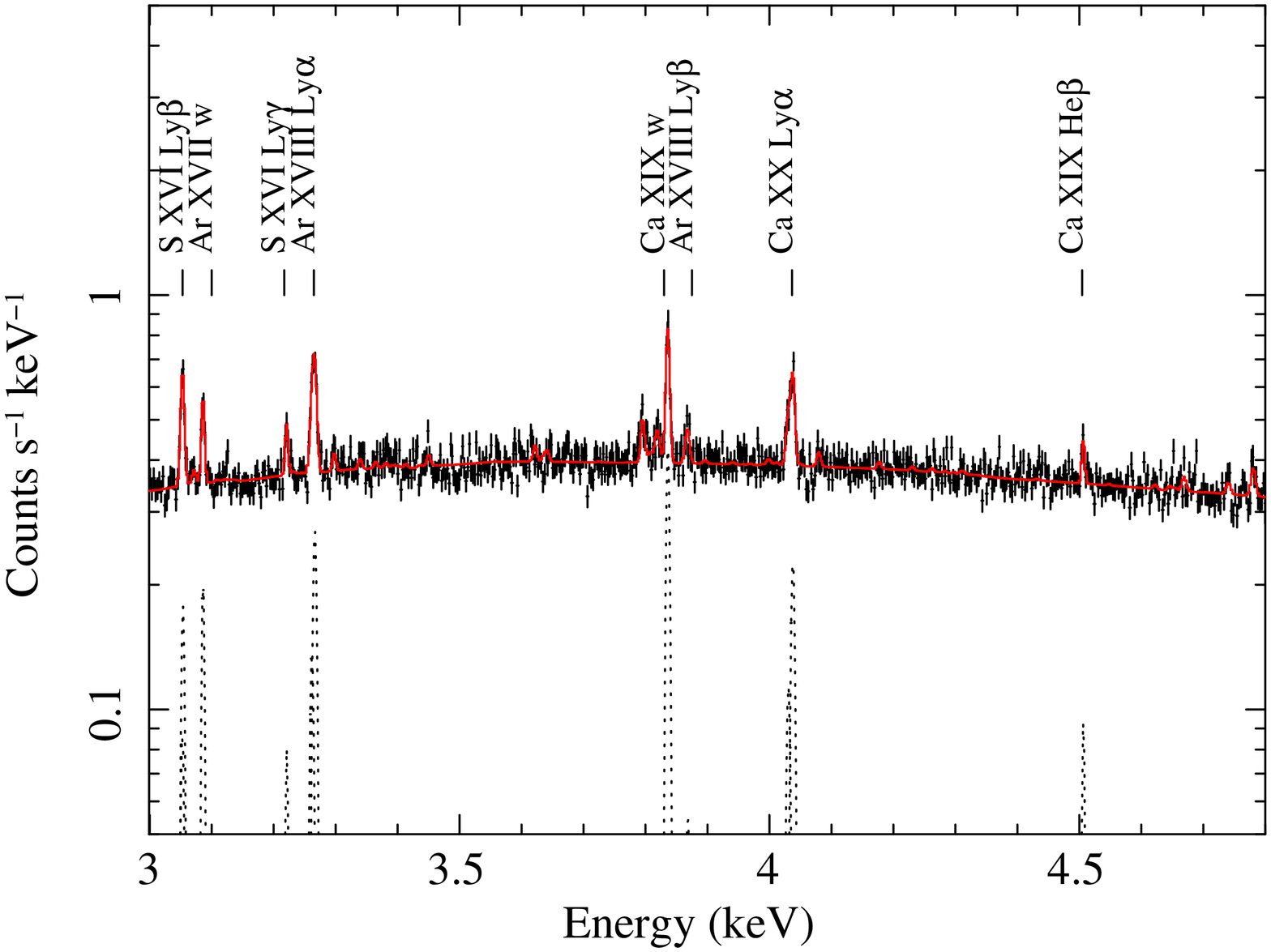}}
\vspace{0.2cm}
\subfloat{\includegraphics[width=7.5cm]{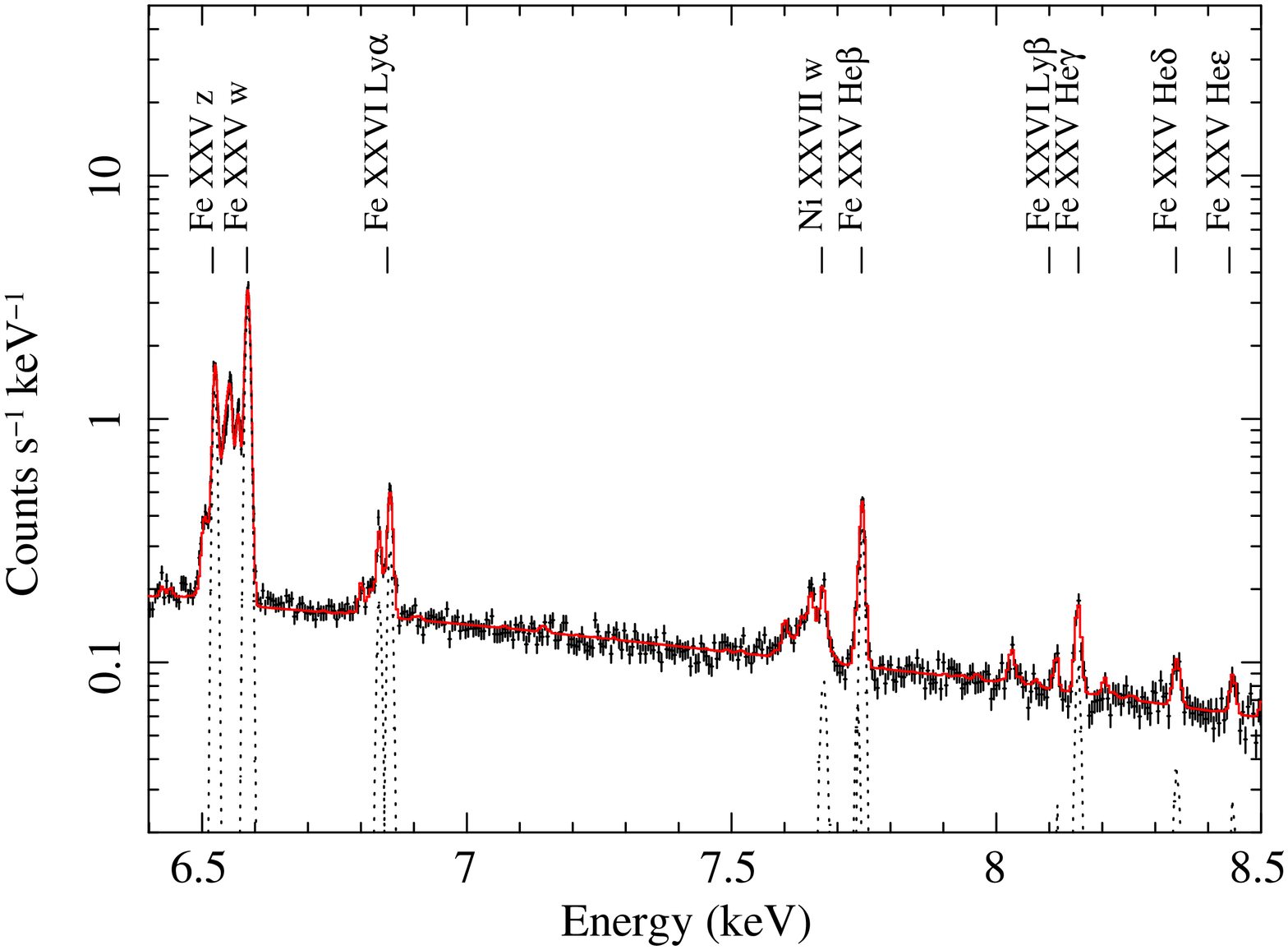}}
\end{center}
\caption{SXS spectra extracted from the Entire core region in the 1.8--3.0~keV (top left), 3.0--4.8~keV (top right), and 6.4--8.5~keV (bottom) bands.
The fitted phenomenological models are shown by the red solid curves.
The Gaussians included in the model are also plotted by the black dotted lines.
}
\label{fig:fit-gaus}
\end{figure*}

\begin{table}[]
\tbl{List of lines considered for the Gaussian fit.\footnotemark[$*$]}{%
\centering
\scriptsize
\begin{tabular}{l@{\hspace{0.2cm}}ccccc}
\hline
\multicolumn{1}{c}{Line name}                                  & $E_0$\footnotemark[$\dagger$]  & \multicolumn{4}{c}{Constraints}                                                                           \\ \cline{3-6} 
                                                               & (eV)         & Tied to                                                    & Center         & Width        & Flux         \\ \hline
Si \emissiontype{XIII} w                                       & 1865.0     & -                                                          & -              & -            & -            \\
Si \emissiontype{XIV} Ly$\alpha_{1}$                           & 2006.1     & -                                                          & -              & -            & -            \\
Si \emissiontype{XIV} Ly$\alpha_{2}$                           & 2004.3     & \multicolumn{1}{l}{Si \emissiontype{XIV} Ly$\alpha_{1}$}   & $- 1.8$ eV     & $\times 1.0$ & $\times 0.5$ \\
Si \emissiontype{XIV} Ly$\beta_{1}$                            & 2376.6     & -                                                          & -              & -            & -            \\
Si \emissiontype{XIV} Ly$\beta_{2}$                            & 2376.1     & \multicolumn{1}{l}{Si \emissiontype{XIV} Ly$\beta_{1}$}    & $- 0.5$ eV     & $\times 1.0$ & $\times 0.5$ \\
S \emissiontype{XV} w                                          & 2460.6     & -                                                          & -              & -            & -            \\
S \emissiontype{XVI} Ly$\alpha_{1}$                            & 2622.7     & -                                                          & -              & -            & -            \\
S \emissiontype{XVI} Ly$\alpha_{2}$                            & 2619.7     & \multicolumn{1}{l}{S \emissiontype{XVI} Ly$\alpha_{1}$}    & $- 3.0$ eV     & $\times 1.0$ & $\times 0.5$ \\
S \emissiontype{XVI} Ly$\beta_{1}$                             & 3106.7     & -                                                          & -              & -            & -            \\
S \emissiontype{XVI} Ly$\beta_{2}$                             & 3105.8     & \multicolumn{1}{l}{S \emissiontype{XVI} Ly$\beta_{1}$}     & $- 0.9$ eV     & $\times 1.0$ & $\times 0.5$ \\
S \emissiontype{XVI} Ly$\gamma_{1}$                            & 3276.3     & -                                                          & -              & -            & -            \\
S \emissiontype{XVI} Ly$\gamma_{2}$                            & 3275.9     & \multicolumn{1}{l}{S \emissiontype{XVI} Ly$\gamma_{1}$}    & $- 0.4$ eV     & $\times 1.0$ & $\times 0.5$ \\
Ar \emissiontype{XVII} w                                       & 3139.6     & -                                                          & -              & -            & -            \\
Ar \emissiontype{XVIII} Ly$\alpha_{1}$                         & 3323.0     & -                                                          & -              & -            & -            \\
Ar \emissiontype{XVIII} Ly$\alpha_{2}$                         & 3328.2     & \multicolumn{1}{l}{Ar \emissiontype{XVIII} Ly$\alpha_{1}$} & $- 4.8$ eV     & $\times 1.0$ & $\times 0.5$ \\
Ar \emissiontype{XVIII} Ly$\beta_{1}$                          & 3935.7     & -                                                          & -              & -            & -            \\
Ar \emissiontype{XVIII} Ly$\beta_{2}$                          & 3934.3     & \multicolumn{1}{l}{Ar \emissiontype{XVIII} Ly$\beta_{1}$}  & $- 1.4$ eV     & $\times 1.0$ & $\times 0.5$ \\
Ca \emissiontype{XIX} w                                        & 3902.4     & -                                                          & -              & -            & -            \\
Ca \emissiontype{XIX} He$\beta_{1}$\footnotemark[$\ddagger$]    & 4583.5     & -                                                          & -              & -            & -            \\
Ca \emissiontype{XX} Ly$\alpha_{1}$                            & 4107.5     & -                                                          & -              & -            & -            \\
Ca \emissiontype{XX} Ly$\alpha_{2}$                            & 4100.1     & \multicolumn{1}{l}{Ca \emissiontype{XX} Ly$\alpha_{1}$}    & $- 7.4$ eV     & $\times 1.0$ & $\times 0.5$ \\
Fe \emissiontype{XXV} z                                        & 6636.6     & -                                                          & -              & -            & -            \\
Fe \emissiontype{XXV} w                                        & 6700.4     & -                                                          & -              & -            & -            \\
Fe \emissiontype{XXV} He$\beta_{1}$                            & 7881.5     & -                                                          & -              & -            & -            \\
Fe \emissiontype{XXV} He$\beta_{2}$                            & 7872.0     & \multicolumn{1}{l}{Fe \emissiontype{XXV} He$\beta_{1}$}    & $- 9.5$ eV     & $\times 1.0$ & -            \\
Fe \emissiontype{XXV} He$\gamma_{1}$\footnotemark[$\ddagger$]   & 8295.5     & -                                                          & -              & -            & -            \\
Fe \emissiontype{XXV} He$\delta_{1}$\footnotemark[$\ddagger$]   & 8487.4     & -                                                          & -              & -            & -            \\
Fe \emissiontype{XXV} He$\epsilon_{1}$\footnotemark[$\ddagger$] & 8588.5     & -                                                          & -              & -            & -            \\
Fe \emissiontype{XXVI} Ly$\alpha_{1}$                          & 6973.1     & -                                                          & -              & -            & -            \\
Fe \emissiontype{XXVI} Ly$\alpha_{2}$                          & 6951.9     & \multicolumn{1}{l}{Fe \emissiontype{XXVI} Ly$\alpha_{1}$}  & -              & $\times 1.0$ & -            \\
Fe \emissiontype{XXVI} Ly$\beta_{1}$                           & 8252.6     & -                                                          & -              & -            & -            \\
Fe \emissiontype{XXVI} Ly$\beta_{2}$                           & 8248.4     & \multicolumn{1}{l}{Fe \emissiontype{XXVI} Ly$\beta_{1}$}   & $- 6.2$ eV     & $\times 1.0$ & -            \\
Ni \emissiontype{XXVII} w                                      & 7805.6     & -                                                          & -              & -            & -            \\ \hline
\multicolumn{6}{c}{Constraints only on the Rim region}                                                                                                                                  \\
Si \emissiontype{XIII} w                                       & 1865.0     & \multicolumn{1}{l}{Si \emissiontype{XIV} Ly$\alpha_{1}$}   & fixed at $E_0$ & $\times 1.0$ & -            \\
Ca \emissiontype{XIX} He$\beta_{1}$                            & 4583.5     & \multicolumn{1}{l}{Ca \emissiontype{XX} Ly$\alpha_{1}$}    & fixed at $E_0$ & $\times 1.0$ & -            \\ \hline
\multicolumn{6}{c}{Constraints only on the Outer region}                                                                                                                                \\
Si \emissiontype{XIII} w                                       & 1865.0     & \multicolumn{1}{l}{Si \emissiontype{XIV} Ly$\alpha_{1}$}   & fixed at $E_0$ & $\times 1.0$ & -            \\
S \emissiontype{XV} w                                          & 2460.6     & \multicolumn{1}{l}{S \emissiontype{XVI} Ly$\alpha_{1}$}    & fixed at $E_0$ & $\times 1.0$ & -            \\
Ca \emissiontype{XIX} He$\beta_{1}$                            & 4583.5     & \multicolumn{1}{l}{Ca \emissiontype{XX} Ly$\alpha_{1}$}    & fixed at $E_0$ & $\times 1.0$ & -            \\
Fe \emissiontype{XXV} He$\gamma_{1}$                           & 8295.5     & \multicolumn{1}{l}{Fe \emissiontype{XXV} He$\beta_{1}$}    & fixed at $E_0$ & $\times 1.0$ & -            \\
Fe \emissiontype{XXV} He$\delta_{1}$                           & 8487.4     & \multicolumn{1}{l}{Fe \emissiontype{XXV} He$\beta_{1}$}    & fixed at $E_0$ & $\times 1.0$ & -            \\
Fe \emissiontype{XXV} He$\epsilon_{1}$                         & 8588.5     & \multicolumn{1}{l}{Fe \emissiontype{XXV} He$\beta_{1}$}    & fixed at $E_0$ & $\times 1.0$ & -            \\ \hline
\end{tabular}
}\label{tab:linelist}
\begin{tabnote}
\hangindent6pt\noindent
\hbox to6pt{\footnotemark[$*$]\hss}\unskip%
Free parameters are denoted by the hyphen (-).
\par
\hangindent6pt\noindent
\hbox to6pt{\footnotemark[$\dagger$]\hss}\unskip%
Fiducial energies of the emission lines at the rest frame in \texttt{AtomDB}~3.0.9
\par
\hangindent6pt\noindent
\hbox to6pt{\footnotemark[$\ddagger$]\hss}\unskip%
Ca \emissiontype{XIX} He$\beta_{2}$, Fe \emissiontype{XXV} He$\gamma_{2}$, Fe \emissiontype{XXV} He$\delta_{2}$, and Fe \emissiontype{XXV} He$\epsilon_{2}$ were omitted because their fluxes are too small to constrain from the SXS spetra.
\end{tabnote}
\end{table}

\begin{table}[]
\tbl{Observed line fluxes derived from  Gaussian fits.\footnotemark[$*$]}{%
\centering\scriptsize
\begin{tabular}{lrrrr}
\hline
\multicolumn{1}{c}{Line name}          & \multicolumn{4}{c}{Flux (10$^{-5}$~ph~cm$^ {-2}$~s$^ {-1}$)} \\
\cline{2-5}                            & \multicolumn{1}{c}{Entire Core}                              & \multicolumn{1}{c}{Nebula} & \multicolumn{1}{c}{Rim}     & \multicolumn{1}{c}{Outer}   \\
\hline
Si \emissiontype{XIII} w               & 6.40$^{+4.71}_{-2.67}$                                       & 5.87$^{+3.60}_{-2.54}$     & \multicolumn{1}{c}{$<$5.45} & \multicolumn{1}{c}{$<$4.54} \\
Si \emissiontype{XIV} Ly$\alpha_{1}$   & 32.43$^{+2.29}_{-2.23}$                                      & 20.11$^{+1.92}_{-1.83}$    & 21.83$^{+2.64}_{-2.52}$     & 4.09$^{+2.27}_{-1.72}$      \\
Si \emissiontype{XIV} Ly$\beta_{1}$    & 6.96$^{+0.91}_{-0.87}$                                       & 5.03$^{+0.74}_{-0.70}$     & 3.93$^{+1.04}_{-0.98}$      & 1.21$^{+0.82}_{-0.58}$      \\
S \emissiontype{XV} w                  & 9.38$^{+1.13}_{-1.11}$                                       & 7.26$^{+0.98}_{-0.99}$     & 3.91$^{+1.03}_{-0.94}$      & \multicolumn{1}{c}{$<$1.08} \\
S \emissiontype{XVI} Ly$\alpha_{1}$    & 22.71$^{+0.73}_{-0.72}$                                      & 15.81$^{+0.64}_{-0.63}$    & 12.46$^{+0.77}_{-0.76}$     & 2.70$^{+0.67}_{-0.64}$      \\
S \emissiontype{XVI} Ly$\beta_{1}$     & 3.83$^{+0.29}_{-0.29}$                                       & 2.55$^{+0.25}_{-0.24}$     & 2.49$^{+0.35}_{-0.33}$      & 0.62$^{+0.27}_{-0.22}$      \\
S \emissiontype{XVI} Ly$\gamma_{1}$    & 1.20$^{+0.20}_{-0.19}$                                       & 0.74$^{+0.15}_{-0.17}$     & 0.92$^{+0.25}_{-0.24}$      & 0.32$^{+0.22}_{-0.17}$      \\
Ar \emissiontype{XVII} w               & 3.72$^{+0.37}_{-0.36}$                                       & 2.82$^{+0.31}_{-0.30}$     & 1.87$^{+0.51}_{-0.47}$      & 1.20$^{+0.41}_{-0.34}$      \\
Ar \emissiontype{XVIII} Ly$\alpha_{1}$ & 5.47$^{+0.29}_{-0.29}$                                       & 3.85$^{+0.25}_{-0.25}$     & 3.15$^{+0.32}_{-0.30}$      & 0.94$^{+0.32}_{-0.30}$      \\
Ar \emissiontype{XVIII} Ly$\beta_{1}$  & 0.77$^{+0.15}_{-0.15}$                                       & 0.51$^{+0.12}_{-0.12}$     & 0.63$^{+0.20}_{-0.18}$      & 0.26$^{+0.15}_{-0.11}$      \\
Ca \emissiontype{XIX} w                & 5.20$^{+0.27}_{-0.27}$                                       & 3.66$^{+0.23}_{-0.23}$     & 2.94$^{+0.30}_{-0.28}$      & 0.93$^{+0.29}_{-0.26}$      \\
Ca \emissiontype{XIX} He$\beta_{1}$    & 0.66$^{+0.16}_{-0.10}$                                       & 0.46$^{+0.13}_{-0.10}$     & 0.67$^{+0.29}_{-0.35}$      & 0.21$^{+0.16}_{-0.12}$      \\
Ca \emissiontype{XX} Ly$\alpha_{1}$    & 2.80$^{+0.18}_{-0.18}$                                       & 1.85$^{+0.16}_{-0.15}$     & 1.81$^{+0.20}_{-0.19}$      & 0.77$^{+0.20}_{-0.19}$      \\
Fe \emissiontype{XXV} w                & 33.14$^{+0.43}_{-0.34}$                                      & 21.09$^{+0.32}_{-0.31}$    & 22.13$^{+0.49}_{-0.35}$     & 9.49$^{+0.45}_{-0.44}$      \\
Fe \emissiontype{XXV} z                & 13.26$^{+0.27}_{-0.25}$                                      & 8.72$^{+0.21}_{-0.22}$     & 8.41$^{+0.28}_{-0.27}$      & 3.03$^{+0.28}_{-0.27}$      \\
Fe \emissiontype{XXV} He$\beta_{1}$    & 4.73$^{+0.12}_{-0.24}$                                       & 2.80$^{+0.12}_{-0.15}$     & 3.35$^{+0.19}_{-0.18}$      & 1.49$^{+0.21}_{-0.20}$      \\
Fe \emissiontype{XXV} He$\beta_{2}$    & 1.04$^{+0.10}_{-0.18}$                                       & 0.73$^{+0.14}_{-0.08}$     & 0.55$^{+0.13}_{-0.13}$      & \multicolumn{1}{c}{$<$0.17} \\
Fe \emissiontype{XXV} He$\gamma_{1}$   & 1.75$^{+0.13}_{-0.13}$                                       & 1.04$^{+0.10}_{-0.10}$     & 1.32$^{+0.14}_{-0.13}$      & 0.25$^{+0.13}_{-0.12}$      \\
Fe \emissiontype{XXV} He$\delta_{1}$   & 0.88$^{+0.12}_{-0.12}$                                       & 0.55$^{+0.10}_{-0.10}$     & 0.63$^{+0.13}_{-0.12}$      & 0.27$^{+0.13}_{-0.11}$      \\
Fe \emissiontype{XXV} He$\epsilon_{1}$ & 0.54$^{+0.10}_{-0.10}$                                       & 0.34$^{+0.08}_{-0.08}$     & 0.43$^{+0.12}_{-0.12}$      & 0.15$^{+0.12}_{-0.10}$      \\
Fe \emissiontype{XXVI} Ly$\alpha_{1}$  & 3.68$^{+0.16}_{-0.16}$                                       & 2.24$^{+0.13}_{-0.13}$     & 2.68$^{+0.17}_{-0.17}$      & 1.35$^{+0.22}_{-0.21}$      \\
Fe \emissiontype{XXVI} Ly$\alpha_{2}$  & 2.17$^{+0.14}_{-0.13}$                                       & 1.31$^{+0.12}_{-0.11}$     & 1.59$^{+0.14}_{-0.14}$      & 0.99$^{+0.20}_{-0.18}$      \\
Fe \emissiontype{XXVI} Ly$\beta_{1}$   & 0.30$^{+0.06}_{-0.06}$                                       & 0.21$^{+0.06}_{-0.05}$     & 0.18$^{+0.07}_{-0.04}$      & 0.16$^{+0.08}_{-0.06}$      \\
Ni \emissiontype{XXVII} w              & 1.43$^{+0.13}_{-0.13}$                                       & 1.01$^{+0.11}_{-0.11}$     & 0.79$^{+0.13}_{-0.13}$      & 0.43$^{+0.16}_{-0.15}$      \\
\hline
\end{tabular}
}\label{tab:linedata}
\begin{tabnote}
\hangindent6pt\noindent
\hbox to6pt{\footnotemark[$*$]\hss}\unskip%
The Ly$\alpha_2$ lines of Si, S, Ar, and Ca are not shown because their parameter values are tied to Ly$\alpha_1$ (see \Tab{tab:linelist} for details).
\end{tabnote}\end{table}

\Fig{fig:fit-gaus} shows the spectra extracted from the Entire core region, focusing on the 1.8--3.0~keV, 3.0--4.8~keV, and 6.4--8.5~keV bands.
Both He$\alpha$ and Ly$\alpha$ emission lines of Si, S, Ar, Ca, and Fe were detected and resolved.
Furthermore, 
some transitions from higher principal quantum numbers
are also resolved; 
up to $\epsilon$ ($n=6$) from Fe in particular.

In order to derive the observed fluxes of these lines, we fitted the spectra in the three energy bands listed above with a phenomenological model consisting of continuum emission and Gaussian lines. 
We used a CIE plasma model based on \texttt{AtomDB} (the \texttt{apec} model) in which the strong lines listed in \Tab{tab:linelist} were replaced with Gaussians.
In accordance with the First paper, the metal abundance of the plasma model was fixed at 0.62 solar and the line-of-sight velocity dispersion was fixed at 146~km~s$^{-1}$ 
to represent weaker emission lines not listed in \Tab{tab:linelist}. 
Even when these parameters are varied by $\pm$20\% (much higher than statistical errors shown in the Z paper and the V paper), there is no significant impact on our line flux measurements.
Doublets of the Lyman series were not resolved except for Fe~\emissiontype{XXVI}~Ly$\alpha$, 
and hence their centroid energies, line widths, and flux ratios were tied as shown in \Tab{tab:linelist}.
The line centroids and widths for Fe~\emissiontype{XXV}~He$\beta_{1}$ and Fe~\emissiontype{XXV}~He$\beta_{2}$ were also tied as described in \Tab{tab:linelist}.
Unresolved structures in Ca~\emissiontype{XIX}~He$\beta$, Fe~\emissiontype{XXV}~He$\gamma$, Fe~\emissiontype{XXV}~He$\delta$, and Fe~\emissiontype{XXV}~He$\epsilon$ were represented by single Gaussians.
The Gaussian fluxes we obtained are shown in \Tab{tab:linedata}.
The results of the line centroids and width, though not relevant to our analysis, are summarized in  Appendix~\ref{appendix:gaus-parameters}. 
Readers are referred to the V~paper for a detailed discussion of the velocity dispersions and line-of-sight velocity shifts.

\begin{figure*}
\begin{center}
\includegraphics[width=16cm]{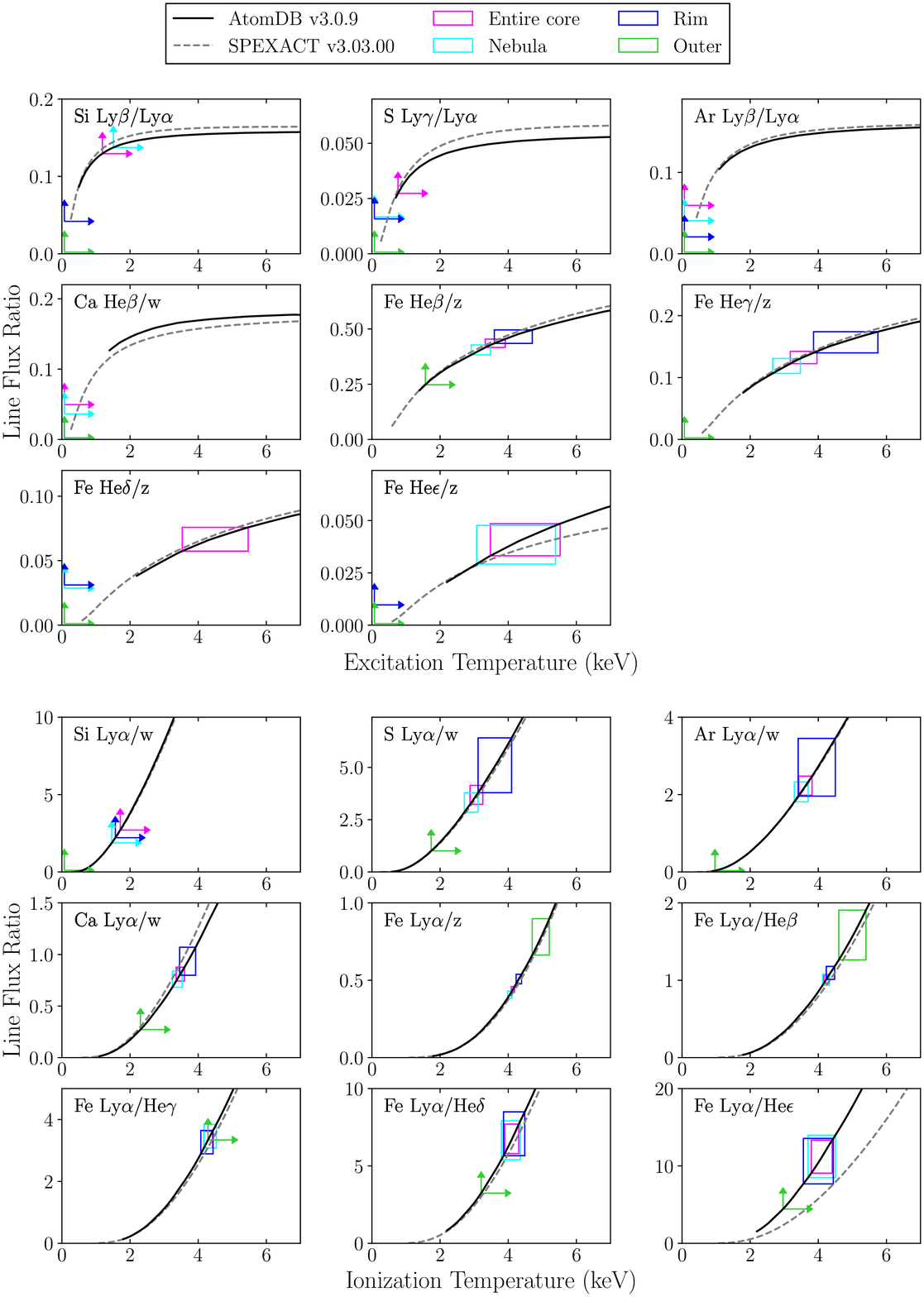} 
\end{center}
\caption{Upper 8 panels show flux ratios of the emission lines as a function of the excitation temperature, calculated from \texttt{AtomDB} (black solid curve) and \texttt{ SPEXACT} (gray dashed curve) assuming a single temperature CIE plasma.
The lines used in the calculations are denoted in each panel.
The color boxes show the ranges of the observed line ratios and the corresponding \texttt{AtomDB} temperatures at the1$\sigma$ confidence level.
Magenta, blue, cyan, and green correspond to the Entire core, Rim, Nebula, and Outer regions, respectively.
When the ranges of the statistical errors of the observed line ratios are outside the models, 3$\sigma$ lower limits are shown instead by the color arrows.
Lower 9 panels are the same as the upper panels but for the ionization temperature.
}
\label{fig:line-ratios}
\end{figure*}

Assuming a single-temperature CIE plasma, 
and employing the \texttt{AtomDB} and \texttt{SPEXACT} databases, we calculated how the line ratios considered here depend on the temperature.
The calculated temperature dependencies are shown in \Fig{fig:line-ratios}.
Line emissivities used in these calculations are given in Appendix~\ref{appendix:emissivities} 
along with measurements of emission measure based on single line fluxes.
Except for He$\epsilon$/z and Ly$\alpha$/He$\epsilon$ ratios, 
the two codes gave consistent values with each other within 5--10\%
for the interesting temperature range, 1--7~keV.
Detailed comparisons of line emissivities between the two codes are discussed in the Atomic paper. 

A line ratio of different transitions in the same ion reflects the kinetic temperature of free electrons in the plasma, and is referred to as ``excitation temperature'' or \telec.
Referring to \Fig{fig:line-ratios}, we calculated the \telec from 
the observed line ratios of Ly$\beta$/Ly$\alpha$ of Si and Ar, Ly$\gamma$/Ly$\alpha$ of S, He$\beta$/w of Ca, and He$\beta$/z,  He$\gamma$/z,  He$\delta$/z, and He$\epsilon$/z of Fe (top three rows of \Fig{fig:line-ratios}).
S Ly$\beta$ is not used because it is not separated from Ar z, whose energy is 3102~eV (see \Fig{fig:fit-gaus}).
Fe Ly$\beta$ is not used because of the low observed flux.
Fluxes of Ly$\alpha_{1}$ and Ly$\alpha_{2}$ were co-added in this calculation.
In the same manner, the fine structures of Ly$\beta$, Ly$\gamma$, He$\beta$, He$\gamma$, He$\delta$ and He$\epsilon$ were also summed.
The interval of the observed line ratios and the corresponding temperature ranges are overlaid on \Fig{fig:line-ratios} as color boxes. 

Separately from the \telec diagnostics, 
we used line ratios of different ionization species to measure the ion fraction for each element.
We parameterize these ratios by 
``ionization temperatures'' or \tioni.
When the emission comes from a single component and optically thin plasma under the CIE, 
\tioni from every element should be the same as \telec.
The \tioni were calculated using the line ratios of Ly$\alpha$/w of Si, S, Ar, and Ca and Ly$\alpha$/z, Ly$\alpha$/He$\beta$, Ly$\alpha$/He$\gamma$, Ly$\alpha$/He$\delta$, and Ly$\alpha$/He$\epsilon$ of Fe (bottom three rows of \Fig{fig:line-ratios}).
The temperature range derived from the observed line ratios are shown in \Fig{fig:line-ratios}. 

\begin{figure*}
\begin{center}
\subfloat[Entire core]{\includegraphics[width=8cm]{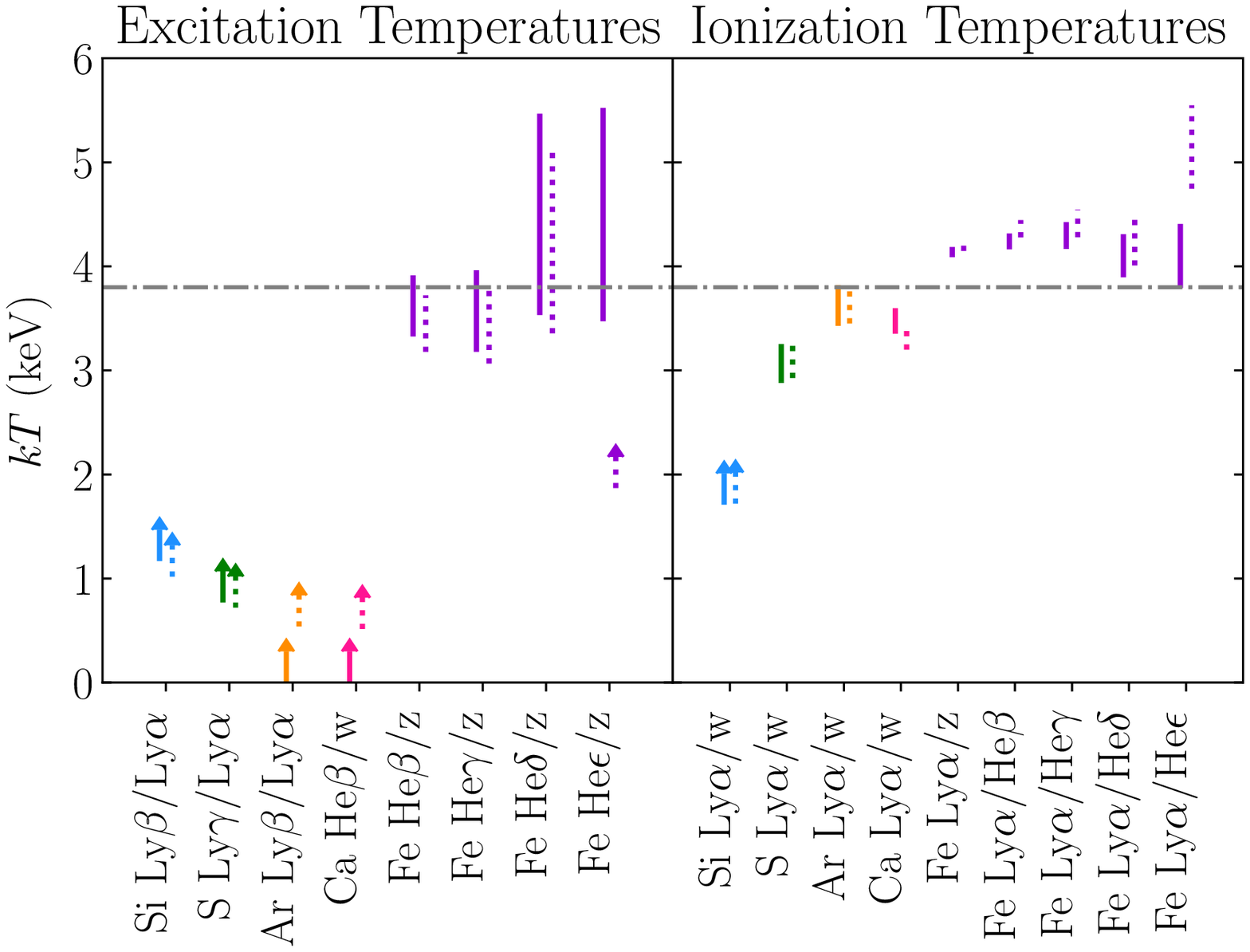}}
\vspace{0.0cm}
\subfloat[Nebula]{\includegraphics[width=8cm]{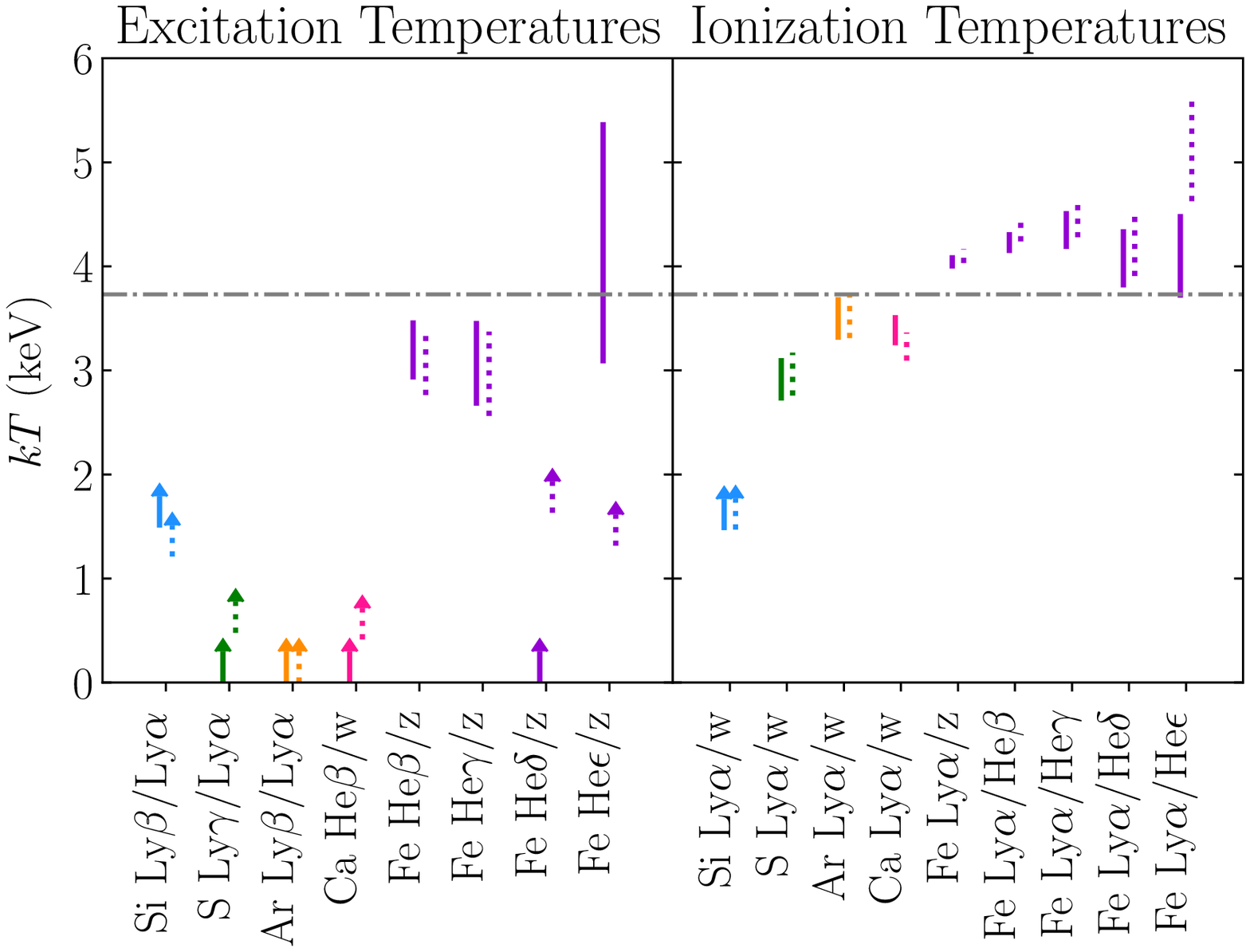}}
\vspace{0.0cm}
\subfloat[Rim]{\includegraphics[width=8cm]{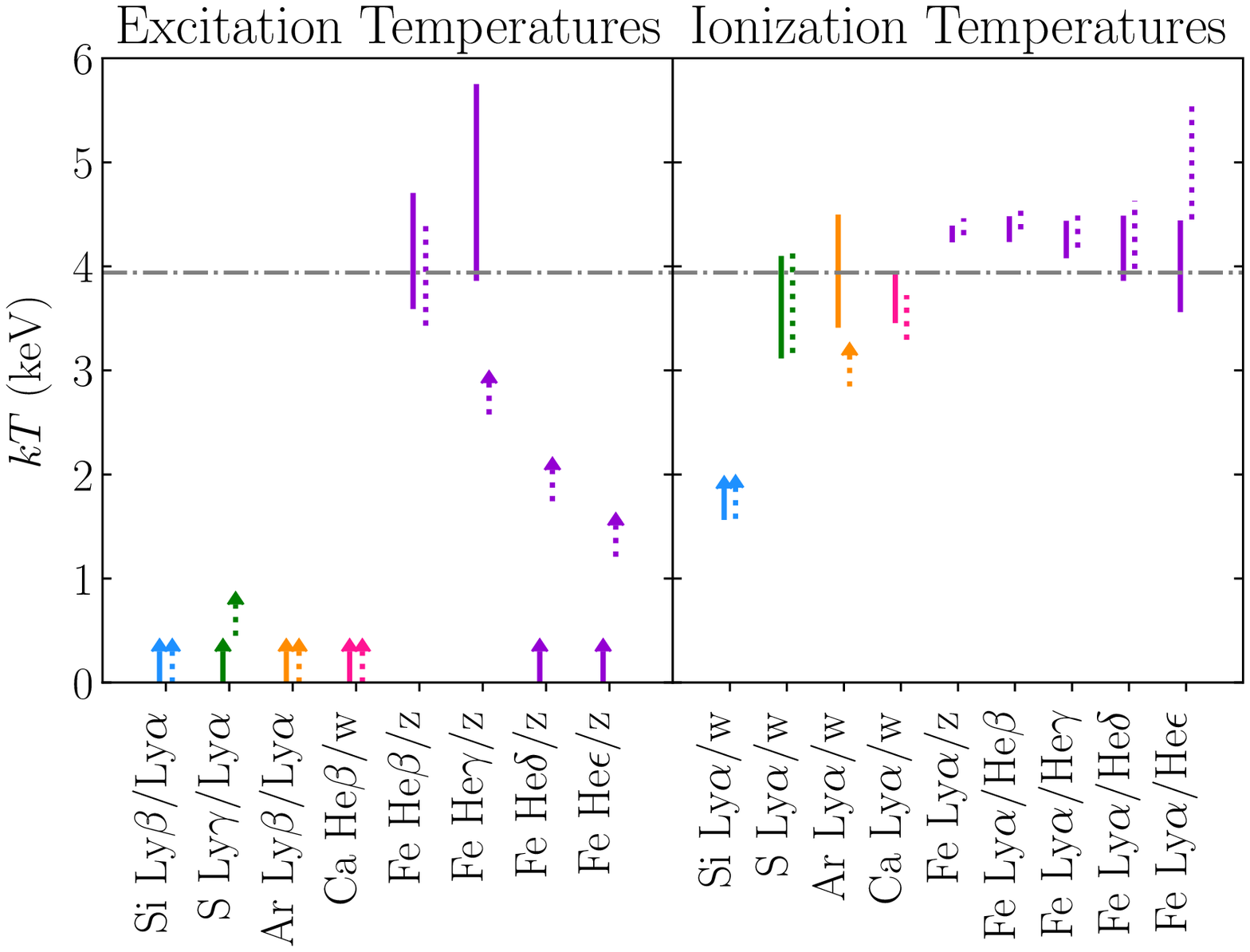}}
\vspace{0.0cm}
\subfloat[Outer]{\includegraphics[width=8cm]{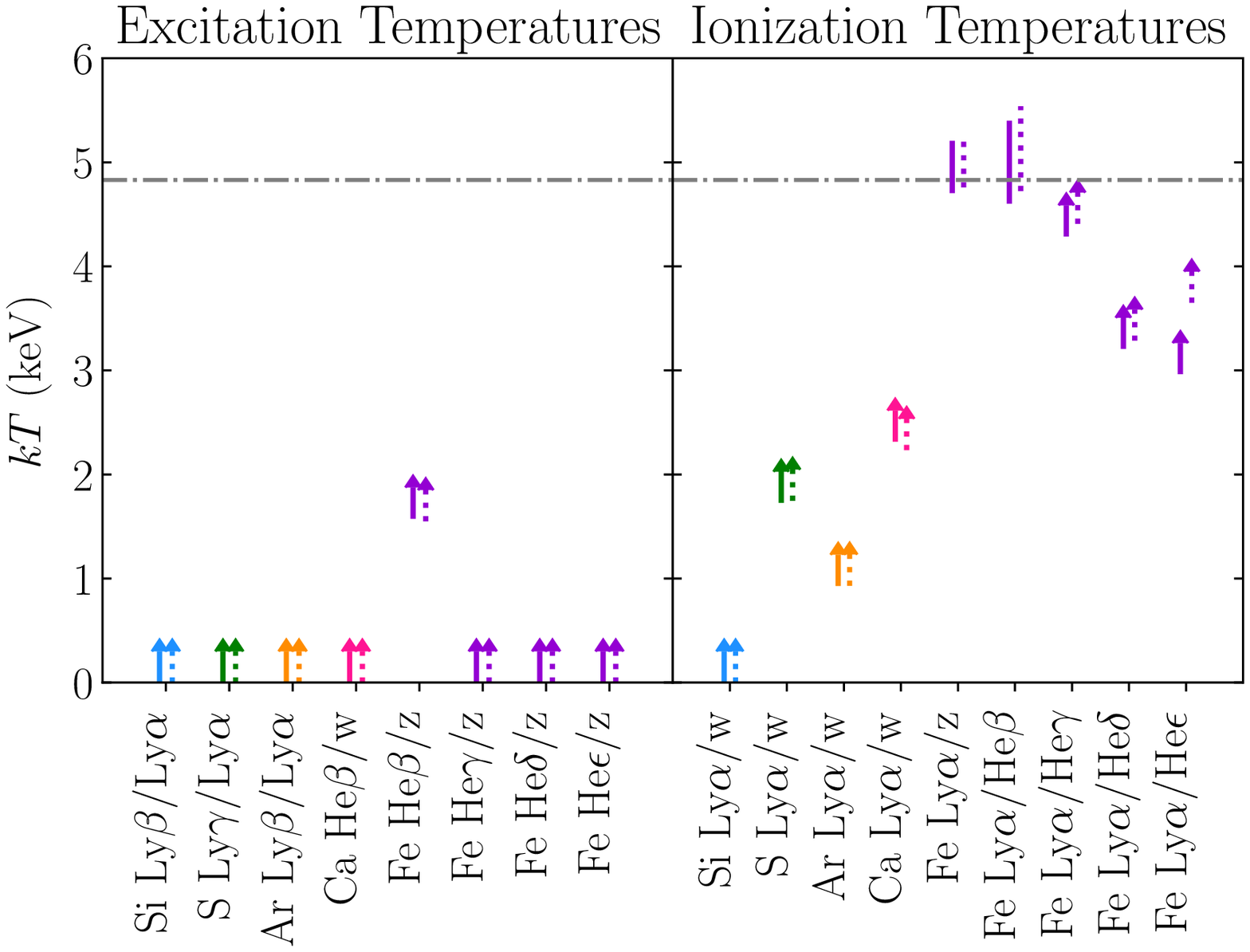}}
\end{center}
\caption{Excitation temperatures and ionization temperatures derived from individual line ratios in (a) the Entire core, (b) Nebula, (c) Rim, and (d) Outer regions.
Cyan, green, orange, pink, and purple indicate Si, S, Ar, Ca, and Fe, respectively.
The results based on \texttt{AtomDB} and \texttt{SPEXACT} are shown by the solid and dotted lines, respectively.
The horizontal dash-dotted lines show the best-fit $kT_\mathrm{line}$ of the modified 1T model described in \S\ref{ana:single-t} and \S\ref{ana:devided-1t}. 
}
\label{fig:line-temperatures}
\end{figure*}

We summarize the derived \telec and \tioni in \Fig{fig:line-temperatures}.
\tioni from Fe, which is determined with the smallest statistical uncertainties,  has typical values of 4--5~keV.
\tioni from the Entire core and Nebula regions are clearly different among elements; namely there is a tendency of increasing \tioni with increasing atomic number. 
These results indicate deviation from a single temperature CIE model.
\tioni from the Rim also suggests a slight deviation from a single temperature model.
The results of the Outer region are consistent with a single temperature approximation.

\telec from Fe for the Nebula and Rim are about 3 and 4~keV, respectively.
In the Nebula and Entire core regions,  \telec from Fe are lower than \tioni at the 2--3~$\sigma$ level, providing further evidence for deviation from the single temperature approximation. 
For Si, S, Ar, and Ca, the line ratios which are sensitive to \telec are all consistent with the CIE prediction with the temperature of 2--4~keV within the statistical 1--2$\sigma$ errors, however, the corresponding \telec are not constrained.


\subsection{Modelling of the Broad-band Spectrum in the Entire Core Region}
\label{ana:entire}

We then tried to reproduce the broad-band (1.8--20.0~keV) spectrum with optically-thin thermal plasma models based on \texttt{AtomDB} and \texttt{SPEXACT}.
In the analysis of this section, we focused on the spectrum of the Entire core region in order to ignore the contamination of photons scattered due to the point spread function (PSF) of the telescope, and to investigate uncertainties due to the atomic codes and the effective area calibration. 

\subsubsection{Single temperature plasma model}
\label{ana:single-t}

\begin{table}[]
\tbl{Best fit parameters for the Entire core region}{%
\centering
\begin{tabular}{lcc}
\hline
Model/Parameter                                            & AtomDB v3.0.9           & SPEXACT v3.03.00        \\
\hline
\hline \multicolumn{3}{l}{1CIE model}                      \\
~~~$kT_{\mathrm{1CIE}}$~(keV)                              & 3.95$^{+0.01}_{-0.01}$  & 3.94$^{+0.01}_{-0.01}$  \\
~~~$N$~(10$^{12}$~cm$^{-5}$)                               & 23.20$^{+0.05}_{-0.05}$ & 22.78$^{+0.04}_{-0.04}$ \\
~~~C-statistics/dof                                        & 13123.6/12979           & 13181.7/12979           \\
\hline \multicolumn{3}{l}{Modified 1CIE model}             \\
~~~$kT_{\mathrm{cont}}$~(keV)                              & 4.01$^{+0.01}_{-0.01}$  & 3.95$^{+0.01}_{-0.01}$  \\
~~~$kT_{\mathrm{line}}$~(keV)                              & 3.80$^{+0.02}_{-0.02}$  & 3.89$^{+0.02}_{-0.02}$  \\
~~~$N$~(10$^{12}$~cm$^{-5}$)                               & 22.77$^{+0.04}_{-0.04}$ & 22.67$^{+0.05}_{-0.05}$ \\
~~~C-statistics/dof                                        & 13085.9/12978           & 13178.7/12978           \\
\hline \multicolumn{3}{l}{2CIE model (modified CIE + CIE)} \\
~~~$kT_{\mathrm{cont}1}$~(keV)                             & 3.66$^{+0.01}_{-0.02}$  & 3.40$^{+0.02}_{-0.01}$  \\
~~~$kT_{\mathrm{line}1}$~(keV)                             & 3.06$^{+0.04}_{-0.03}$  & 2.92$^{+0.03}_{-0.03}$  \\
~~~$kT_2$~(keV)                                            & 4.51$^{+0.02}_{-0.03}$  & 4.73$^{+0.02}_{-0.02}$  \\
~~~$N_1$~(10$^{12}$~cm$^{-5}$)                             & 12.98$^{+0.05}_{-0.05}$ & 13.27$^{+0.13}_{-0.09}$ \\
~~~$N_2$~(10$^{12}$~cm$^ {-5}$)                            & 9.71$^{+0.06}_{-0.05}$  & 9.45$^{+0.07}_{-0.05}$  \\
~~~C-statistics/dof                                        & 13058.5/12976           & 13093.9/12976           \\
\hline \multicolumn{3}{l}{Power-law DEM model}             \\
~~~index                                                   & 10.92$^{+0.11}_{-0.11}$ & 4.68$^{+0.03}_{-0.03}$  \\
~~~$kT_{\mathrm{max}}$~(keV)                               & 4.01$^{+0.06}_{-0.01}$  & 4.29$^{+0.01}_{-0.01}$  \\
~~~$N$~(10$^{12}$~cm$^{-5}$)                               & 21.38$^{+0.24}_{-0.24}$ & 15.39$^{+0.04}_{-0.04}$ \\
~~~C-statistics/dof                                        & 13123.4/12978           & 13147.6/12978           \\
\hline \multicolumn{3}{l}{Gaussian DEM model}              \\
~~~$kT_{\mathrm{mean}}$~(keV)                              & 3.94$^{+0.01}_{-0.01}$  & 3.89$^{+0.01}_{-0.01}$  \\
~~~$\sigma$~(keV)                                          & 0.60$^{+0.08}_{-0.11}$  & 1.01$^{+0.05}_{-0.05}$  \\
~~~$N$~(10$^{12}$~cm$^{-5}$)                               & 11.65$^{+0.02}_{-0.02}$ & 11.67$^{+0.03}_{-0.03}$ \\
~~~C-statistics/dof                                        & 13121.1/12978           & 13138.7/12978           \\
\hline
\end{tabular}
}\label{tab:fit-entirecore}
\end{table}

\begin{figure*}
\begin{center}
\subfloat[]{\includegraphics[width=7.5cm]{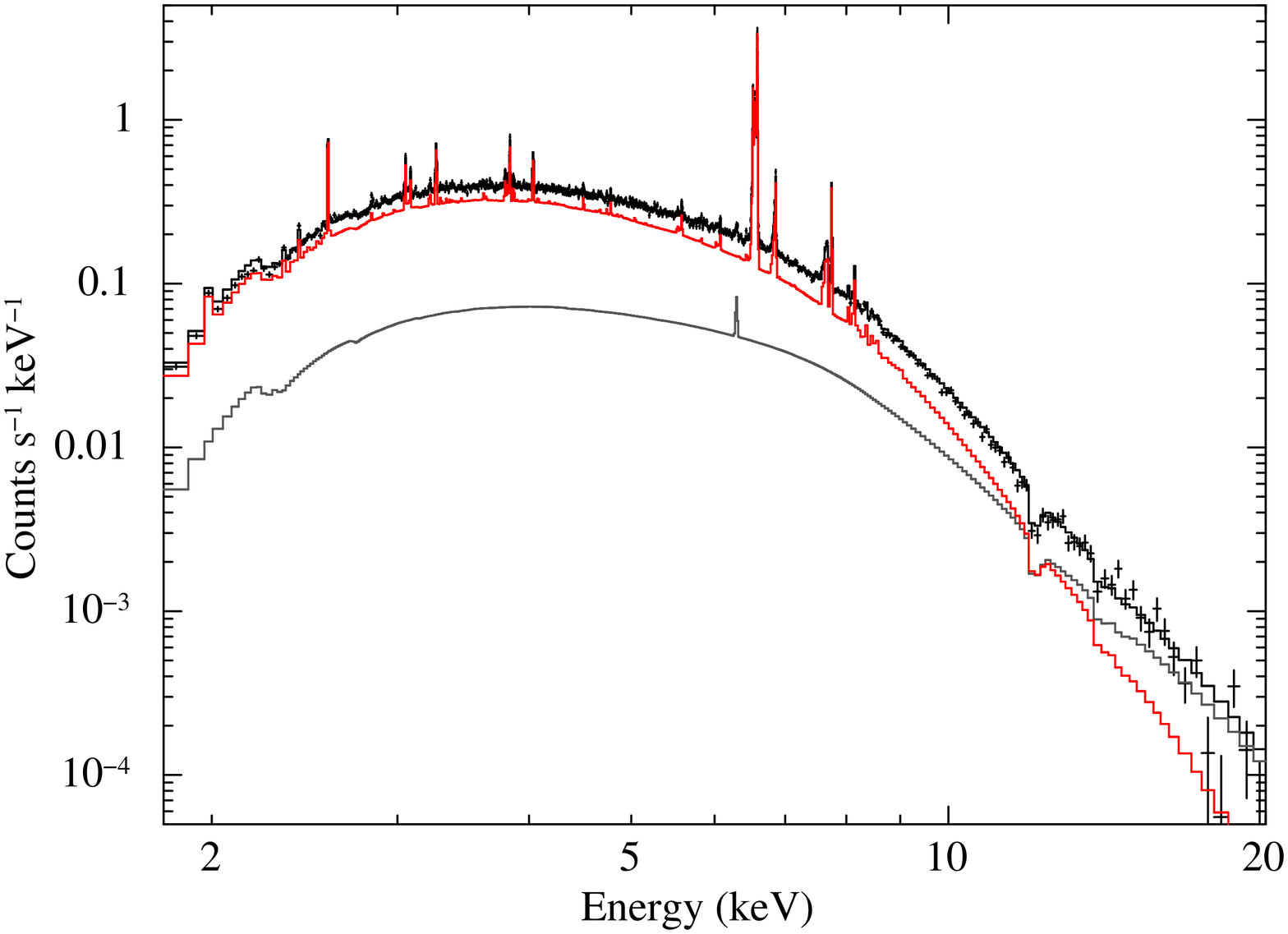}}
\hspace{0.5cm}
\subfloat[]{\includegraphics[width=7.5cm]{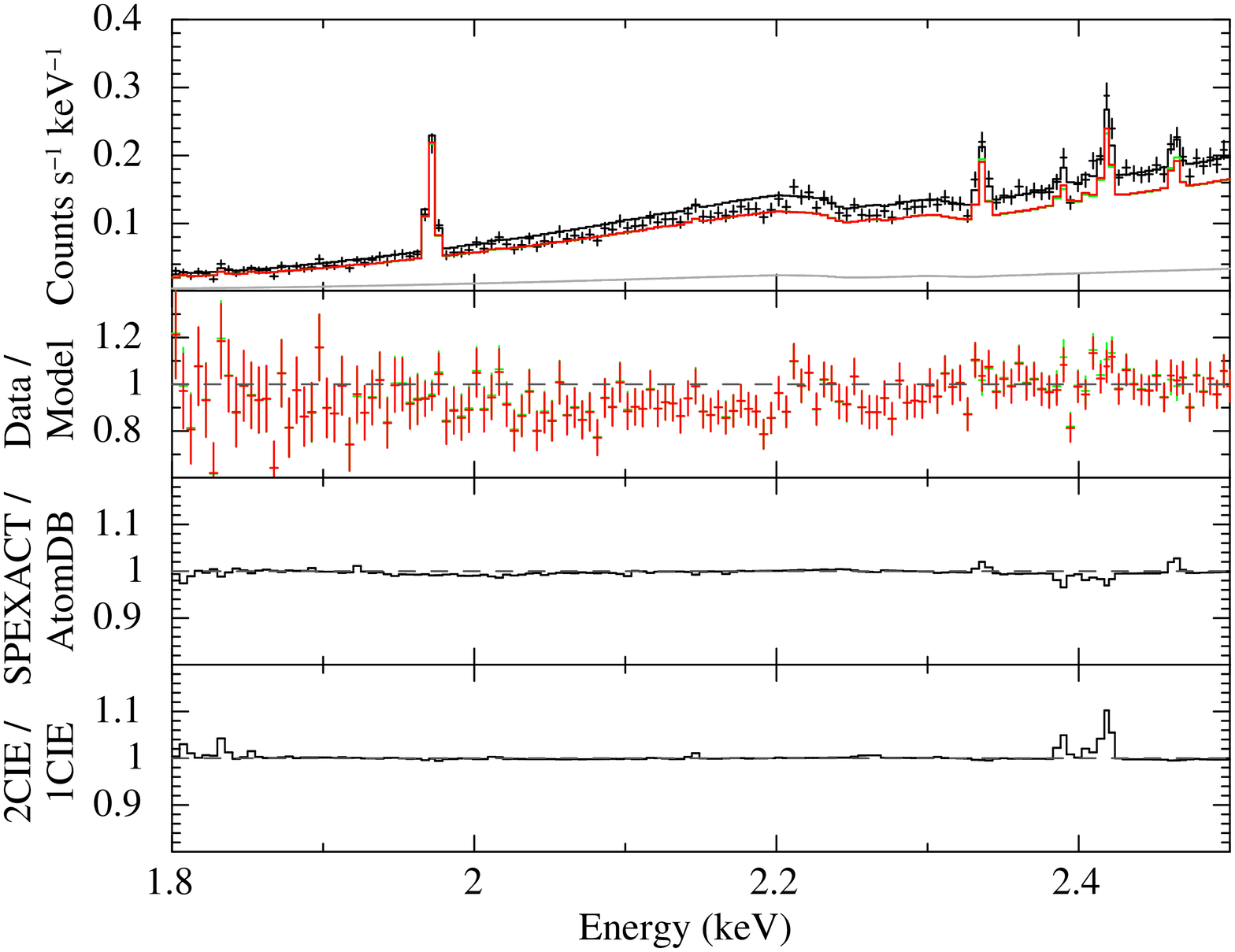}}
\vspace{0.2cm}
\subfloat[]{\includegraphics[width=7.5cm]{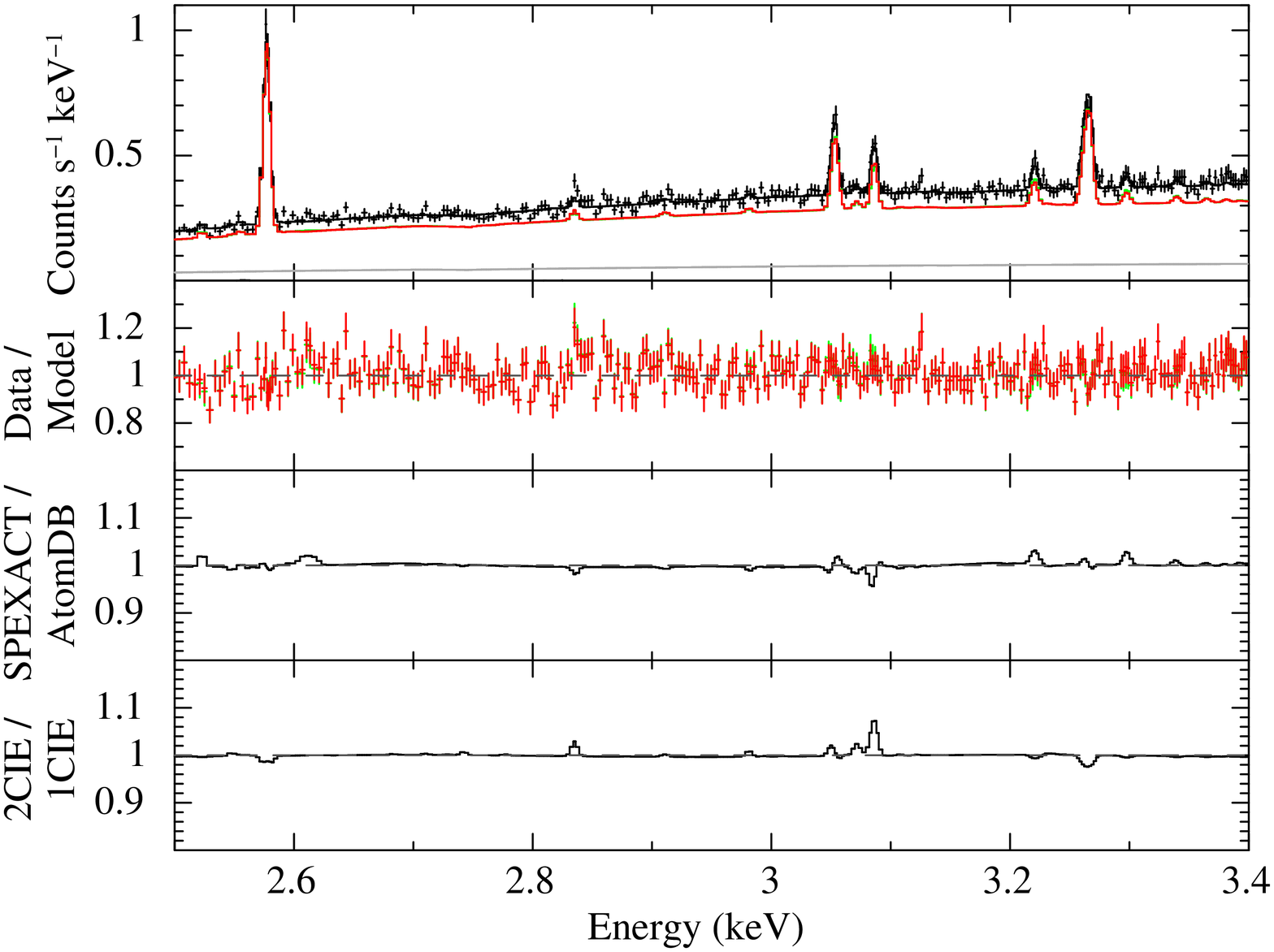}}
\hspace{0.5cm}
\subfloat[]{\includegraphics[width=7.5cm]{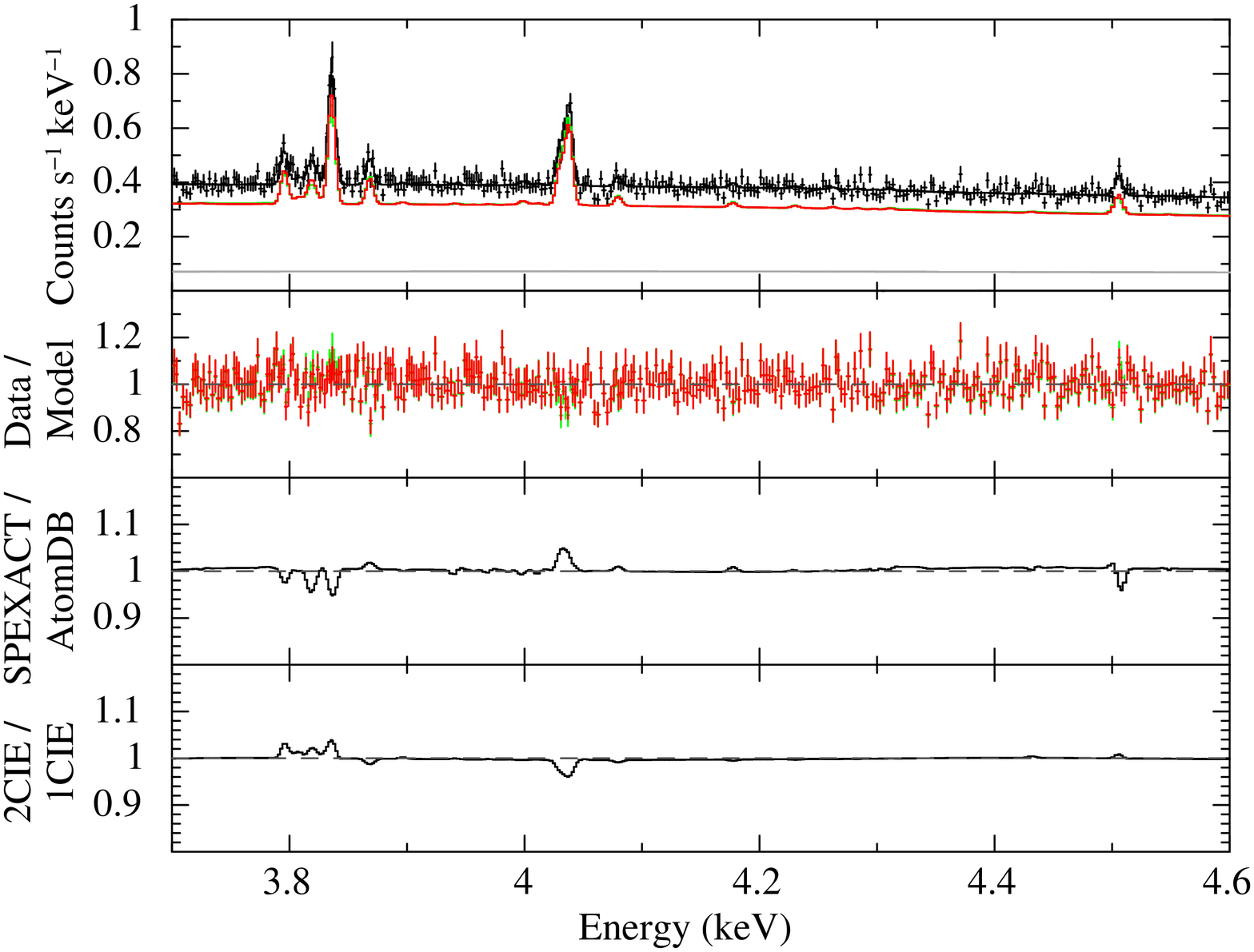}}
\vspace{0.2cm}
\subfloat[]{\includegraphics[width=7.5cm]{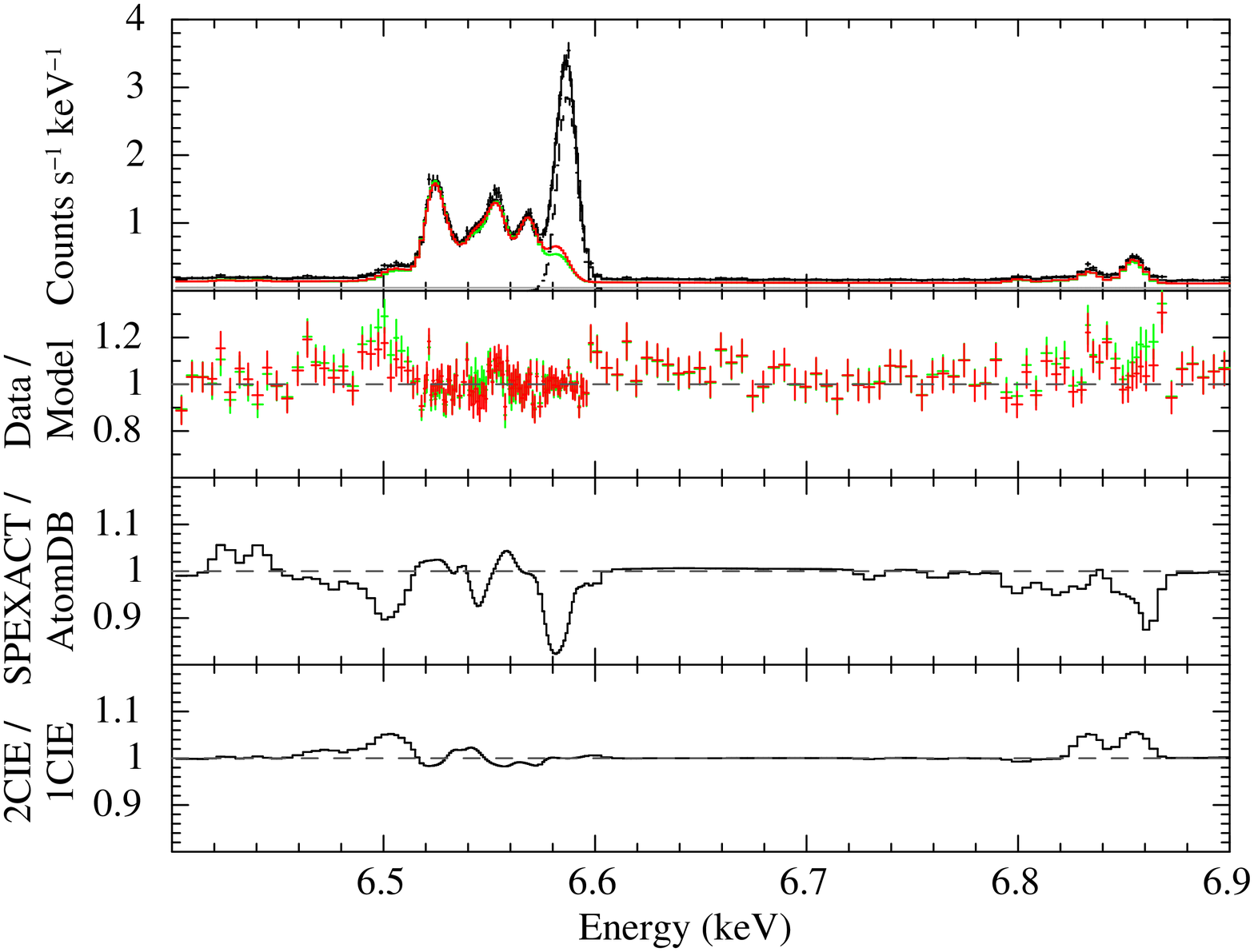}}
\hspace{0.5cm}
\subfloat[]{\includegraphics[width=7.5cm]{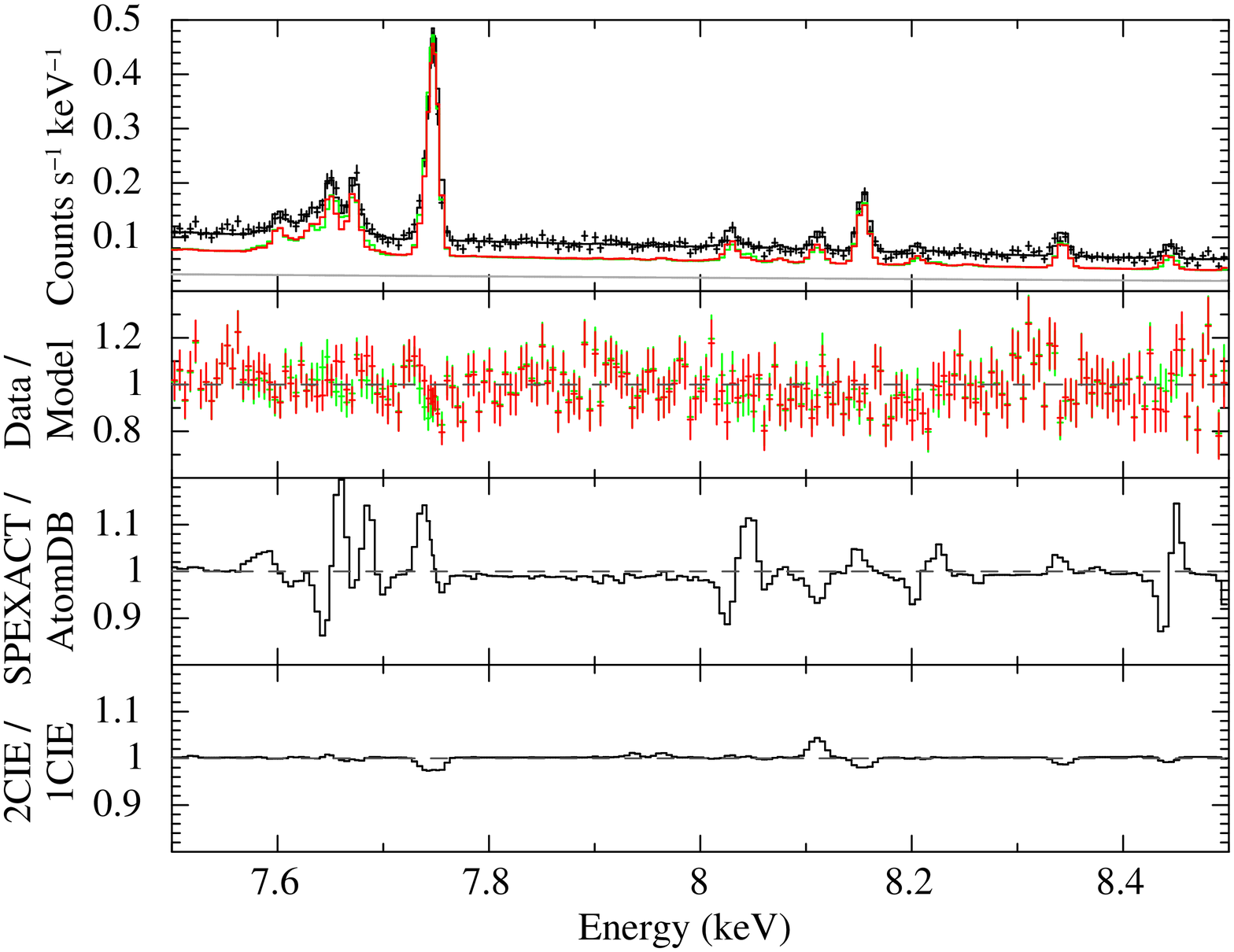}}
\end{center}
\caption{The spectra in the Entire core region fitted with the modified-1CIE model.
The entire energy band of 1.8--20.0~keV is shown in (a), and narrower energy bands of 1.8--2.5~keV, 2.5--3.4~keV, 3.7--4.6~keV,  6.4--6.9~keV, and 7.5--8.5~keV are shown in (b)--(f).
The black solid curve is the total model flux, and the red and gray curves indicate the ICM component based on \texttt{AtomDB} and the AGN component, respectively.
(b)--(f) include the green lines indicating the ICM component based on \texttt{SPEXACT}.
The figure (e), covering the 6.4--6.9~keV band, shows also the Gaussian (black dashed curve) which substitutes Fe~\emissiontype{XXV}~w in the plasma model.
All the spectra are rebinned after the fitting just for display purposes.
The second panels in (b)--(f) are the ratio of the data to the model of \texttt{AtomDB} (red) and \texttt{SPEXACT} (green).
The third panels in (b)--(f) are the comparison of \texttt{SPEXACT} and \texttt{AtomDB} in the modified 1CIE model. 
The bottom panels in (b)--(f) shows the ratio of the 2CIE model to the modified 1CIE model based on \texttt{AtomDB}.
}
\label{fig:fit-1cie}
\end{figure*}

\begin{figure}
\begin{center}
\includegraphics[width=8cm]{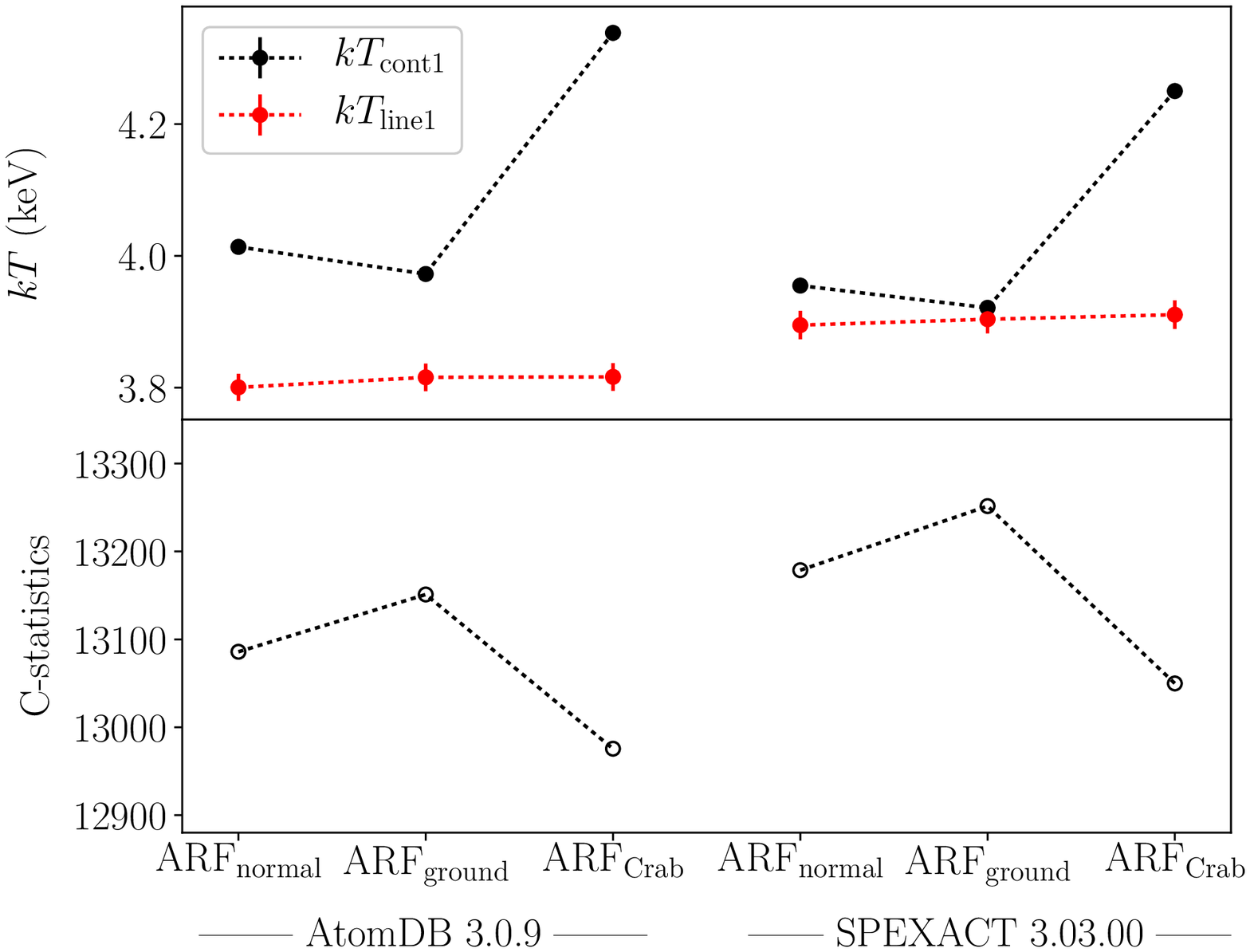}
\end{center}
\caption{Comparison of the best-fit temperatures and C-statistics among different ARFs and atomic databases for the modified 1CIE model.
}
\label{fig:compare-arf-1cie}
\end{figure}
Although the SXS spectra indicate multi-temperature conditions, we begin by fitting the data with the simplest model, that is, a single temperature CIE plasma model (hereafter the 1CIE model), with the  temperature ($kT_{\mathrm{1CIE}}$), the abundances of Si, S, Ar, Ca, Cr, Mn, Fe, and Ni, the line-of-sight velocity dispersion, and the normalization ($N$) as free parameters.
The abundances of other elements from Li through Zn were tied to that of Fe. 
Since the resonance line of He-like Fe (Fe~\emissiontype{XXV}~w) is subject to the resonance scattering effect (see the RS~paper), we replaced it by a single Gaussian so that it does not affect the parameters we obtained.
The best-fit parameters are shown in \Tab{tab:fit-entirecore}; \texttt{AtomDB} and \texttt{SPEXACT} give consistent temperatures of $3.95\pm0.01$~keV and $3.94\pm0.01$~keV, respectively.
The C-statistics are within the expected range that is calculated according to \cite{2017arXiv170709202K}, and hence the fits are acceptable even in these simple models.

In the 1CIE fit, both the continuum shape and the emission-line fluxes participate in the temperature determination.
In order to fully utilize the line resolving power of the SXS, we then modeled the continuum and lines separately and determined the continuum temperatures ($kT_\mathrm{cont}$) and the line temperatures ($kT_\mathrm{line}$) (hereafter the modified 1CIE model).
In this model, $kT_\mathrm{cont}$ and $kT_\mathrm{line}$ were independently allowed to vary whereas the other parameters were common (implemented as the \texttt{bvvtapec} model in Xspec). 
The best-fit parameters we obtained are shown in \Tab{tab:fit-entirecore}.
Both \texttt{AtomDB} and \texttt{SPEXACT} provide a reasonably good fit to the observed spectrum as shown in \Fig{fig:fit-1cie}.
Compared to $kT_{\mathrm{1CIE}}$, $kT_\mathrm{cont}$ and $kT_\mathrm{line}$ become slightly higher and lower, respectively, for both \texttt{AtomDB} and \texttt{SPEXACT}.
Since $kT_\mathrm{cont}$ is closer to $kT_{\mathrm{1CIE}}$ than $kT_\mathrm{line}$, the continuum shape most likely determines the temperature of the 1CIE model, rather than the line fluxes, even with high-resolution spectroscopy measurements. 
The temperature differences between \texttt{AtomDB} and \texttt{SPEXACT} are formally statistically significant, but are less than 0.1~keV.

The difference between $kT_\mathrm{cont}$ and $kT_\mathrm{line}$ is at most 0.23~keV but statistically significant.
As we found the multi-temperature structure from the line ratio diagnostics (\S\ref{ana:line}), that difference is possible.
However, an uncertainty in the effective area also might affects the results; the in-flight calibration of Hitomi was not completed because of its short life time.
We therefore assessed this uncertainty using the modified ARF based on the ground telescope calibration (ARF$_\mathrm{ground}$) and the actual Crab data (ARF$_\mathrm{Crab}$).  See Appendix~\ref{appendix:arf} for the detailed correction method.
We fitted the modified 1CIE model using ARF$_\mathrm{ground}$ and ARF$_\mathrm{Crab}$.
The correction of the ARF slightly affects the parameters of the AGN components as well (see Table~4 of  the AGN~paper).
Even though the differences are very small, we used the specific AGN parameter values corresponding to each assumed ARF in our fits.
The temperatures and C-statistics we obtained are summarized in \Fig{fig:compare-arf-1cie}.
$kT_\mathrm{cont}$ varies depending on ARF because the continuum shape is subject to the effective area shape.
On the other hand, the values of $kT_\mathrm{line}$ measured with different assumptions for the ARF remain consistent with each other.
Therefore, $kT_\mathrm{line}$ provides the most robust estimate of the temperature from the SXS spectrum assuming a single-phase model.
In terms of the C-statistics, the ARF$_\mathrm{Crab}$ gives the best-fit, but this choice of ARF also results in the largest difference between $kT_\mathrm{cont}$ and $kT_\mathrm{line}$.
This illustrates the difficulty of effective area calibration with the limited amount of available data.  

Even though the AGN~paper carefully modeled the AGN emission, the uncertainty of its model parameters and their impact on the best-fit temperature structure should also be considered.
If the AGN model is slightly changed, $kT_\mathrm{cont}$ would again change, while $kT_\mathrm{line}$ would be less affected as demonstrated in the comparison of the ARFs.

\subsubsection{Two temperature plasma models}
\label{ana:two-t}

\begin{figure}
\begin{center}
\includegraphics[width=8cm]{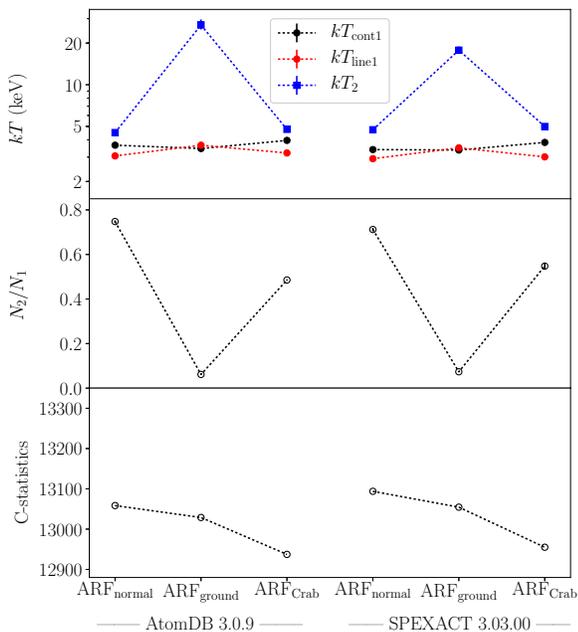}
\end{center}
\caption{Comparison of the best-fit temperatures and C-statistics among different ARFs and atomic databases for the 2CIE model.
}
\label{fig:compare-arf-2cie}
\end{figure}
The line ratio diagnostics in \S\ref{ana:line} actually indicate the presence of multi-temperature structure in the Perseus cluster core.
As a simple approximation of the deviations from a single thermal phase, we first used a two-temperature model where another CIE model was added to the modified 1CIE model (hereafter the 2CIE model).
The free parameters of the additional CIE component were the temperature and the normalization, while the abundances and the line-of-sight velocity dispersion were tied to those of the primary component.
The results are shown in \Tab{tab:fit-entirecore}.
As expected, the C-statistics are significantly improved from those of the modified 1CIE model ($\Delta C=$~30--91).
However, as shown in the bottom panels of \Fig{fig:fit-1cie} (b)--(f), the continuum is almost the same and the difference of line emissivities are at most 10$\%$ compared to the 1CIE model.
The temperatures and normalizations obtained with the two spectral codes are in reasonably good agreement, although some differences are statistically significant. 
The dominant component now has a temperature of $kT_\mathrm{line}=3.06\pm0.03$~keV from \texttt{AtomDB}, which is fully consistent with $kT_\mathrm{line}=3.06^{+0.03}_{-0.08}$~keV from \texttt{SPEXACT}. 
The second thermal component is from hotter gas with $kT_2\sim5$ keV; for this component, \texttt{SPEXACT} gives a $\sim10$\% higher temperature than \texttt{AtomDB}, and a somewhat lower relative normalization ($N_2/(N_1+N_2)$ of 31\% with \texttt{SPEXACT} and 43\% with \texttt{AtomDB}). 
The temperatures derived from the 2CIE fit are consistent with the line ratio diagnostics shown in \Fig{fig:line-temperatures}; the ionization temperature of S is $\sim 3$~keV and that of Fe is $\sim4.5$~keV.
We also checked difference of the line-of-sight velocity dispersion between the lower and higher temperature components, but no significant difference was found (see Appendix~\ref{appendix:sigmav} for details). 

In the same manner as for the modified 1CIE model (\S\ref{ana:single-t}), we examined the effect of different ARFs (ARF$_\mathrm{ground}$ and ARF$_\mathrm{Crab}$) for the 2CIE model.
\Fig{fig:compare-arf-2cie} shows the resulting temperatures, the ratio of the normalizations ($N_{2}/N_{1}$), and the C-statistics for each ARF.
The best-fit parameter values vary significantly depending on the choice of ARF, but the temperatures of ARF$_\mathrm{normal}$ and ARF$_\mathrm{Crab}$ are very close to each other ($\sim$3~keV plus $\sim$5~keV).
Only ARF$_\mathrm{ground}$ shows the presence of a $>$20~keV component, which seems physically less well motivated. 
The different trend in ARF$_\mathrm{ground}$ is likely caused by an incomplete modeling of the continuum; as shown in the middle panel of \Fig{fig:compare-residuals} (Appendix~\ref{appendix:arf}), downward convex residuals are seen in the 2--7~keV band for ARF$_\mathrm{ground}$.
In any case, the trend where the dominant component has a temperature of 3--4~keV and the sub-dominant additional phase has a higher temperature, is robust. 

\subsubsection{Other combinations of collisional plasma models}
\label{ana:other-t}
We also tried to add one more CIE component to the 2CIE model (i.e., 3CIE model), but no significant improvements of the C-statistics are found.
Therefore, the 2CIE model is sufficient to reproduce the observed spectrum.

The actual temperature structure of the ICM might not consist of discrete temperature components but rather of a continuous temperature distribution. 
Indeed, some hints of a power-law  or a Gaussian temperature distributions were reported in the literature (e.g., \cite{2004A&A...413..415K,2009A&amp;A...493..409S}).
We therefore applied these simple differential emission measure (DEM) models to the SXS spectrum.  
The emission measure profile, $EM(kT)$, is proportional to $(kT/kT_\mathrm{max})^{\alpha}$ for the power-law DEM model and to $\exp(-(kT - kT_\mathrm{mean})^2/2\sigma)$ for the Gaussian DEM model. 
The best-fit parameters of the models are summarized in \Tab{tab:fit-entirecore}.
Both the power-law and the Gaussian DEM models show steep temperature distributions peaked at $\sim4$~keV, even though the distributions based on \texttt{SPEXACT} are slightly wider (smaller index or larger $\sigma$) than those based on \texttt{AtomDB}.
In any case, we found no significant improvements from the 2CIE model.
Further investigation of the multi-temperature model is shown in \Sec{ana:multi-t} and \Fig{fig:norm-profile}.

Another possible cause of the deviation from a single temperature model shown in the line ratio diagnostics is the NEI state, which is often observed in supernova remnants. We thus tried to fit the spectrum with a NEI model (the possibilities of both an ionizing and a recombining plasma are considered).
However, the obtained ionization parameter becomes $nt >1\times10^{12}$~cm$^{-3}$~s$^{-1}$, and the temperature is almost the same as the 1T model; therefore the model is consistent with a CIE state, and we find no significant signature of the NEI. 

\subsection{Spatial Variation of the Temperature Structure}
\label{ana:devided-1t}

\begin{table*}[]
\tbl{Best fit parameters for Neubla, Rim and Outer}{%
\centering
\begin{tabular}{@{\extracolsep{4pt}}lcccccc}
\hline
Model/Parameter                                & \multicolumn{3}{c}{AtomDB v3.0.9} & \multicolumn{3}{c}{SPEXACT v3.03.00} \\
\cline{2-4}\cline{5-7}                         & Nebula                            & Rim                     & Outer                  & Nebula                               & Rim                     & Outer                  \\
\hline
\hline \multicolumn{7}{l}{Modified 1CIE model} \\
\hspace{0.3cm}$kT_{\mathrm{cont}}$~(keV)       & 3.96$^{+0.01}_{-0.01}$            & 4.02$^{+0.01}_{-0.01}$  & 4.93$^{+0.10}_{-0.10}$ & 3.90$^{+0.02}_{-0.02}$               & 3.97$^{+0.01}_{-0.01}$  & 4.85$^{+0.09}_{-0.09}$ \\
\hspace{0.3cm}$kT_{\mathrm{line}}$~(keV)       & 3.73$^{+0.03}_{-0.03}$            & 3.94$^{+0.04}_{-0.04}$  & 4.83$^{+0.15}_{-0.16}$ & 3.82$^{+0.03}_{-0.07}$               & 4.07$^{+0.04}_{-0.04}$  & 4.97$^{+0.16}_{-0.14}$ \\
\hspace{0.3cm}$N$~(10$^{12}$~cm$^{-5}$)        & 14.75$^{+0.05}_{-0.05}$           & 15.31$^{+0.04}_{-0.04}$ & 5.22$^{+0.10}_{-0.11}$ & 14.68$^{+0.08}_{-0.08}$              & 15.23$^{+0.04}_{-0.04}$ & 5.21$^{+0.09}_{-0.09}$ \\
~~~C-statistics/dof                            & 11948.0/12200                     & 10168.9/10300           & 6323.8/6929            & 12013.0/12200                        & 10188.6/10300           & 6326.5/6929            \\
\hline \multicolumn{7}{l}{2CIE model}          \\
\hspace{0.3cm}$kT_{\mathrm{cont}1}$~(keV)      & 3.56$^{+0.42}_{-0.07}$            & 3.65$^{+0.02}_{-0.02}$  & $\cdots$               & 3.39$^{+0.07}_{-0.09}$               & 3.40$^{+0.02}_{-0.02}$  & $\cdots$               \\
\hspace{0.3cm}$kT_{\mathrm{line}1}$~(keV)      & 2.78$^{+0.10}_{-0.37}$            & 3.49$^{+0.05}_{-0.05}$  & $\cdots$               & 2.60$^{+0.24}_{-0.23}$               & 3.27$^{+0.05}_{-0.05}$  & $\cdots$               \\
\hspace{0.3cm}$kT_2$~(keV)                     & 4.32$^{+0.02}_{-0.36}$            & 4.98$^{+0.05}_{-0.05}$  & $\cdots$               & 4.30$^{+0.28}_{-0.16}$               & 4.99$^{+0.04}_{-0.03}$  & $\cdots$               \\
\hspace{0.3cm}$N_1$~(10$^{12}$~cm$^{-5}$)      & 6.91$^{+0.45}_{-3.40}$            & 11.16$^{+0.06}_{-0.06}$ & $\cdots$               & 6.24$^{+2.14}_{-2.21}$               & 9.94$^{+0.06}_{-0.06}$  & $\cdots$               \\
\hspace{0.3cm}$N_2$~(10$^{12}$~cm$^ {-5}$)     & 7.73$^{+3.46}_{-0.45}$            & 4.28$^{+0.04}_{-0.04}$  & $\cdots$               & 8.32$^{+1.56}_{-1.97}$               & 5.47$^{+0.05}_{-0.05}$  & $\cdots$               \\
~~~C-statistics/dof                            & 11926.0/12198                     & 10163.2/10298           & $\cdots$               & 11958.2/12198                        & 10173.3/10298           & $\cdots$               \\
\hline \multicolumn{7}{l}{PSF corrected model} \\
\hspace{0.3cm}$kT_{\mathrm{cont1}}$~(keV)      & 3.64$^{+0.03}_{-0.03}$            & 3.92$^{+0.02}_{-0.02}$  & 5.11$^{+0.05}_{-0.05}$ & 3.46$^{+0.03}_{-0.03}$               & 3.82$^{+0.02}_{-0.02}$  & 5.01$^{+0.06}_{-0.05}$ \\
\hspace{0.3cm}$kT_{\mathrm{line1}}$~(keV)      & 2.68$^{+0.04}_{-0.05}$            & 3.88$^{+0.05}_{-0.05}$  & 5.00$^{+0.16}_{-0.16}$ & 2.66$^{+0.04}_{-0.04}$               & 3.97$^{+0.06}_{-0.05}$  & 5.19$^{+0.17}_{-0.16}$ \\
\hspace{0.3cm}$kT_{2}$~(keV)                   & 4.27$^{+0.03}_{-0.03}$            & 5.37$^{+0.28}_{-0.30}$  & $\cdots$               & 4.53$^{+0.03}_{-0.03}$               & 6.80$^{+0.56}_{-0.46}$  & $\cdots$               \\
\hspace{0.3cm}$N_1$~(10$^{12}$~cm$^{-5}$)      & 5.54$^{+0.04}_{-0.04}$            & 10.18$^{+0.05}_{-0.05}$ & 4.51$^{+0.04}_{-0.04}$ & 6.63$^{+0.05}_{-0.21}$               & 10.35$^{+0.05}_{-0.05}$ & 4.52$^{+0.04}_{-0.04}$ \\
\hspace{0.3cm}$N_2$~(10$^{12}$~cm$^{-5}$)      & 5.86$^{+0.03}_{-0.04}$            & 0.70$^{+0.04}_{-0.03}$  & $\cdots$               & 4.72$^{+0.04}_{-0.04}$               & 0.52$^{+0.04}_{-0.03}$  & $\cdots$               \\
~~~C-statistics/dof                            & ~                                 & 28404.6/29425           & ~                      & ~                                    & 28444.1/29425           & ~                      \\
\hline
\end{tabular}
}\label{tab:fitothers}
\end{table*}

\begin{table}[]
\tbl{The fraction of integrated photons coming from each sky region.}{%
\centering
\begin{tabular}{lccc}
\hline
 & \multicolumn{3}{c}{Sky regions}\\
\cline{2-4}
Integrated regions & Nebula & Rim & Outer \\
\hline
Nebula & 0.800 & 0.192 & 0.008 \\
Rim    & 0.273 & 0.719 & 0.007 \\
Outer  & 0.034 & 0.111 & 0.855 \\
\hline
\end{tabular}
}\label{tab:psfratio}
\end{table}

We next modeled the broad-band spectra in the Nebula, Rim, and Outer regions in order to look for spatial trends in the temperature distribution. 
The fit results obtained with the modified 1CIE model are shown in the top rows of \Tab{tab:fitothers}.
Compared to the result from the Entire core region, the temperature in the Nebula region is slightly lower, while that in the Rim region is slightly higher. The temperature continues to increase at larger radii, reaching 5~keV in the Outer region.
These results are consistent with the temperature map obtained from XMM-Newton and Chandra observations \citep{2003ApJ...590..225C,2007MNRAS.381.1381S}.

The line ratio diagnostics show a deviation from the single temperature approximation in the Nebula and Rim regions. 
We thus applied the 2CIE model to the spectra of those regions.
The best-fit parameters are also shown in the middle rows of \Tab{tab:fitothers}.
The C-statistics were improved from the modified 1CIE model ($\Delta C =$~6--59).
Both the Nebula and Rim regions show the same composition as the Entire core (roughly 3~keV plus 5~keV), but with different normalization ratios (the relative contribution of the hotter component is lower in the Rim region, although significant differences between the two spectral codes are also found). 
Large asymmetrical errors of the normalizations in the Nebula region are likely due to the
comparable normalization values of the two components and the limited energy band ($>1.8$ keV). 
In the Nebula region, the discrepancy between $kT_\mathrm{cont}$ and $kT_\mathrm{line}$ becomes large ($\sim$1.0~keV), and $kT_\mathrm{line}$ shows the lowest temperature of $\sim$2.7~keV among the different spatial regions considered.
We also checked the 2CIE model in the Outer region, but no improvements from the modified 1CIE model were found ($\Delta C < 1$), as expected from the line ratio diagnostics. 
The systematic uncertainty of the temperature measurements due to the different ARFs has a similar trend as the analysis of the Entire core region (see Appendix~\ref{appendix:arf}).

The sizes of the regions used for spatially resolved spectroscopy are comparable to the angular resolution of the telescope.
Therefore, photons scattered from the adjacent regions due to the telescope's PSF tail might affect the fitting results. 
We calculated the expected fraction of scattered photons with ray-tracing simulations, and show the results in \Tab{tab:psfratio}; the fractions reach up to 30\%, and are not negligible.
We thus performed a ``PSF corrected'' analysis, in which all the regions were simultaneously fitted taking into account the expected fluxes of photons scattered between regions. 
We used the 2CIE model for the Nebula and Rim regions and the 1 CIE model for the Outer region according to the results presented above.
The best-fit parameters of the PSF corrected model are shown in the bottom rows of \Tab{tab:fitothers}.
After the PSF correction, the ratios of the normalizations are changed but the temperatures we obtained are almost consistent with those derived from the PSF ``uncorrected'' analysis.

\subsection{Comparison with Multi-temperature Models from Previous Observations}
\label{ana:multi-t}

\begin{figure*}
\begin{center}
\subfloat{\includegraphics[width=8cm]{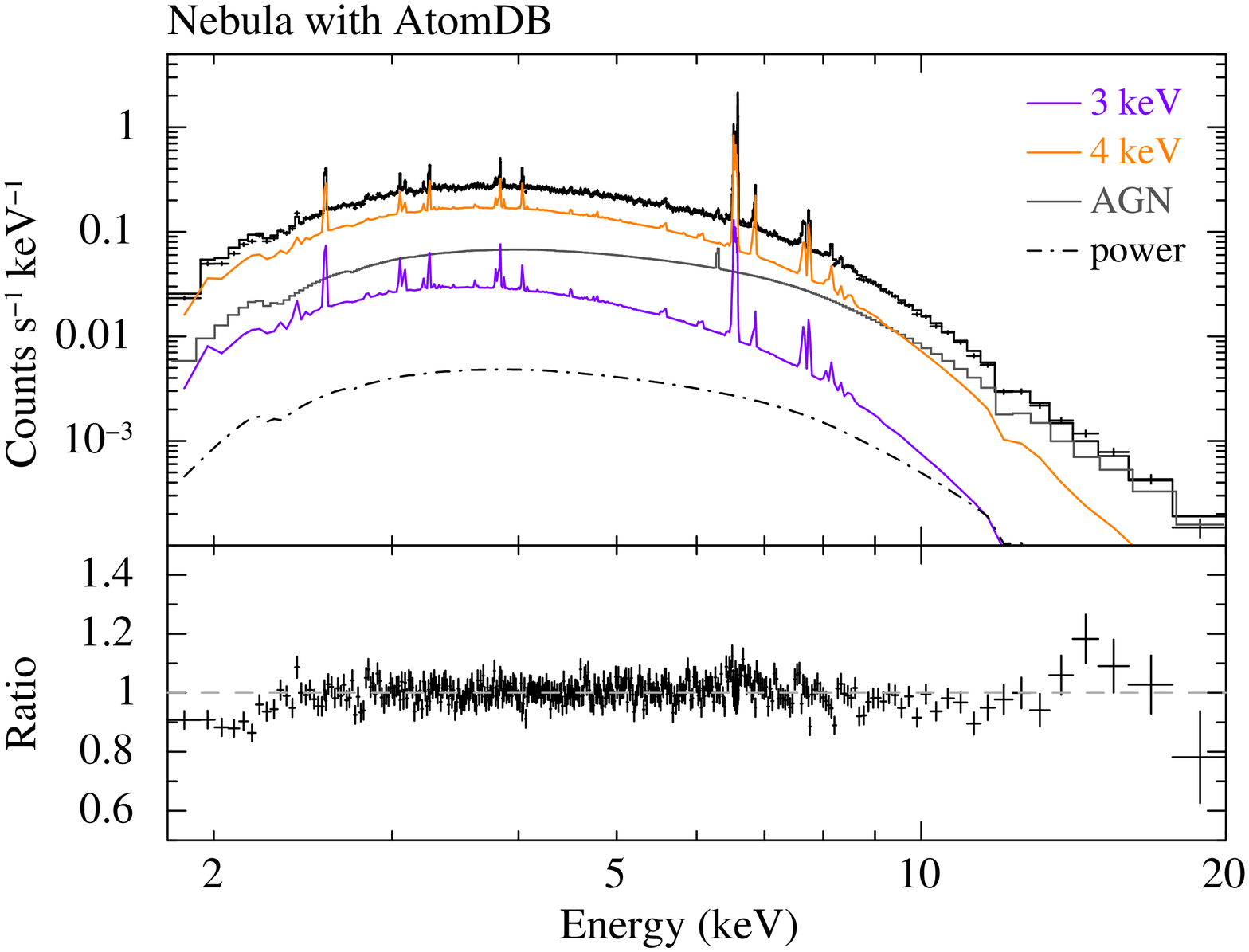}}
\hspace{0.1cm}
\subfloat{\includegraphics[width=8cm]{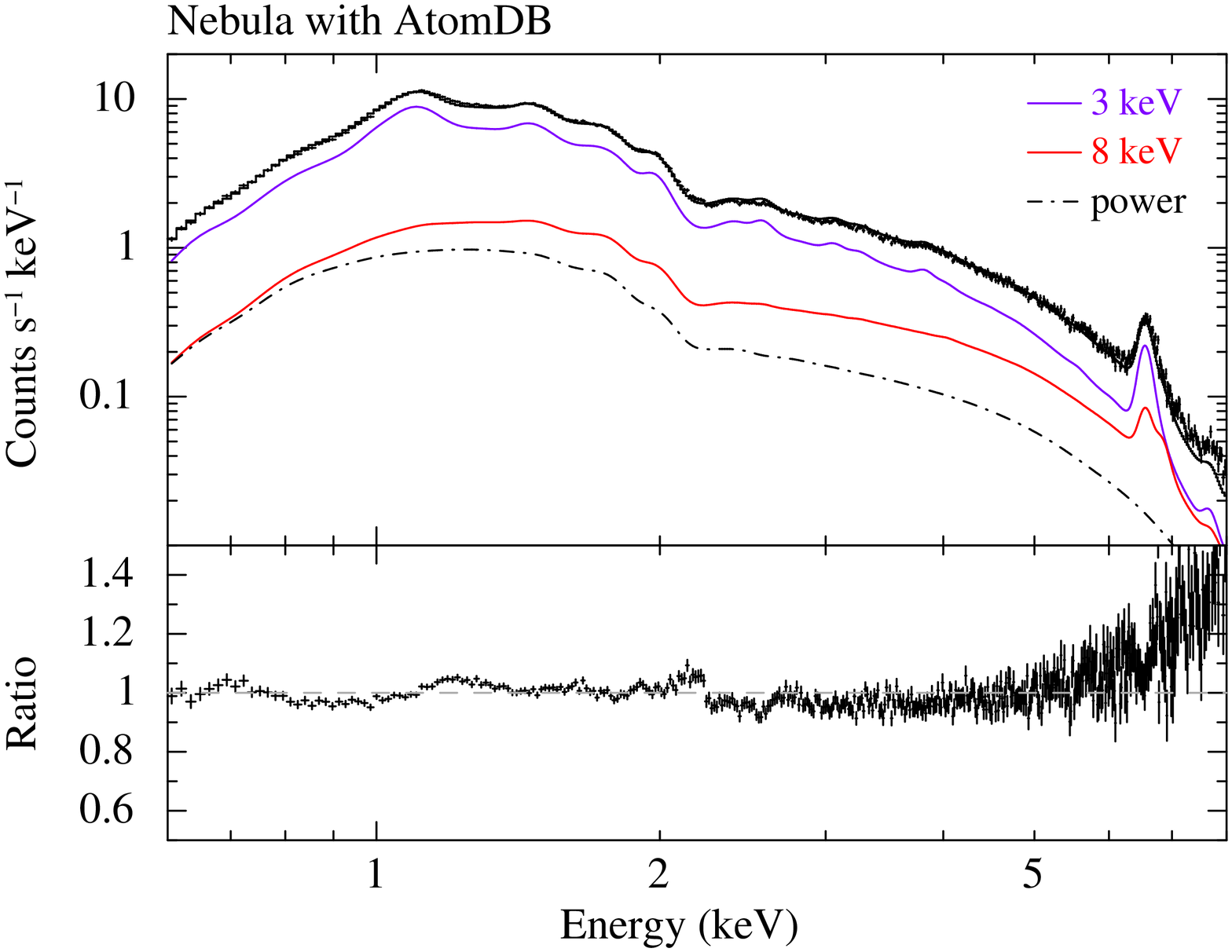}}
\vspace{0.1cm}
\subfloat{\includegraphics[width=8cm]{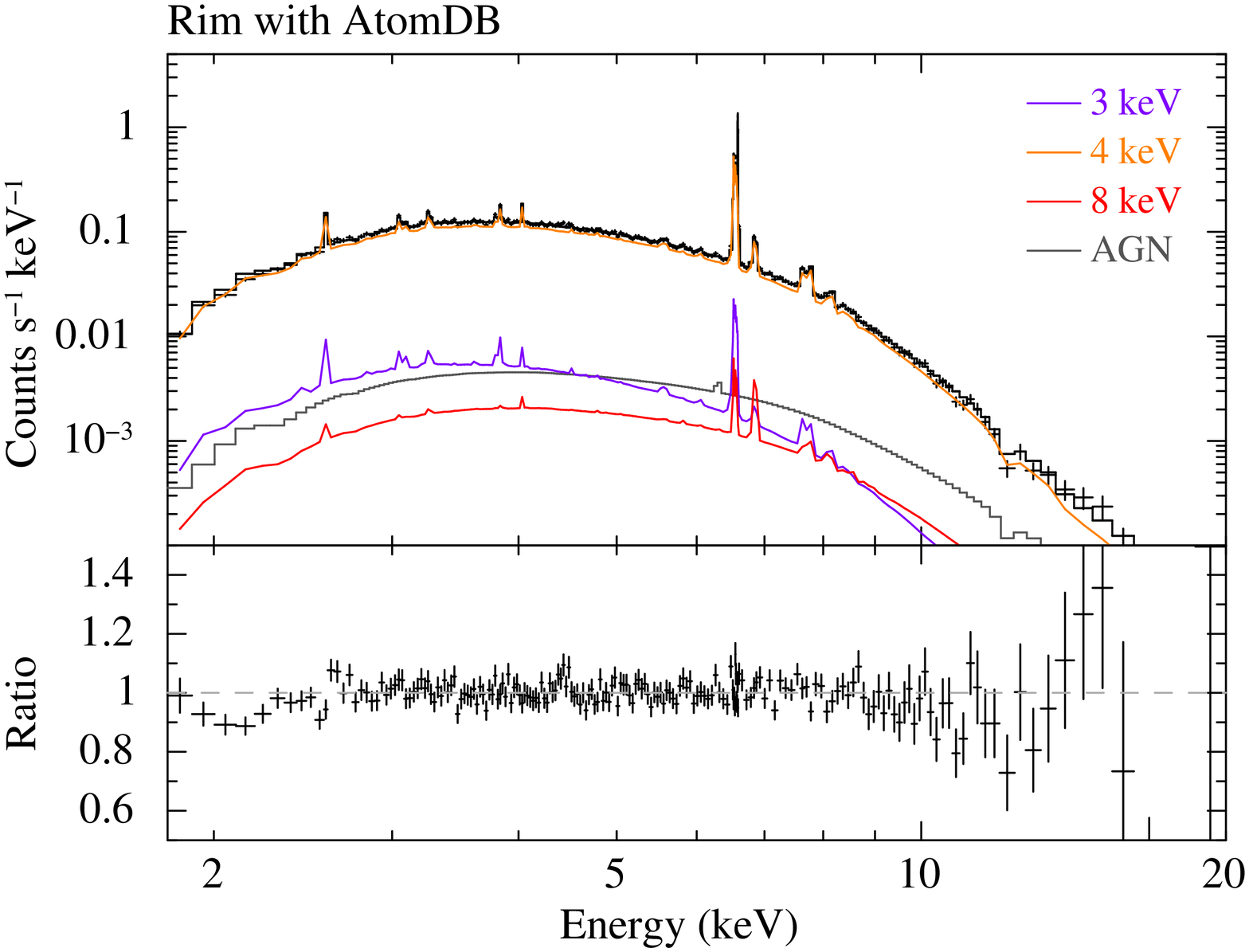}}
\hspace{0.1cm}
\subfloat{\includegraphics[width=8cm]{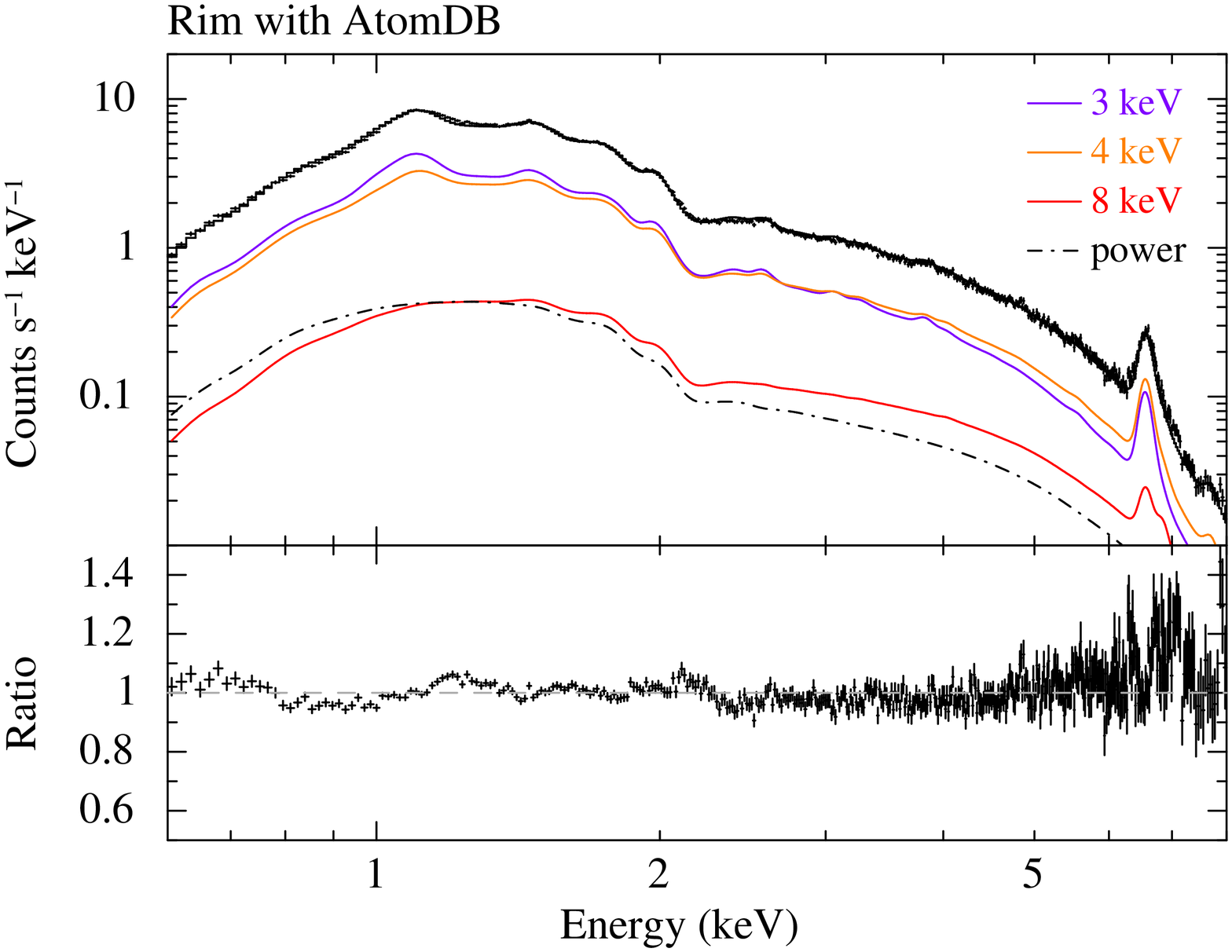}}
\end{center}
\caption{
Spectra fitted with the six-temperature model (upper panels) and ratios between the data and the model (lower panels).
Left and right columns show the Hitomi/SXS and Chandra/ACIS spectra, respectively.
The top and bottom rows correspond to the Nebula and Rim regions, respectively.
The components included in the model are shown in color lines as denoted in each panel.
The components whose normalizations are very low are not shown. 
}
\label{fig:fit-multi-t}
\end{figure*}

\begin{figure*}
\begin{center}
\subfloat{\includegraphics[width=8cm]{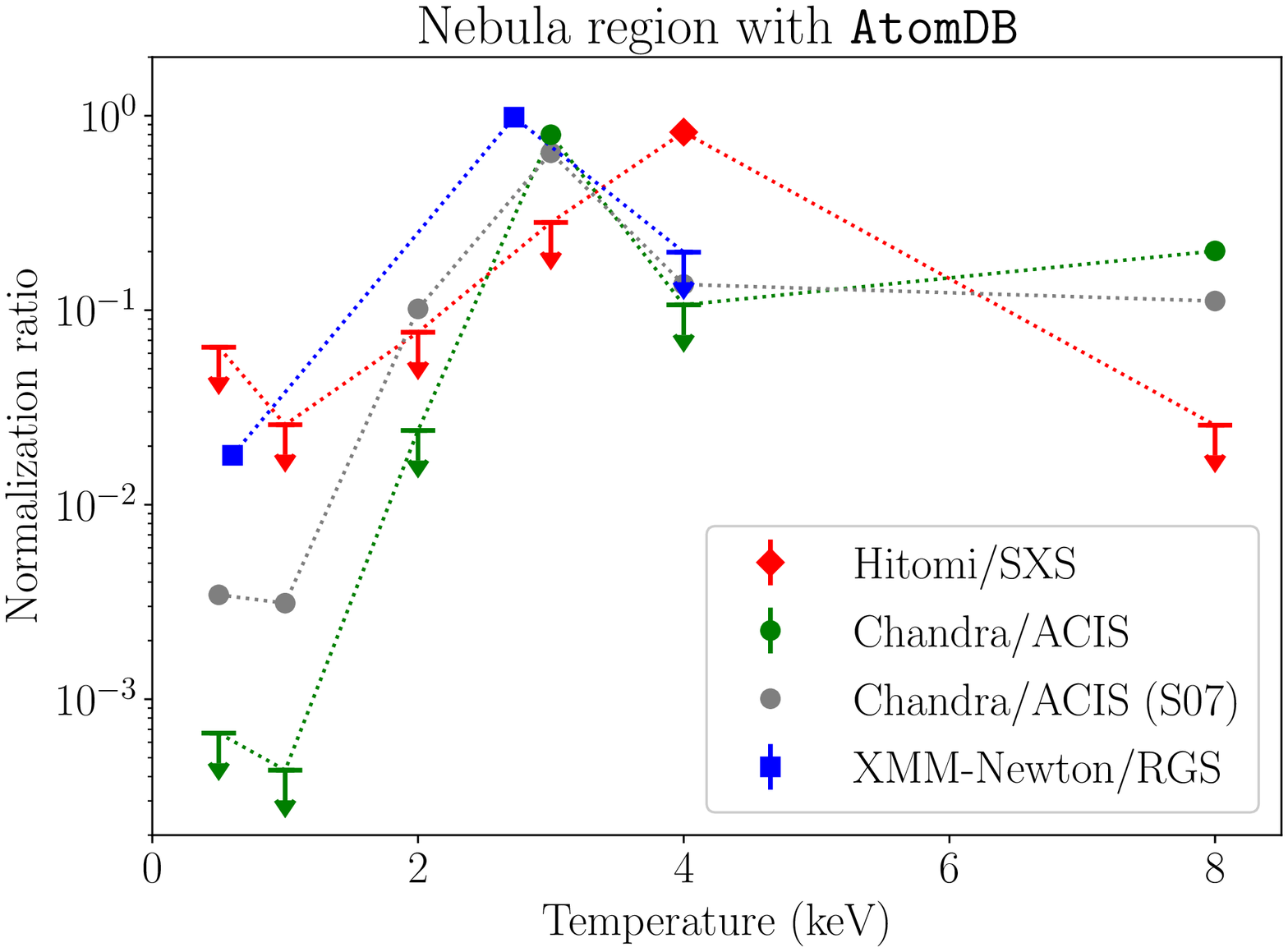}}
\hspace{0.1cm}
\subfloat{\includegraphics[width=8cm]{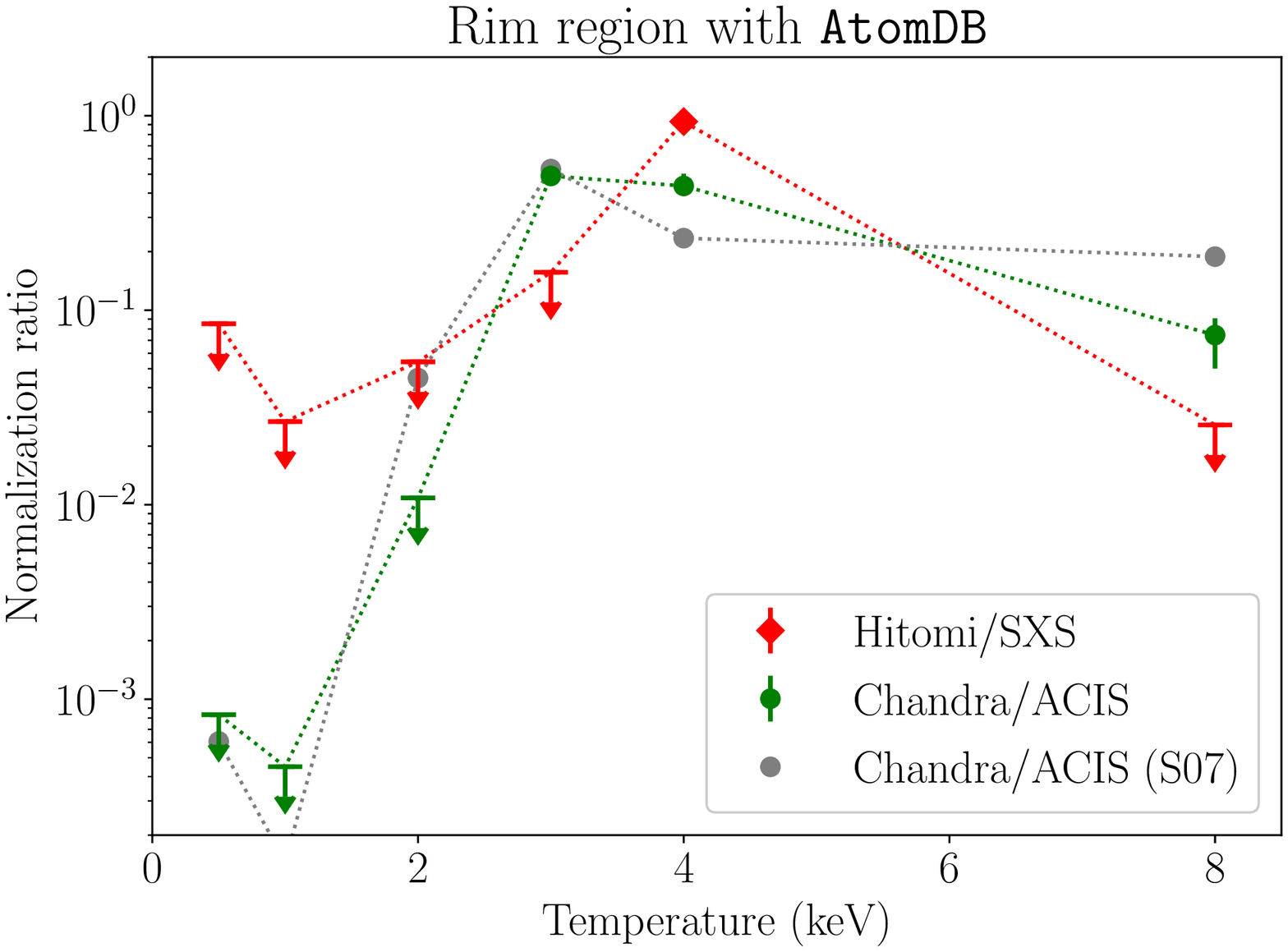}}
\vspace{0.1cm}
\subfloat{\includegraphics[width=8cm]{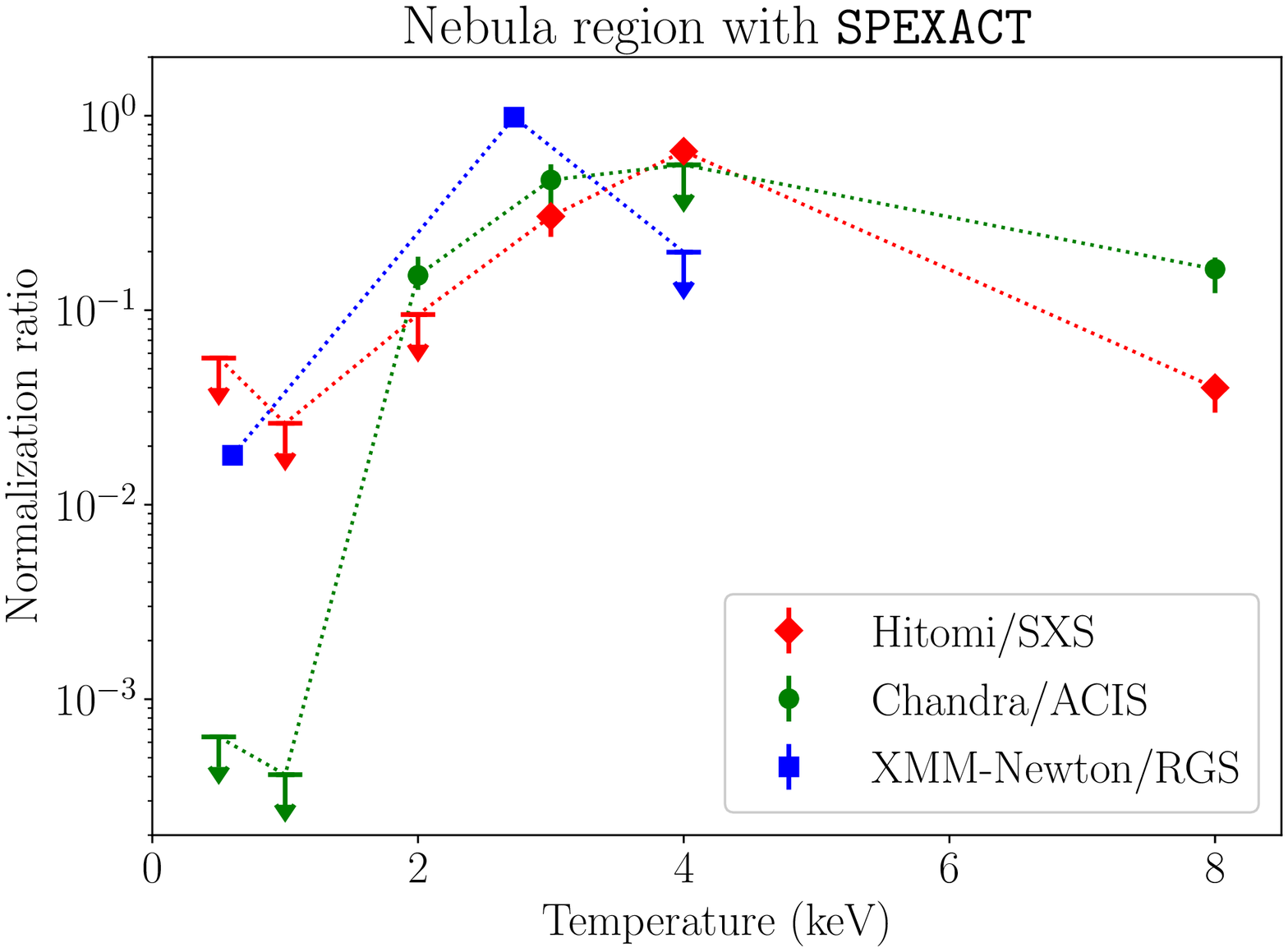}}
\hspace{0.1cm}
\subfloat{\includegraphics[width=8cm]{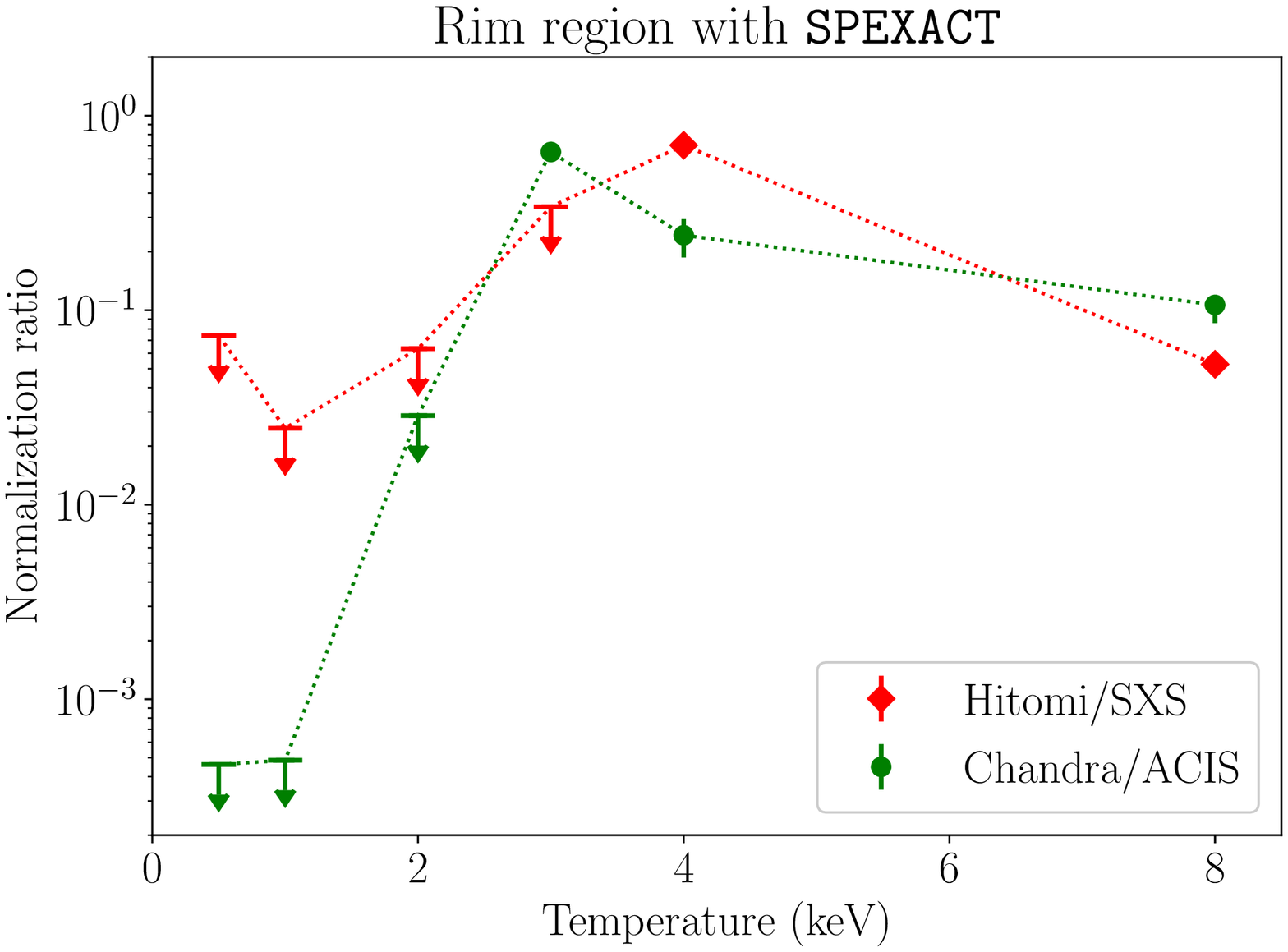}}
\end{center}
\caption{Normalization ratios of each temperature component derived from the multi-temperature models. The top rows are the results from \texttt{AtomdB} and the bottom rows are those from \texttt{SPEXACT}. The left and right columns correspond to Nebula and Rim, respectively. The red diamonds are Hitomi/SXS, the black circles are Chandra/ACIS, the gray circles are Chandra/ACIS of \citet{2007MNRAS.381.1381S}, and the blue squares are XMM-Newton/RGS. The results of XMM-Newton/RGS are shown only in the Nebula region because the RGS data does not cover the Rim region.
}
\label{fig:norm-profile}
\end{figure*}

\begin{table}[]
\tbl{Surface brightness of the power-law component.\footnotemark[$*$]}{%
\centering
\begin{tabular}{@{\extracolsep{4pt}}lcccccc}
\hline
Instrument             & \multicolumn{2}{c}{AtomDB 3.0.9} & \multicolumn{2}{c}{SPEXACT 3.03.00} \\
\cline{2-3}\cline{4-5} & Nebula                              & Rim                       & Nebula                               & Rim      \\
\hline
Hitomi/SXS             & $<$3.4                              & $<$1.2                    & $<1.8$                             & $<0.6$ \\
Chandra/ACIS         & $6.3_{-0.5}^{+0.5}$                 & $2.0_{-0.3}^{+0.3}$       & $6.6_{-0.5}^{+0.5}$                & $1.9_{-0.3}^{+0.3}$ \\
\hline
\end{tabular}
}\label{tab:powerlaw}
\begin{tabnote}
\hangindent6pt\noindent
\hbox to6pt{\footnotemark[$*$]\hss}\unskip%
In the unit of 10$^{-16}~$~erg~cm$^{-2}$~s$^{-1}$~arcsec$^{-2}$ (2--10~keV band)
\end{tabnote}\end{table}

Chandra/ACIS and XMM-Newton/RGS observations revealed a multi-temperature structure ranging between 0.5--8.0~keV in the core of the Perseus cluster \citep{2007MNRAS.381.1381S,2016MNRAS.461.2077P}.
Here we use a similar multi-temperature analysis to check the consistency between Hitomi/SXS and these previous measurements. 

We fitted the SXS spectra extracted from the Nebula and Rim regions with a six-temperature CIE model consisting of 0.5~keV, 1~keV, 2~keV, 3~keV, 4~keV and 8~keV components following  \citet{2007MNRAS.381.1381S}.
The temperature of each component was fixed, and the abundance and line-of-sight velocity dispersion were common to all the components.
The power-law component that was found in \citet{2007MNRAS.381.1381S} and interpreted as a possible inverse-Compton emission was also included in our model with a fixed photon index of $\Gamma = 2$.
The spectra and the best-fit models in the Nebula and the Rim regions are shown in the left column of \Fig{fig:fit-multi-t}.
The normalizations we obtained for each temperature were scaled to sum to unity, and the results are plotted in \Fig{fig:norm-profile} as red diamonds.
The profile of the scaled normalizations are very similar between \texttt{AtomDB} and \texttt{SPEXACT}, except for the 8~keV component which is detected with \texttt{SPEXACT} in both the Nebula and Rim regions while only its upper limit was obtained for \texttt{AtomDB}.
The results indicate that the combination of the 3~keV, 4~keV, and 8~keV components approximates the 2CIE model obtained in \S\ref{ana:devided-1t} (roughly 3--4~keV plus 5~keV).

We also reanalyzed the Chandra/ACIS data because the effective area calibration was significantly improved during 2007--2009 \citep{2010A&amp;A...523A..22N} and the atomic codes have been updated since the original work of \citet{2007MNRAS.381.1381S}.
We fitted the spectra of the Nebula and Rim regions with the same six-temperature model as the SXS spectrum.
The abundances and the velocity dispersion were fixed at the value obtained from the SXS analysis because Chandra's  energy resolution is not sufficient to determine these parameters.
The AGN model was not included because we excluded the AGN from the ACIS spectral extraction region. 
When the absorption column density was fixed at $1.38 \times 10^{21}$~cm$^{-2}$, we found a significant excess of the model over the data below 1~keV. 
We therefore allowed the absorption column density to vary to compensate for these residuals.
The best-fit column density is $\sim2.0 \times 10^{21}$~cm$^{-2}$ for both the Nebula and Rim regions.
The fitted spectra are shown in the right column of \Fig{fig:fit-multi-t}.
Large residuals can be seen above 5~keV in the Nebula region.
Fitting these residuals with an additional power-law would require this to have a negative photon index.
Therefore, we suspect these residuals are due to an instrumental effect rather than true astrophysical emission.
The fact that no such residuals are seen in the SXS spectrum supports this inference. 
In addition, we see the wavy residuals in the entire energy band, which is probably due to the systematic uncertainty of the detector responses. 
The scaled normalizations we obtained are plotted in \Fig{fig:norm-profile} as black circles.
The two spectral codes show similar trends, except for the 2~keV component in the Nebula region that is seen with \texttt{SPEXACT} but not with \texttt{AtomDB}.

Compared to the results of \citet{2007MNRAS.381.1381S}, our ACIS analysis shows a similar trend, but $\leq 2$~keV components are not detected (\Fig{fig:norm-profile}).
That is probably because  the analysis of \citet{2007MNRAS.381.1381S} used much smaller regions and could detect the lower temperature component that is concentrated in the cluster core and the filamentary structures.
In the spectra of our analysis, which is taken from a much larger region, the lower temperature components could be smeared out by the dominant higher temperature component.   

\begin{figure}
\begin{center}
\includegraphics[width=8cm]{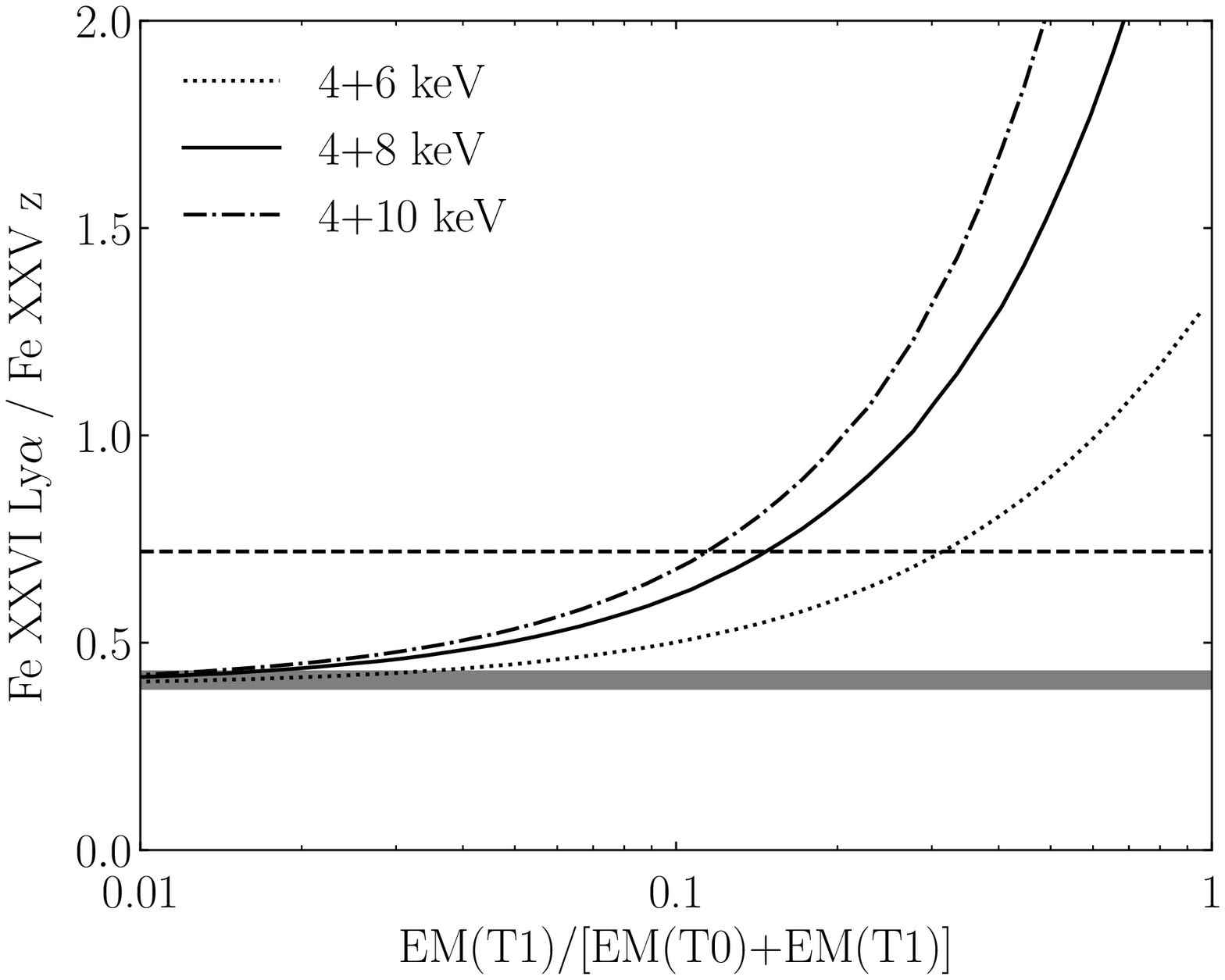}
\end{center}
\caption{
The model line flux ratio of Fe~\emissiontype{XXVI}~Ly$\alpha$ to Fe~\emissiontype{XXV}~z as a function of the fractional emission measure of the secondary component in a two CIE model based on \texttt{AtomDB}. 
The temperature of the main component ($T0$) is 4.0~keV. 
The solid, dotted, and two dashed lines show the model for a second component with a temperature ($T1$) of 8.0, 6.0, and 10.0~keV, respectively. 
The ratio of SXS measurement in the Nebula region (0.41$\pm$0.02) is shown by the gray shaded area, and that of the Chandra model in the same region (0.8) is shown by the horizontal dashed line.
}
\label{fig:felineratio}
\end{figure}

The Hitomi/SXS upper limits of the $\le$2~keV components are consistent with the Chandra/ACIS results.
However, the distribution of the higher temperature components seems somewhat different. The 4~keV component has the highest normalization in the SXS analysis, while the 3~keV component seems dominant in the ACIS fit. 
When the lower end of the energy band for the ACIS analysis was changed to 1.8~keV as same as the SXS analysis, we found that the peak of the normalization ratio became at 4~keV.
It suggests that a derived normalization ratio of the ACIS analysis is affected by the fitted energy band, especially the band of Fe L-shell lines.  
The normalization of the 8~keV component for the SXS is lower than that for the ACIS by a factor of 2--10.
To show sensitivity of the line ratio, Fe~\emissiontype{XXVI}~Ly$\alpha$/Fe~\emissiontype{XXV}~z, to the 8~keV component emission, we calculate the line ratio as a function of its fractional emission measure in \Fig{fig:felineratio}.
The SXS observed line ratio ($\sim$0.4) and the expected line ratio derived from the best-fit ACIS multi-temperature model is also shown in the same figure. 
This indicates that the line flux of Fe~\emissiontype{XXVI}~Ly$\alpha$ primarily limits the hotter component emission.
This SXS spectroscopic constraint is more robust and less dependent on the modeling of the continuum components, compared with previous continuum-based analysis.

Although we employed this particular six-temperature model 
just to examine consistency with the Chandra result 
(\cite{2007MNRAS.381.1381S}, and our own analysis),
we admit that the assumed six temperatures are not necessarily appropriate,
because no emission measure is considered between the temperatures of 4 keV and 8 keV.
This condition is inconsistent with the very likely presence  of 
a component with $\sim 5$~keV temperature in the present SXS spectra,
as indicated in \Tab{tab:fitothers} by the 2T fit to the Nebula and Rim spectra.
In addition, the outer region of Perseus is known to have 
a typical temperature of 6--7~keV (e.g., \cite{2003ApJ...590..225C}),
and such a component must contribute to the SXS spectra at least due to projection.
Given these, we repeated the multi-temperature fitting 
by adding a 7th component with its temperature fixed at 6 keV.
As a result, the Nebula spectrum constrained the normalization (the same as in \Fig{fig:norm-profile})
of this 6~keV component as $< 0.05$ with \texttt{AtomDB} and $< 0.13$ with \texttt{SPEXACT},
which are lower than those for the 4~keV emission.
At the same time, the SXS normalization of the 8~keV component is $<0.03$ with \texttt{AtomDB} and $<0.07$ with \texttt{SPEXACT}, and is consistent with the six-temperature results.
Therefore, the additional 6~keV component has no significant effect on the normalizations of the other temperature components.

The fluxes of the additional power-law component are shown in \Tab{tab:powerlaw}.
The SXS detected no significant power-law component, while the ACIS data clearly require it in both the Nebula and Rim regions.
These differences are discussed in \S\ref{dis:power-law}.

The RGS data covers the energy band below 2~keV with high spectral resolution, and is complementary to the Hitomi/SXS data. Indeed, a very low temperature component with $kT<1$~keV was reported from the XMM-Newton/RGS observations \citep{2016MNRAS.461.2077P}. Here, we fitted the RGS spectrum with a three-temperature plasma model by adding a fixed-temperature 4~keV component to the two-temperature model used in \citet{2016MNRAS.461.2077P}.
For the RGS analysis, we have used the SPEX fitting package, because accounting for the line broadening due to the spatial extent of the source is not easily implemented in Xspec. A \texttt{user} model that calls Xspec externally and returns the model calculation to SPEX is used to implement fitting the RGS data with \texttt{AtomDB}.

The obtained best-fit temperatures are $0.60_{-0.02}^{+0.02}$~keV and $2.7_{-0.05}^{+0.08}$~keV for \texttt{AtomDB} and $0.55_{-0.06}^{+0.03}$~keV and $2.4_{-0.10}^{+0.08}$~keV for \texttt{SPEXACT}. The relative normalizations of each component are over-plotted on \Fig{fig:norm-profile}.
The profile peaks at $kT\sim2.5$~keV, lower than both Chandra and Hitomi, and gives a significant detection of gas with $kT\sim0.6$~keV. The normalization of this low temperature component measured with RGS is consistent with the Hitomi, but not with the Chandra upper limits for the 0.5~keV gas included in the six-temperature model; the upper limits for the 4~keV gas in both RGS and Chandra are lower than the Hitomi measurement for this temperature.

A simultaneous fitting of the SXS, ACIS, and RGS might provide a more complete picture of the temperature distribution. 
However, cross-instruments issues, such as the different spectral extraction regions and cross calibration of the effective areas,
require more detailed analysis on the systematic errors. We therefore consider such analysis as a future work.

\section{Discussion}
\subsection{Origin of the Deviations from a Single-temperature Model}
\begin{figure*}
\begin{center}
\subfloat{\includegraphics[width=8cm]{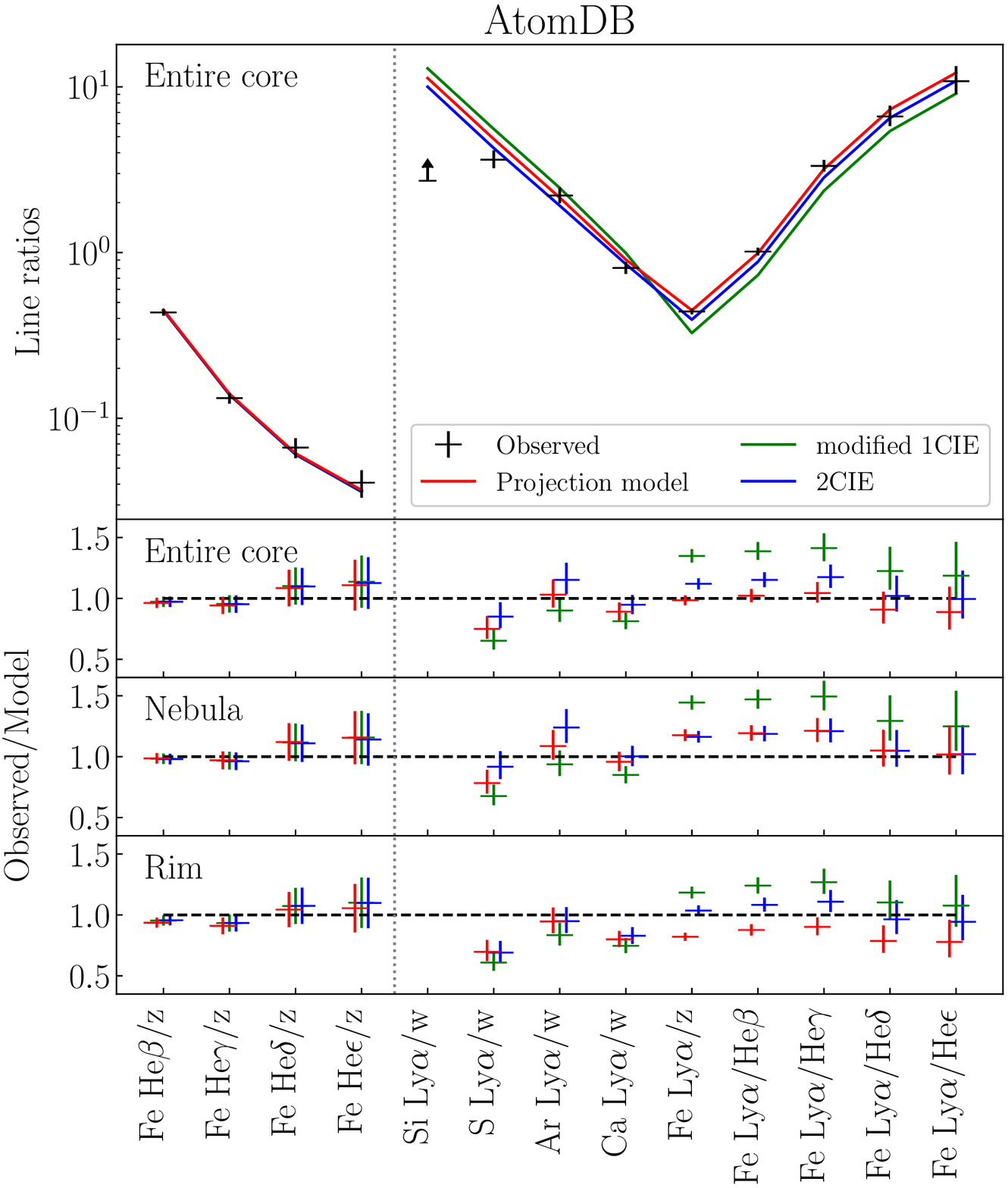}}
\hspace{0.1cm}
\subfloat{\includegraphics[width=8cm]{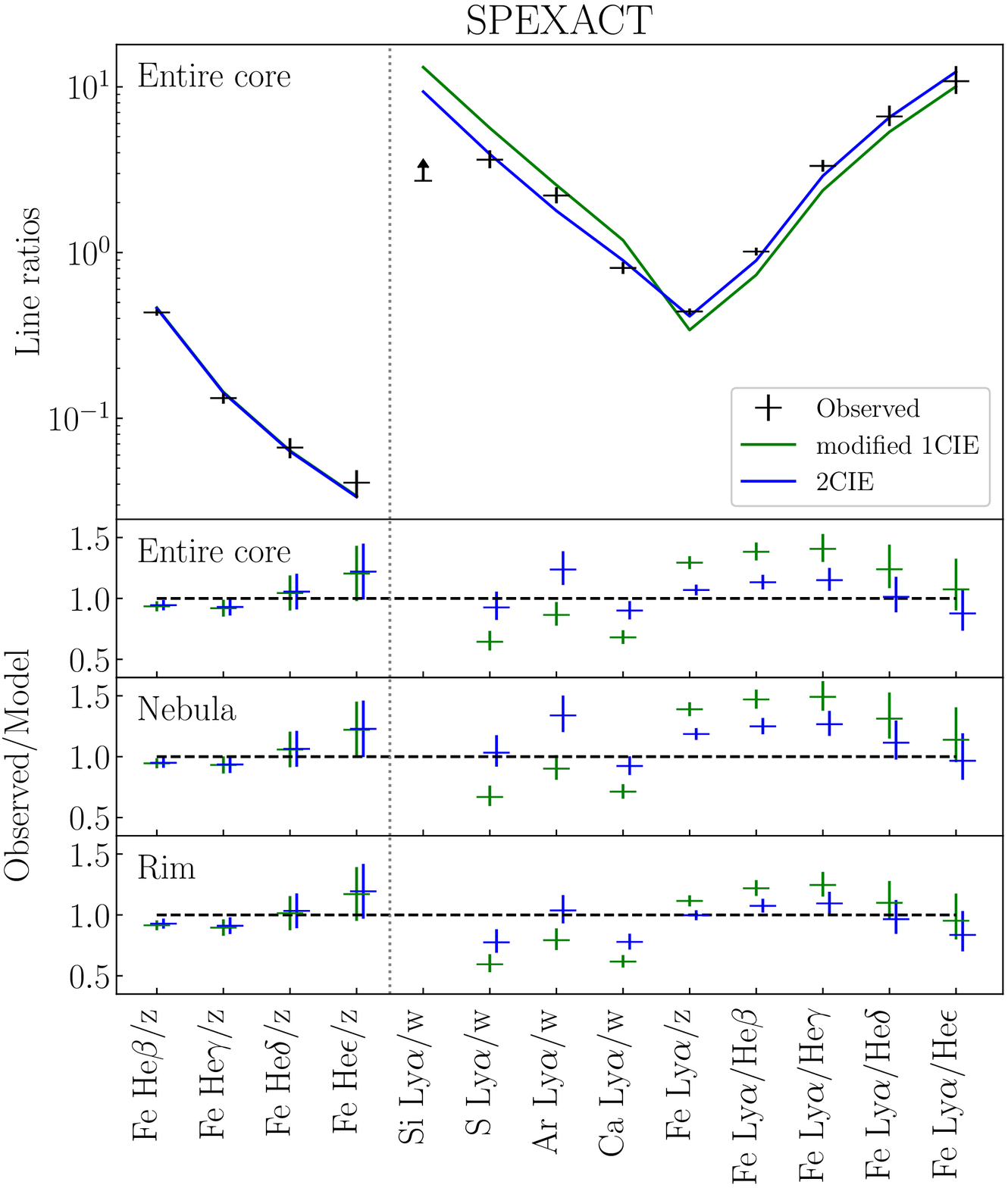}}
\end{center}
\caption{Comparison between the observed line ratios (black crosses),  the modified 1CIE model (green line), the 2CIE model (blue line), and the projection model (red line).
The left and right figures show the calculations based on \texttt{AtomDB} and \texttt{SPEXACT}, respectively. 
The projection model is not shown in the right (\texttt{SPEXACT}) figure.
The lower sub-panels show the ratios of the observation to the model in the Entire core, Nebula, and Rim regions from top thorough bottom.
The vertical dotted line in the panels separates the ratios measuring the excitation temperature (left side) and the ionization temperature (right side).
}
\label{fig:compare-line-ratio}
\end{figure*}
The line ratio diagnostics presented in \S\ref{ana:line} show that, with the exception of the Outer region, the derived ionization temperatures are different for each element (\Fig{fig:line-temperatures}), and indicate multi-temperature structure.

In \S\ref{ana:entire}, we modeled the spectrum of the Entire core, Nebula, and Rim regions with single- and two-temperature plasma models. In \Fig{fig:compare-line-ratio}, we compare the observed line ratios and the line ratios predicted by the modified 1CIE and the 2CIE models, in order to investigate how these model approximations are able to reproduce the observations, and where the biggest discrepancies are found.
This figure includes not only the line ratios that allow us to estimate the ionization temperatures, but also those sensitive to the excitation temperatures (Fe He$\beta$/z, He$\gamma$/z, He$\delta$/z, and He$\epsilon$/z).
As expected from the C-statistics shown in \Tab{tab:fit-entirecore} and \Tab{tab:fitothers}, the line ratios of the 2CIE models are closer to the observed ones than those of the modified 1CIE models in all the region and in both \texttt{AtomDB} and \texttt{SPEXACT}.
We then calculate chi-squared values ($\chi^2$) of the 2CIE models with respect to the observed line ratios in the Entire core region.
The results are 22.0 and 19.6 in \texttt{AtomDB} and \texttt{SPEXACT}, respectively for 12 considered line ratio measurements.
The major line ratios are reproduced better by the 2CIE model of \texttt{SPEXACT}, even though the broad-band fitting with the 2CIE model gives larger C-statistics in \texttt{SPEXACT} than \texttt{AtomDB} (\Tab{tab:fit-entirecore}).

One of the physical origins of the multi-temperature structure is the projection effect; the radial temperature gradient from the core to the outskirts is accumulated along the line of sight.
In order to check this possibility, we used the azimuthally-averaged radial profiles of the temperature, density, and abundances derived from the de-projection analysis of the Chandra data using \texttt{AtomDB} (see Figure~7 in the RS~paper).
In this radial profile model, the temperatures vary from 3~keV to 6.5~keV with increasing radius from the cluster center in the range of 3--1000~kpc. 
We integrated this model along the line of sight and calculated the model line ratios shown in red in \Fig{fig:compare-line-ratio} (left).
The projection model of \texttt{SPEXACT} is not shown because the radial profile is derived based on \texttt{AtomDB}.
In the Entire core region, $\chi^2$ of the projection model is 13.0, and is considerably better than that of the 2CIE model ($\chi^2=22.0$).
On the other hand, in the Nebula and Rim regions, the projection model is almost the same or even worse compared to the 2CIE model.
Azimuthal variation in temperature is probably significant in the Nebula and Rim regions as observed with Chandra (e.g., \cite{2011MNRAS.418.2154F}), and this likely causes the difference from the projection model that is based on the azimuthally averaged radial profile.

A configuration in which the two temperature components determined from the 2CIE model are truly co-spatial cannot be ruled out.
However, the derived temperatures in the 2CIE model ($\sim3$~keV plus $\sim5$~keV) are very close to the temperatures observed in the cluster center and outer.
Therefore, the projection effect naturally explains the observed trend of lower ionization temperatures for ions with lower atomic number.
As the equivalent widths of spectral lines from lighter elements are generally larger for lower temperatures, in the presence of a radial gradient along the line of sight, the emission-measure weighted average fluxes of these lines will naturally be biased towards the cooler gas, while the emission-measure weighted average fluxes of elements with higher atomic numbers will be biased towards values more typical of the hotter gas.

The low-temperature gas components ($kT<3$~keV) reported in previous Chandra and XMM-Newton observations are not seen in \Fig{fig:compare-line-ratio}.
That is simply because the Hitomi/SXS energy band is restricted to energies above 1.8 keV and so is not sensitive to such low-temperature components; the derived upper limits from Hitomi for these thermal phases are not in conflict with previous results. 

\subsection{Uncertainties in Modeling the Multi-temperature Plasma}
\label{discussion:multi-t}

In \S\ref{ana:multi-t}, we compared the Hitomi/SXS results with the Chandra/ACIS and the XMM-Newton/RGS results.
The best-fit emission measure distribution as a function of temperature is  different among instruments (\Fig{fig:norm-profile}): the normalization peaks at temperatures of 4~keV, 3~keV, and 2~keV for the SXS, ACIS, and RGS, respectively.
The uncertainty of the detector response also affects the results, as shown by the discrepancy of the 8~keV component in the Hitomi/SXS and Chandra/ACIS.

Even in the single- or two-temperature model, the best-fit parameters are sensitive to the effective area calibration as demonstrated in \Fig{fig:compare-arf-1cie}, \Fig{fig:compare-arf-2cie}, and Appendix~\ref{appendix:arf}.
In the single-temperature modeling, we can robustly determine temperatures using only the line fluxes.
However, in the two-temperature modeling, it is difficult to determine both temperatures and normalizations exactly. 

Furthermore, we found that a small change in the atomic code significantly affects the result.
As shown in Appendix~\ref{appendix:atomdb}, the \texttt{AtomDB}~3.0.8 gives the temperatures of 1.7~keV and 4.1~keV in the 2CIE model, which is completely different from the results based on \texttt{AtomDB}~3.0.9.
The difference between the two codes is only the emissivity of the dielectric-recombination satellite lines that is significantly lower than that of the transitions in the He-like ions (w, x, y, and z). 

Therefore, we demonstrated that quantifying deviations from a single temperature model is a complex problem.
As \Fig{fig:fit-1cie} shows, the spectral differences between a two-temperature model consisting of a mixture of 3 and 5 keV plasma and a single-temperature model with $kT=4$~keV are very small, and thus the results of the 2CIE fit are sensitive to a large number of factors. 
These factors include the analysis energy band, the energy resolution, the calibration of the effective area, and atomic codes. 
For accurate analysis of the multi-temperature, non-dispersive high-resolution spectroscopy, a broad spectral band (0.5--10 keV), as will be achieved by XARM and Athena, is necessary.

\subsection{Upper Limit for the power-law component}
\label{dis:power-law}
The diffuse radio emission is thought to be generated by Synchrotron mechanism of relativistic energy electrons with a 0.1--10~$\mu$G magnetic field in the ICM~\citep{2014IJMPD..2330007B}. 
These electrons scatter the CMB photons via the inverse-Compton scattering, which allows us to investigate the magnetic field in the ICM~\citep{2012RAA....12..973O}. Based on an assumption that inverse-Compton emission is generated by the same population of
relativistic electrons, the volume-integrated magnetic field strength can be derived from intensities/upper-limits of IC
emission~\citep{1979rpa..book.....R}.

\citet{2007MNRAS.381.1381S} reported the detection of a diffuse power-law component in the core of the Perseus cluster, which was not confirmed by the XMM-Newton analysis \citep{2009A&amp;A...493...13M}.
The corresponding surface brightness in the 2--10~keV band measured by \citet{2007MNRAS.381.1381S} is $\sim15\times10^{-16}$ and $\sim8\times10^{-16}$~erg~cm$^{-2}$~s$^{-1}$~arcsec$^{-2}$ in Nebula and Rim, respectively.
Our reanalysis of the Chandra/ACIS data also suggests the presence of such a power-law component, but with observed fluxes lower by a factor of 2--4 in better agreement with the upper limit of $5\times10^{-16}$~erg~cm$^{-2}$~s$^{-1}$~arcsec$^{-2}$ reported by \cite{2009A&amp;A...493...13M}.
That is likely caused by the update of the calibration database described in \citet{2010A&amp;A...523A..22N}; the response for the higher energy band is improved and significantly reduces the flux in that energy band.

In contrast, the Hitomi/SXS results show upper limits for this power-law component that are significantly lower than the fluxes measured with Chandra/ACIS..
As shown in \Fig{fig:fit-multi-t}, large systematic residuals are present above 5~keV in the Chandra spectra even after the update of the effective area calibration, and they likely bias the power-law fluxes. A similar discrepancy was also reported in the comparison with the XMM-Newton/EPIC results \citep{2009A&amp;A...493...13M}.
Since the Hitomi/SXS covers a wider energy range up to 20~keV, the obtained upper limits would be robust at least in the Rim region, in which the level of the AGN contamination is low. 

Assuming the power-law component is due to inverse-Compton scattering, we can estimate the strength of the magnetic field ($B$) as discussed in \citet{2005MNRAS.360..133S}.
Using the SXS upper limit of $<1.2\times10^{-16}$~erg~cm$^{-2}$~s$^{-1}$~arcsec$^{2}$ in the Rim region, we obtained the lower limit of $B>0.4$~$\mu$G.
This value is consistent with the results of other observations performed at other wavelengths ($\sim$7--25~$\mu$G) as discussed in \citet{2009A&amp;A...493...13M}.

\section{Conclusion}

Compared to the intricate structures revealed by the deep Chandra image of the core of the Perseus Cluster (e.g, \cite{2011MNRAS.418.2154F}), at first glance the high-quality Hitomi SXS spectra of this source, which are sensitive to the temperature range of $\gtrsim 3$~keV, present a surprisingly quiescent view: the velocity dispersions are rather small (the First~paper, the V~paper), the chemical composition is remarkably similar to the solar neighborhood (the Z~paper), and the spectra between 1.8--20 keV are largely well approximated by a single temperature model. The diffuse power-law component reported from previous Chandra measurements is also not required by the Hitomi data.


We have resolved line emission from various ions.
This provides the first direct measurements of the electron temperature and ionization degree separately from different transitions of He-like and H-like ions of Si, S, Ar, Ca, and Fe.
Compared with previous temperature measurements mostly based on the continuum shape, the new diagnostics are more sensitive to excitation processes and plasma conditions.
We found that all observed ratios are broadly consistent with the CIE approximation.
However, there are two signs of small deviation from a single temperature model.
Firstly there is a trend of increasing ionization temperature with increasing atomic mass, 
particularly in the Nebula (central) region and possibly also in the surrounding Rim region.
Secondly, the excitation temperature from Fe ($\sim 3$~keV) is lower than  the corresponding ionization temperature and than the electron temperature determined from the spectral continuum ($\sim 4$~keV) for the Nebula.
In the Nebula and Rim regions, the best-fit two temperature models suggest a mix of roughly 3 and 5 keV plasma, both of which are expected to be present based on deprojected temperature profiles previously measured with Chandra.
On the other hand, the Outer region, corresponding to the farthest observation from the cluster core performed by Hitomi, shows no significant deviation from single temperature.
No additional third temperature component, Gaussian nor power-law DEM model, nor significant emission from non-equilibrium ionization plasma are required to describe the spectra.  

Even though we can not rule out a true multi-phase structure in which different temperature components are co-spatial, the projection effect is a natural explanation for the observed deviations from single temperature.

Best-fit models of lower-resolution spectra that include the energy band below 2 keV and the RGS spectra seem to present a contrasting picture, requiring a multi-phase thermal structure that the Hitomi observations are currently not sensitive to. 
It is clear that the dominant thermal component in the spectral fit depends on the energy band observed, and that detectors able to cover simultaneously the emission lines from all phases of the ICM are needed in order for a reliable temperature structure to be pinned down. High-resolution, non-dispersive spectroscopy with XARM or Athena will thus be crucial in order to assess the origins and robustness of the multi-temperature structure reported by CCD studies, and verify to what extent the complexity of cluster cores revealed by high-spatial resolution images corresponds to an equally complex picture along the energy axis.

\begin{contribution}
S. Nakashima led this study in data analysis and wrote the manuscript.
K. Matsushita, A. Simionescu, and T. Tamura reviewed the manuscript fully.
K. Sato and T. Tamura performed a cross-check of the Hitomi/SXS analysis.
Y. Kato analyzed the Chandra/ACIS archival data.
A. Simionescu and N. Werner performed the XMM-Newton/RGS analysis.
M. Furukawa and K. Sato constructed the projection model and checked the effect of the resonance scattering. 
M. Bautz, H. Akamatsu, M. Tsujimoto, H. Yamaguchi, K. Makishima, C. Pinto, Y. Fukazawa, R. Mushotzky, and J. de Plaa provided valuable comments that improved the draft.
S. Nakashima, K, Sato, K. Nakazawa, T. Okajima, and N. Yamasaki contributed to fabrication of the instruments and performed the in-orbit operation and calibration.
\end{contribution}

\begin{ack}
We are thankful for the support from the JSPS Core-to-Core Program.
We acknowledge all the JAXA members who have contributed to the ASTRO-H (Hitomi)
project.
All U.S. members gratefully acknowledge support through the NASA Science Mission
Directorate. Stanford and SLAC members acknowledge support via DoE contract to SLAC
National Accelerator Laboratory DE-AC3-76SF00515. Part of this work was performed under
the auspices of the U.S. DoE by LLNL under Contract DE-AC52-07NA27344.
Support from the European Space Agency is gratefully acknowledged.
French members acknowledge support from CNES, the Centre National d'\'{E}tudes Spatiales.
SRON is supported by NWO, the Netherlands Organization for Scientific Research.  Swiss
team acknowledges support of the Swiss Secretariat for Education, Research and
Innovation (SERI).
The Canadian Space Agency is acknowledged for the support of Canadian members.  
We acknowledge support from JSPS/MEXT KAKENHI grant numbers 15J02737,
15H00773, 15H00785, 15H02090, 15H03639, 15H05438, 15K05107, 15K17610,
15K17657, 16J00548, 16J02333, 16H00949, 16H06342, 16K05295, 16K05296,
16K05300, 16K13787, 16K17672, 16K17673, 21659292, 23340055, 23340071,
23540280, 24105007, 24244014, 24540232, 25105516, 25109004, 25247028,
25287042, 25400236, 25800119, 26109506, 26220703, 26400228, 26610047,
26800102, JP15H02070, JP15H03641, JP15H03642, JP15H06896,
JP16H03983, JP16K05296, JP16K05309, JP16K17667, and JP16K05296.
The following NASA grants are acknowledged: NNX15AC76G, NNX15AE16G, NNX15AK71G,
NNX15AU54G, NNX15AW94G, and NNG15PP48P to Eureka Scientific.
H. Akamatsu acknowledges support of NWO via Veni grant.  
C. Done acknowledges STFC funding under grant ST/L00075X/1.  
A. Fabian and C. Pinto acknowledge ERC Advanced Grant 340442.
P. Gandhi acknowledges JAXA International Top Young Fellowship and UK Science and
Technology Funding Council (STFC) grant ST/J003697/2. 
Y. Ichinohe, K. Nobukawa, and H. Seta are supported by the Research Fellow of JSPS for Young
Scientists.
N. Kawai is supported by the Grant-in-Aid for Scientific Research on Innovative Areas
``New Developments in Astrophysics Through Multi-Messenger Observations of Gravitational
Wave Sources''.
S. Kitamoto is partially supported by the MEXT Supported Program for the Strategic
Research Foundation at Private Universities, 2014-2018.
B. McNamara and S. Safi-Harb acknowledge support from NSERC.
T. Dotani, T. Takahashi, T. Tamagawa, M. Tsujimoto and Y. Uchiyama acknowledge support
from the Grant-in-Aid for Scientific Research on Innovative Areas ``Nuclear Matter in
Neutron Stars Investigated by Experiments and Astronomical Observations''.
N. Werner is supported by the Lend\"ulet LP2016-11 grant from the Hungarian Academy of
Sciences.
D. Wilkins is supported by NASA through Einstein Fellowship grant number PF6-170160,
awarded by the Chandra X-ray Center, operated by the Smithsonian Astrophysical
Observatory for NASA under contract NAS8-03060.

We are grateful for contributions by many companies, including in particular, NEC, Mitsubishi Heavy
Industries, Sumitomo Heavy Industries, and Japan Aviation Electronics Industry. Finally,
we acknowledge strong support from the following engineers.  JAXA/ISAS: Chris Baluta,
Nobutaka Bando, Atsushi Harayama, Kazuyuki Hirose, Kosei Ishimura, Naoko Iwata, Taro
Kawano, Shigeo Kawasaki, Kenji Minesugi, Chikara Natsukari, Hiroyuki Ogawa, Mina Ogawa,
Masayuki Ohta, Tsuyoshi Okazaki, Shin-ichiro Sakai, Yasuko Shibano, Maki Shida, Takanobu
Shimada, Atsushi Wada, Takahiro Yamada; JAXA/TKSC: Atsushi Okamoto, Yoichi Sato, Keisuke
Shinozaki, Hiroyuki Sugita; Chubu U: Yoshiharu Namba; Ehime U: Keiji Ogi; Kochi U of
Technology: Tatsuro Kosaka; Miyazaki U: Yusuke Nishioka; Nagoya U: Housei Nagano;
NASA/GSFC: Thomas Bialas, Kevin Boyce, Edgar Canavan, Michael DiPirro, Mark Kimball,
Candace Masters, Daniel Mcguinness, Joseph Miko, Theodore Muench, James Pontius, Peter
Shirron, Cynthia Simmons, Gary Sneiderman, Tomomi Watanabe; ADNET Systems: Michael
Witthoeft, Kristin Rutkowski, Robert S. Hill, Joseph Eggen; Wyle Information Systems:
Andrew Sargent, Michael Dutka; Noqsi Aerospace Ltd: John Doty; Stanford U/KIPAC: Makoto
Asai, Kirk Gilmore; ESA (Netherlands): Chris Jewell; SRON: Daniel Haas, Martin Frericks,
Philippe Laubert, Paul Lowes; U of Geneva: Philipp Azzarello; CSA: Alex Koujelev, Franco
Moroso.

\end{ack}

\bibliographystyle{aasjournal}
\bibliography{bibdesk}


\appendix 
\section{Gain Correction}
\label{appendix:gain-correction}
We checked uncertainties of the gain scale of the SXS and corrected them using the Perseus data themselves, because the gain scale calibration is limited due to the short life of Hitomi.
The procedure described in this section is essentially the same as that used in the Z paper and Atomic paper, except that the reference redshift was changed from 0.01756 to 0.017284 according to the V paper.

We first applied a linear gain shift for each pixel in each observation so that the apparent energies of the Fe \emissiontype{XXV} lines agree with a redshift of 0.017284.
The resulting amount of the energy shift at 6.5~keV is 1.0$\pm$1.9~eV (mean and standard deviation). 
This pixel-by-pixel redshift correction removes not only the remaining gain errors among pixels but also the spatial variation of the Doppler shift for the ICM.
Our results are not affected by this possible ``over-correction''.

\begin{figure*}
\begin{center}
\subfloat[obs1]{\includegraphics[width=7.5cm]{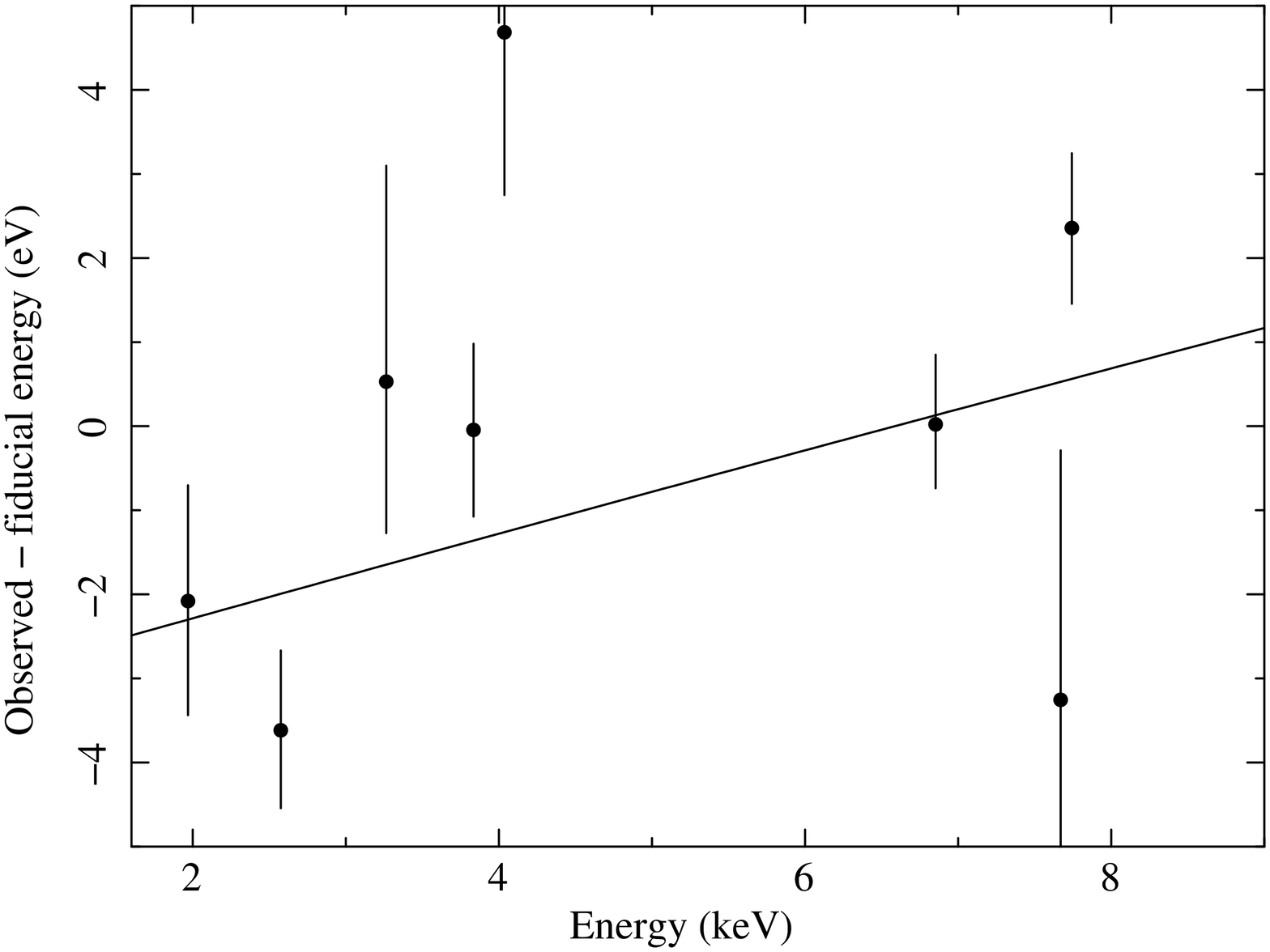}}
\hspace{1.0cm}
\subfloat[obs2]{\includegraphics[width=7.5cm]{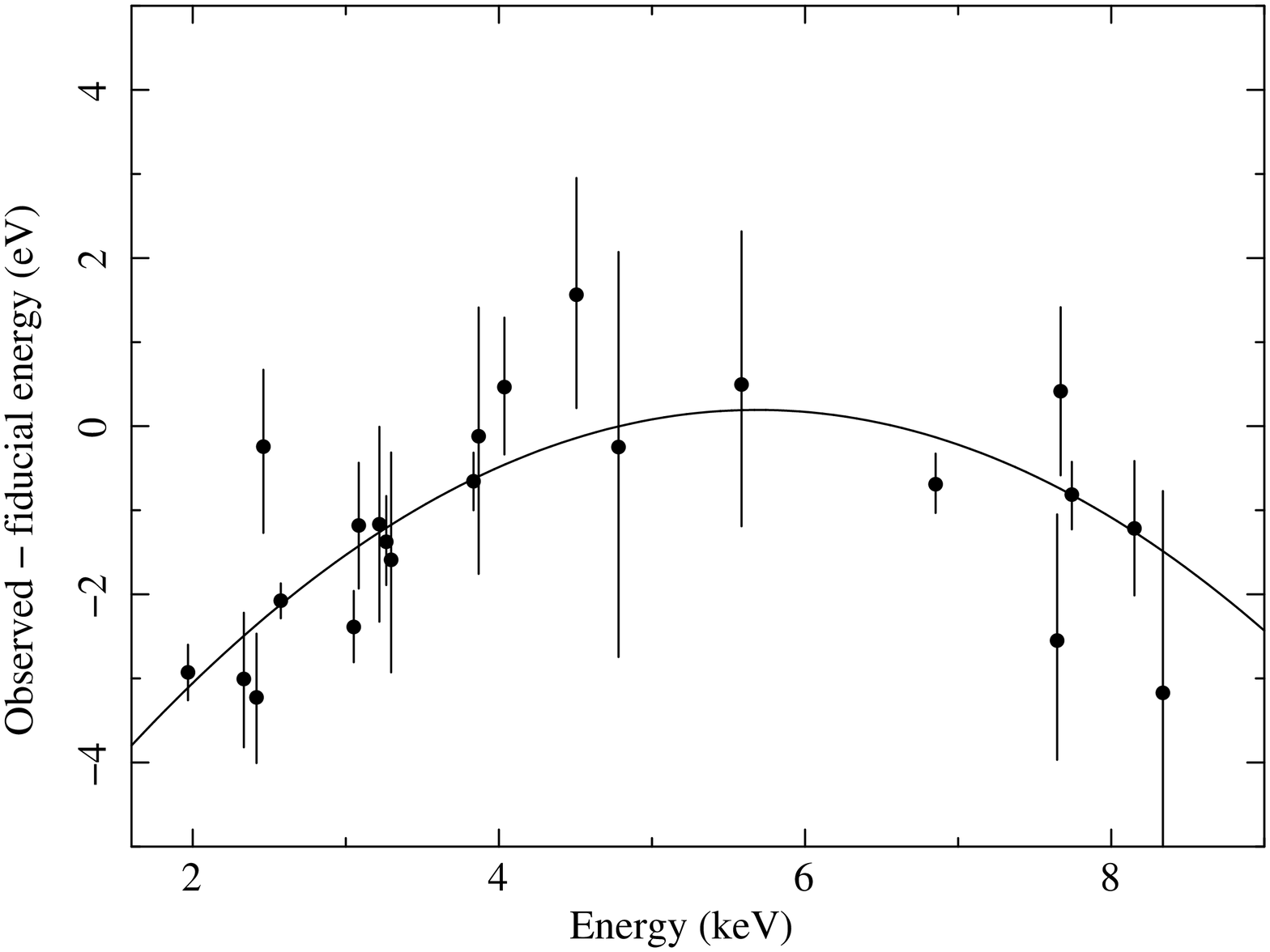}}
\vspace{0.5cm}
\subfloat[obs3]{\includegraphics[width=7.5cm]{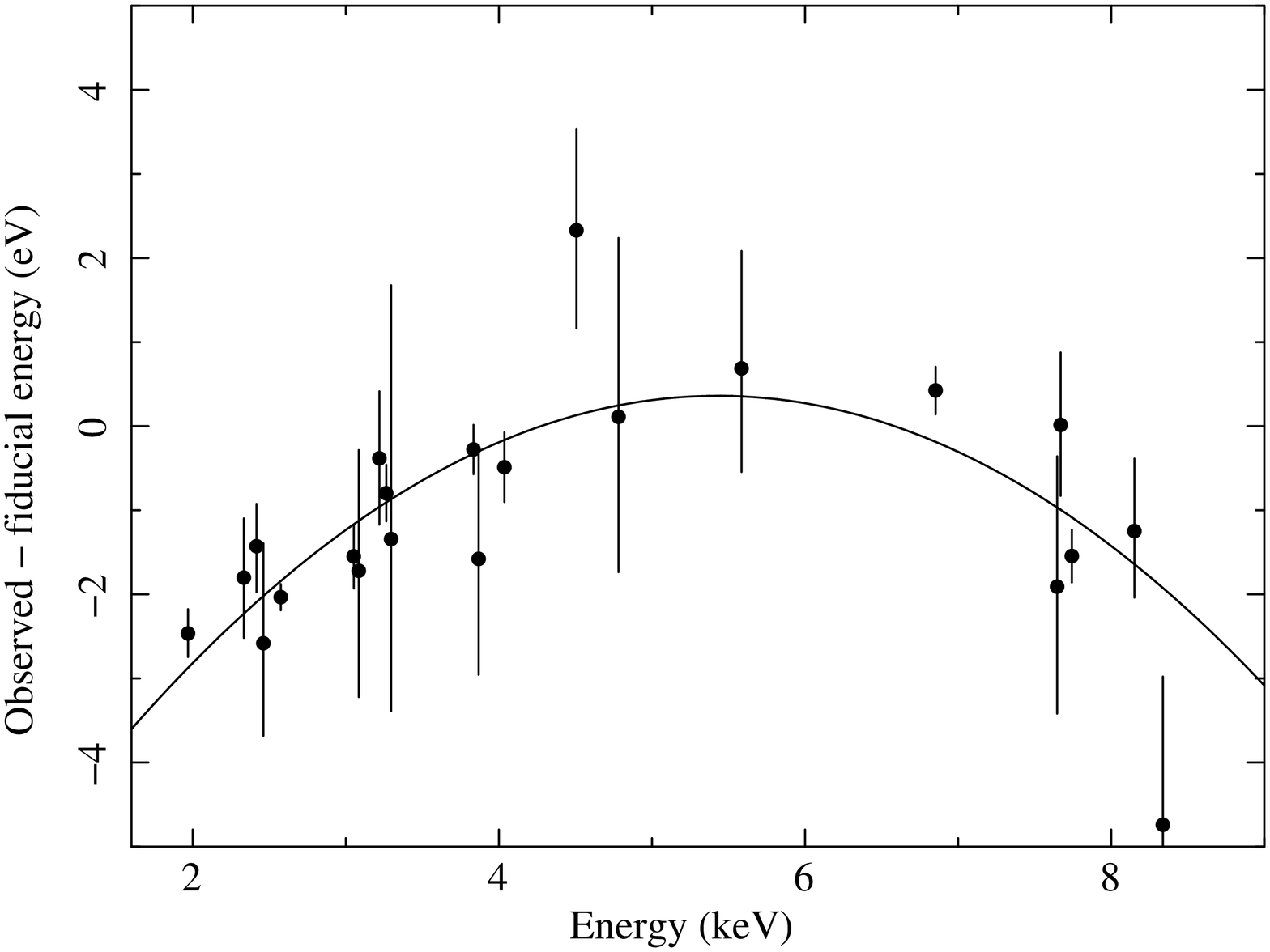}}
\hspace{1.0cm}
\subfloat[obs4]{\includegraphics[width=7.5cm]{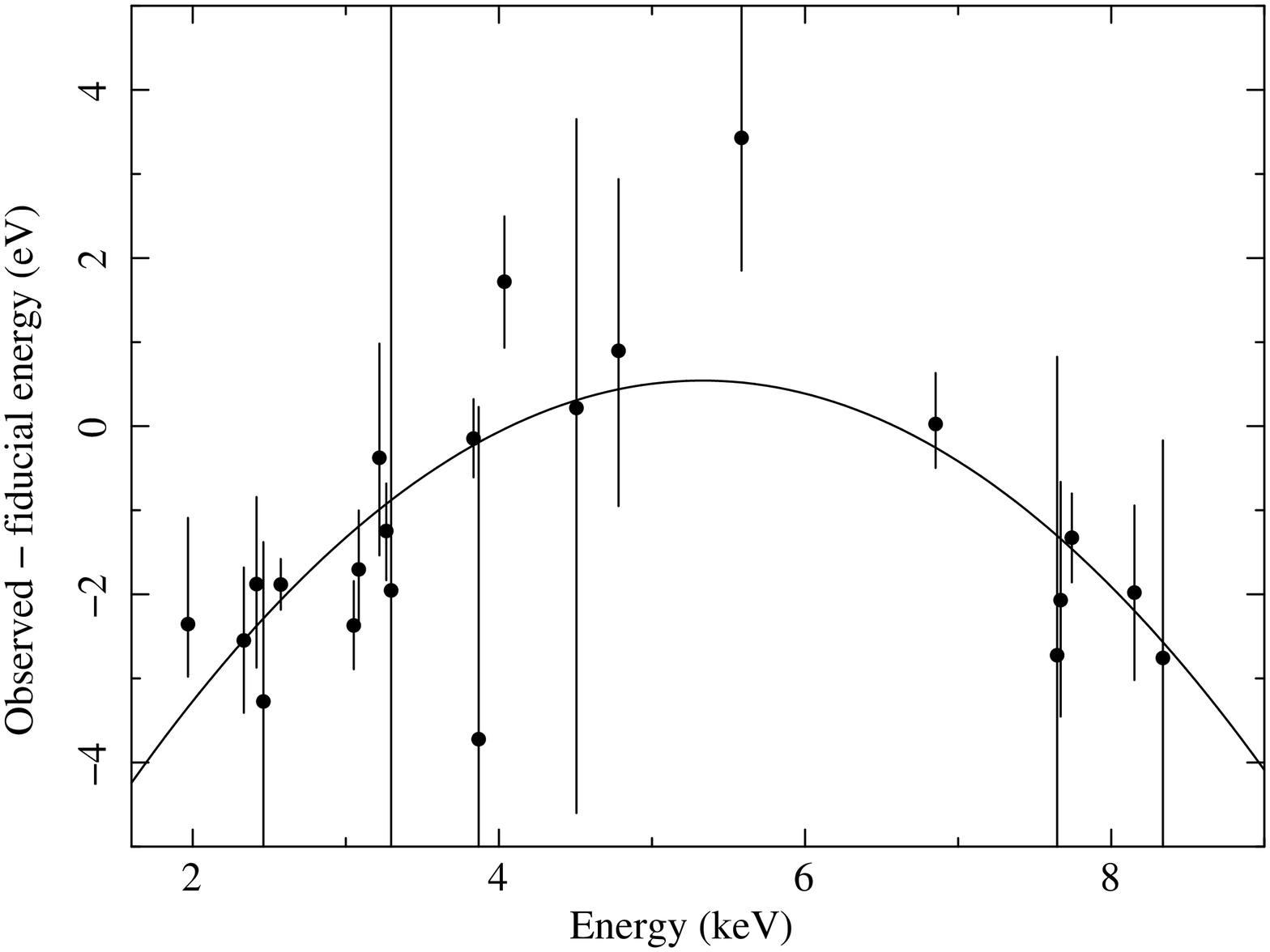}}
\end{center}
\caption{Parabolic functions for the gain correction in each observation. }
\label{fig:gain-correction}
\end{figure*}

We then co-added spectra of all the pixels for each observation and investigated the energy shifts of each line in the 1.8--9.0 keV band.
\Fig{fig:gain-correction} summarizes the differences of line energies from the fiducial values assuming a redshift of 0.017284. 
We modeled these data by the parabolic function shown below, in which $\Delta E$ at $E=6586.5$~eV was constrained to zero:
\begin{eqnarray}
E_{\rm cor} = E + a \times(E-6586.5) + b \times(E-6586.5)^{2} {\rm\; eV}.
\end{eqnarray}
The obtained parameters are 
\begin{eqnarray}
a &= - 4.888\times10^{-4} &{\rm \;and\;}  b = 2.035\times10^{-6}, \\
a &= 4.303\times10^{-4}  &{\rm \;and\;} b = 2.390\times10^{-4}, \\
a &= 6.250\times10^{-4} &{\rm \;and\;} b = 2.702\times10^{-4}, \\
a &= 8.640\times10^{-4} &{\rm \;and\;} b = 3.439\times10^{-4}, 
\end{eqnarray}
for obs1, obs2, obs3, and obs4, respectively.
We applied these corrections to all the pixels.
As confirmed in Figure 6 of the RS~paper, these gain corrections have no impact on the line flux measurement, which is crucial for the temperature measurements. 

\section{Detailed best-fit parameters of the Gaussian fits}
\label{appendix:gaus-parameters}
The centers and widths derived from the Gaussian fits in \S\ref{ana:line} are shown in \Tab{tab:linedata2}.
The obtained line widths are consistent with the results described in V paper.

\begin{table*}[]
\tbl{Observed line centers and widths derived from the Gaussian fits..\footnotemark[$*$]}{%
\centering\scriptsize
\begin{tabular}{llcccclcccc}
\hline
\multicolumn{1}{c}{Line name}          & \multicolumn{1}{c}{$E_0$~(eV)\footnotemark[$\dagger$]} & \multicolumn{4}{c}{Center (eV)} & ~ & \multicolumn{4}{c}{Width (eV)}  \\
\cline{3-6}\cline{8-11}                & ~                                                      & \multicolumn{1}{c}{Entire Core} & \multicolumn{1}{c}{Nebula} & \multicolumn{1}{c}{Rim}     & \multicolumn{1}{c}{Outer}   & ~ & \multicolumn{1}{c}{Entire Core} & \multicolumn{1}{c}{Nebula} & \multicolumn{1}{c}{Rim}    & \multicolumn{1}{c}{Outer}  \\
\hline
Si \emissiontype{XIII} w               & 1865.0                                                 & 1864.5$^{+1.1}_{-1.0}$          & 1864.3$^{+1.9}_{-0.9}$     & \multicolumn{1}{c}{(fixed)} & \multicolumn{1}{c}{(fixed)} & ~ & \multicolumn{1}{c}{$<$1.7}      & \multicolumn{1}{c}{$<$1.7} & \multicolumn{1}{c}{(tied)} & \multicolumn{1}{c}{(tied)} \\
Si \emissiontype{XIV} Ly$\alpha_{1}$   & 2006.1                                                 & 2006.4$^{+0.3}_{-0.2}$          & 2006.5$^{+0.3}_{-0.2}$     & 2006.4$^{+0.4}_{-0.3}$      & 2006.5$^{+1.2}_{-1.3}$      & ~ & 1.4$^{+0.3}_{-0.3}$             & \multicolumn{1}{c}{$<$1.5} & 1.7$^{+0.5}_{-0.5}$        & \multicolumn{1}{c}{$<$2.9} \\
Si \emissiontype{XIV} Ly$\beta_{1}$    & 2376.6                                                 & 2376.5$^{+0.5}_{-0.5}$          & 2376.2$^{+0.5}_{-0.5}$     & 2377.8$^{+1.3}_{-1.4}$      & 2378.5$^{+1.1}_{-1.3}$      & ~ & 2.4$^{+0.5}_{-0.5}$             & 2.2$^{+0.5}_{-0.5}$        & 3.8$^{+1.3}_{-1.2}$        & \multicolumn{1}{c}{$<$2.0} \\
S \emissiontype{XV} w                  & 2460.6                                                 & 2460.7$^{+0.5}_{-0.3}$          & 2460.6$^{+0.6}_{-0.4}$     & 2460.7$^{+0.8}_{-0.6}$      & \multicolumn{1}{c}{(fixed)} & ~ & 2.2$^{+0.6}_{-0.6}$             & 2.6$^{+0.8}_{-0.7}$        & \multicolumn{1}{c}{$<$2.2} & \multicolumn{1}{c}{(tied)} \\
S \emissiontype{XVI} Ly$\alpha_{1}$    & 2622.7                                                 & 2622.6$^{+0.1}_{-0.1}$          & 2622.7$^{+0.1}_{-0.2}$     & 2622.6$^{+0.2}_{-0.2}$      & 2621.6$^{+0.9}_{-0.9}$      & ~ & 1.8$^{+0.2}_{-0.2}$             & 1.9$^{+0.2}_{-0.2}$        & 1.7$^{+0.3}_{-0.3}$        & 2.0$^{+1.0}_{-1.2}$        \\
S \emissiontype{XVI} Ly$\beta_{1}$     & 3106.7                                                 & 3106.4$^{+0.3}_{-0.3}$          & 3106.0$^{+0.4}_{-0.4}$     & 3107.0$^{+0.3}_{-0.5}$      & 3116.1$^{+0.7}_{-1.3}$      & ~ & 2.1$^{+0.4}_{-0.3}$             & 2.4$^{+0.5}_{-0.4}$        & 1.8$^{+0.5}_{-0.7}$        & \multicolumn{1}{c}{$<$1.8} \\
S \emissiontype{XVI} Ly$\gamma_{1}$    & 3276.3                                                 & 3276.8$^{+0.3}_{-0.8}$          & 3276.7$^{+0.5}_{-0.8}$     & 3276.3$^{+1.0}_{-0.8}$      & 3275.7$^{+1.4}_{-1.4}$      & ~ & \multicolumn{1}{c}{$<$2.0}      & \multicolumn{1}{c}{$<$2.1} & 1.8$^{+1.2}_{-1.4}$        & \multicolumn{1}{c}{$<$2.4} \\
Ar \emissiontype{XVII} w               & 3139.6                                                 & 3139.5$^{+0.3}_{-0.3}$          & 3139.5$^{+0.3}_{-0.4}$     & 3139.3$^{+1.2}_{-1.1}$      & 3142.4$^{+0.8}_{-1.0}$      & ~ & 1.5$^{+0.4}_{-0.4}$             & 1.1$^{+0.5}_{-0.6}$        & 3.0$^{+1.4}_{-1.1}$        & \multicolumn{1}{c}{$<$1.9} \\
Ar \emissiontype{XVIII} Ly$\alpha_{1}$ & 3323.0                                                 & 3322.9$^{+0.2}_{-0.3}$          & 3323.1$^{+0.3}_{-0.4}$     & 3322.5$^{+0.5}_{-0.4}$      & 3326.6$^{+2.2}_{-2.2}$      & ~ & 2.7$^{+0.4}_{-0.4}$             & 3.0$^{+0.4}_{-0.4}$        & 2.2$^{+0.6}_{-0.6}$        & 4.1$^{+2.3}_{-1.6}$        \\
Ar \emissiontype{XVIII} Ly$\beta_{1}$  & 3935.7                                                 & 3935.0$^{+0.9}_{-1.0}$          & 3936.2$^{+0.7}_{-1.3}$     & 3931.4$^{+1.5}_{-1.7}$      & 3946.7$^{+0.6}_{-1.2}$      & ~ & 2.8$^{+1.0}_{-0.8}$             & 2.3$^{+1.0}_{-0.9}$        & 3.7$^{+1.7}_{-1.6}$        & \multicolumn{1}{c}{$<$1.2} \\
Ca \emissiontype{XIX} w                & 3902.4                                                 & 3902.4$^{+0.2}_{-0.2}$          & 3902.5$^{+0.2}_{-0.3}$     & 3902.1$^{+0.3}_{-0.4}$      & 3902.4$^{+1.5}_{-1.1}$      & ~ & 2.3$^{+0.2}_{-0.2}$             & 2.4$^{+0.3}_{-0.3}$        & 2.2$^{+0.4}_{-0.4}$        & 2.7$^{+1.5}_{-1.2}$        \\
Ca \emissiontype{XIX} He$\beta_{1}$    & 4583.5                                                 & 4583.9$^{+0.4}_{-0.9}$          & 4583.8$^{+0.2}_{-0.9}$     & \multicolumn{1}{c}{(fixed)} & \multicolumn{1}{c}{(fixed)} & ~ & \multicolumn{1}{c}{$<$1.6}      & \multicolumn{1}{c}{$<$1.4} & \multicolumn{1}{c}{(tied)} & \multicolumn{1}{c}{(tied)} \\
Ca \emissiontype{XX} Ly$\alpha_{1}$    & 4107.5                                                 & 4107.9$^{+0.3}_{-0.5}$          & 4107.1$^{+0.7}_{-0.4}$     & 4108.4$^{+0.5}_{-0.6}$      & 4113.1$^{+1.8}_{-1.5}$      & ~ & 2.6$^{+0.4}_{-0.4}$             & 2.9$^{+0.5}_{-0.5}$        & 2.4$^{+0.7}_{-0.6}$        & 3.0$^{+2.1}_{-1.2}$        \\
Fe \emissiontype{XXV} w                & 6700.4                                                 & 6700.7$^{+0.0}_{-0.1}$          & 6700.7$^{+0.1}_{-0.1}$     & 6700.6$^{+0.1}_{-0.1}$      & 6701.9$^{+0.4}_{-0.2}$      & ~ & 4.2$^{+0.1}_{-0.1}$             & 4.3$^{+0.1}_{-0.1}$        & 4.0$^{+0.1}_{-0.1}$        & 4.3$^{+0.2}_{-0.2}$        \\
Fe \emissiontype{XXV} z                & 6636.6                                                 & 6636.5$^{+0.1}_{-0.1}$          & 6636.4$^{+0.2}_{-0.1}$     & 6636.5$^{+0.1}_{-0.2}$      & 6637.5$^{+0.5}_{-0.5}$      & ~ & 3.4$^{+0.1}_{-0.1}$             & 3.6$^{+0.1}_{-0.1}$        & 3.2$^{+0.2}_{-0.2}$        & 3.9$^{+0.5}_{-0.5}$        \\
Fe \emissiontype{XXV} He$\beta_{1}$    & 7881.5                                                 & 7881.1$^{+0.2}_{-0.2}$          & 7881.3$^{+0.3}_{-0.3}$     & 7881.0$^{+0.3}_{-0.3}$      & 7882.6$^{+0.5}_{-1.8}$      & ~ & 4.2$^{+0.2}_{-0.3}$             & 3.8$^{+0.3}_{-0.4}$        & 4.4$^{+0.3}_{-0.3}$        & 5.0$^{+0.9}_{-0.8}$        \\
Fe \emissiontype{XXV} He$\beta_{2}$    & 7872.0                                                 & \multicolumn{1}{c}{(tied)}      & \multicolumn{1}{c}{(tied)} & \multicolumn{1}{c}{(tied)}  & \multicolumn{1}{c}{(tied)}  & ~ & \multicolumn{1}{c}{(tied)}      & \multicolumn{1}{c}{(tied)} & \multicolumn{1}{c}{(tied)} & \multicolumn{1}{c}{(tied)} \\
Fe \emissiontype{XXV} He$\gamma_{1}$   & 8295.5                                                 & 8295.3$^{+0.4}_{-0.6}$          & 8295.5$^{+0.7}_{-0.7}$     & 8295.1$^{+0.6}_{-1.1}$      & \multicolumn{1}{c}{(fixed)} & ~ & 5.0$^{+0.5}_{-0.5}$             & 4.6$^{+0.7}_{-0.6}$        & 5.6$^{+0.8}_{-0.7}$        & \multicolumn{1}{c}{$<$0.0} \\
Fe \emissiontype{XXV} He$\delta_{1}$   & 8487.4                                                 & 8484.9$^{+1.4}_{-1.2}$          & 8485.7$^{+1.7}_{-1.7}$     & 8483.1$^{+2.1}_{-1.7}$      & \multicolumn{1}{c}{(fixed)} & ~ & 6.7$^{+1.3}_{-1.0}$             & 6.5$^{+1.7}_{-1.3}$        & 7.1$^{+2.8}_{-1.6}$        & \multicolumn{1}{c}{$<$0.0} \\
Fe \emissiontype{XXV} He$\epsilon_{1}$ & 8588.5                                                 & 8592.6$^{+1.3}_{-1.4}$          & 8592.6$^{+1.0}_{-1.9}$     & 8594.3$^{+3.0}_{-3.0}$      & \multicolumn{1}{c}{(fixed)} & ~ & 4.8$^{+1.2}_{-1.0}$             & 3.8$^{+1.4}_{-1.2}$        & 7.8$^{+3.5}_{-2.2}$        & \multicolumn{1}{c}{$<$0.0} \\
Fe \emissiontype{XXVI} Ly$\alpha_{1}$  & 6973.1                                                 & 6973.6$^{+0.1}_{-0.5}$          & 6973.3$^{+0.4}_{-0.4}$     & 6973.6$^{+0.2}_{-0.6}$      & 6973.9$^{+1.1}_{-1.6}$      & ~ & 4.3$^{+0.2}_{-0.2}$             & 4.7$^{+0.3}_{-0.3}$        & 3.8$^{+0.2}_{-0.4}$        & 6.1$^{+1.2}_{-0.9}$        \\
Fe \emissiontype{XXVI} Ly$\alpha_{2}$  & 6951.9                                                 & 6952.8$^{+0.3}_{-0.5}$          & 6952.7$^{+0.5}_{-0.8}$     & 6952.6$^{+0.5}_{-0.5}$      & 6953.1$^{+1.4}_{-1.9}$      & ~ & \multicolumn{1}{c}{(tied)}      & \multicolumn{1}{c}{(tied)} & \multicolumn{1}{c}{(tied)} & \multicolumn{1}{c}{(tied)} \\
Fe \emissiontype{XXVI} Ly$\beta_{1}$   & 8252.6                                                 & 8255.3$^{+1.0}_{-1.3}$          & 8254.6$^{+1.5}_{-1.9}$     & 8254.7$^{+1.5}_{-0.5}$      & 8247.3$^{+2.8}_{-0.9}$      & ~ & \multicolumn{1}{c}{$<$3.6}      & 3.0$^{+2.2}_{-1.9}$        & \multicolumn{1}{c}{$<$1.6} & \multicolumn{1}{c}{$<$3.5} \\
Ni \emissiontype{XXVII} w              & 7805.6                                                 & 7806.9$^{+0.9}_{-0.6}$          & 7807.2$^{+0.9}_{-0.8}$     & 7806.2$^{+1.5}_{-1.2}$      & 7804.0$^{+2.5}_{-4.1}$      & ~ & 5.4$^{+0.8}_{-0.7}$             & 5.4$^{+0.9}_{-0.8}$        & 5.8$^{+1.7}_{-1.4}$        & 6.6$^{+3.6}_{-2.3}$        \\
\hline
\end{tabular}
}\label{tab:linedata2}
\begin{tabnote}
\hangindent6pt\noindent
\hbox to6pt{\footnotemark[$*$]\hss}\unskip%
The Ly$\alpha_2$ lines of Si, S, Ar, and Ca are not shown because all the paramters of them are tied to Ly$\alpha_1$ (see \Tab{tab:linelist} for details).
\par
\hangindent6pt\noindent
\hbox to6pt{\footnotemark[$\dagger$]\hss}\unskip%
Fiducial energies of the lines at the rest frame. 
\end{tabnote}\end{table*}

\section{Single-line based emission measure limits}
\label{appendix:emissivities}
\begin{figure*}
 \begin{center}
 \subfloat[]{\includegraphics[width=7.5cm]{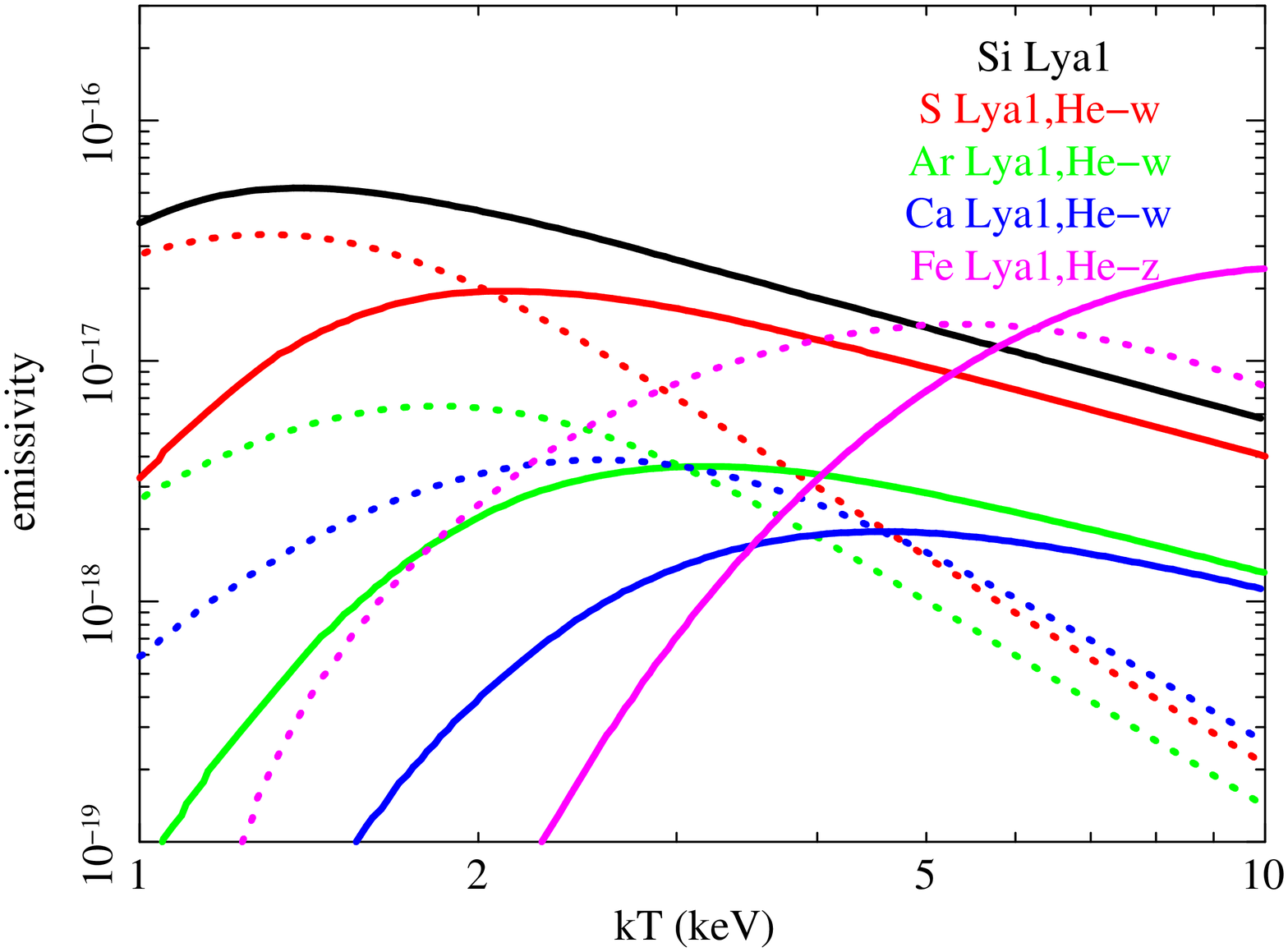}}
\hspace{1.0cm}
\subfloat[]{\includegraphics[width=7.5cm]{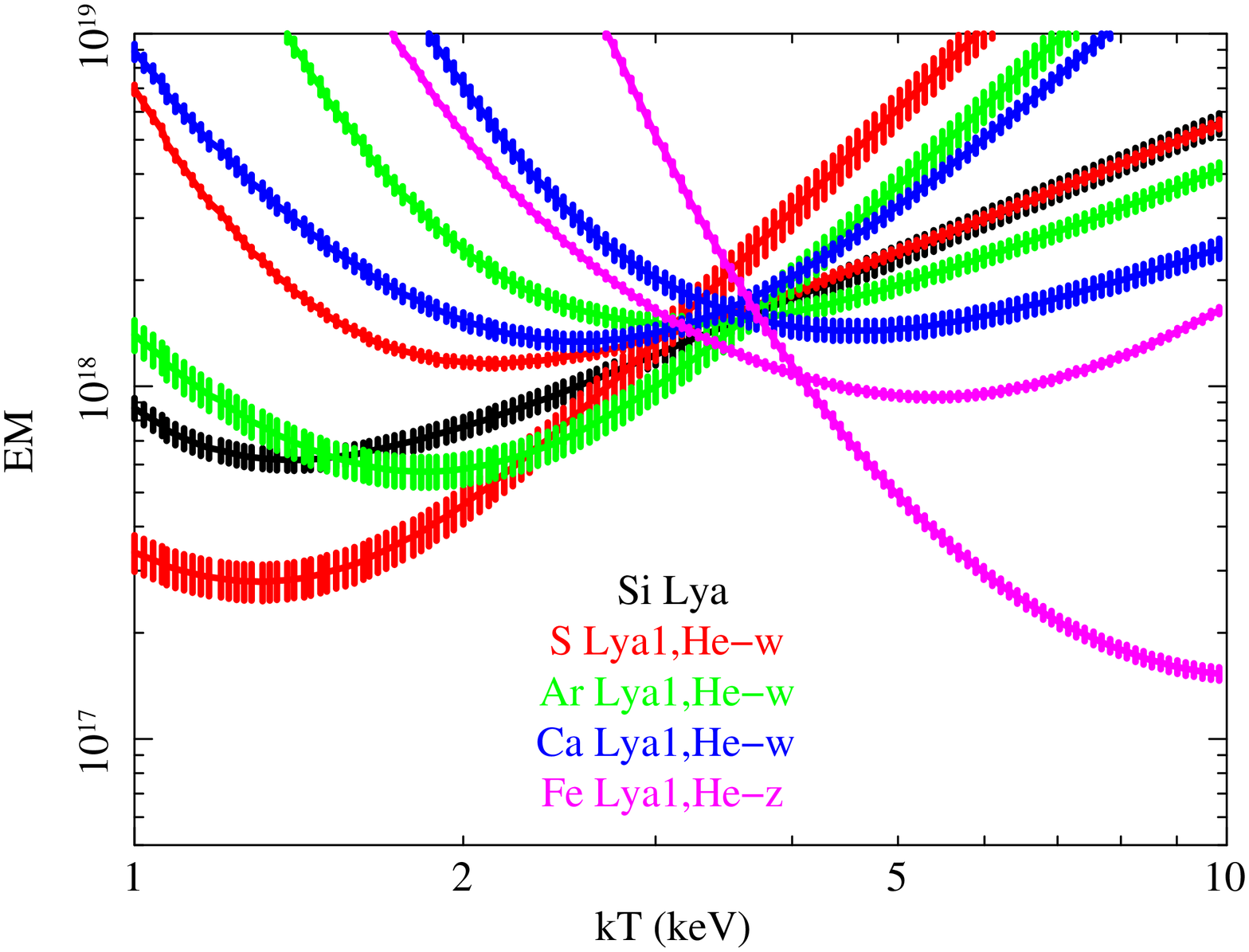}}
 \end{center}
 \caption{
   (a) Line emissivities
   of strong transitions from \texttt{AtomDB}
   for a given emission measure and metal abundances.
   The solid
   and dashed lines
   show H-like and He-like transitions,
   respectively.
   (b)
   Emission measure limits calculated from the \texttt{AtomDB} CIE model
   and the observed fluxes
   from the entire core.
 }
 \label{fig:emissivities}
\end{figure*}

In \Fig{fig:emissivities},
we show theoretical emissivities
for some of observed line transitions
as a function of electron temperature
based on \texttt{AtomDB}.
The peak temperature,
where the emissivity becomes maximum,
from these ions
cover a temperature range
from 1.5~keV (H-like Si)
to 13~keV (H-like Fe).
Therefore
measurements of line fluxes and ratios
are sensitive to emission from plasma at around this temperature range.

Combined with these emissivity values, 
the observed flux for each transition (\Tab{tab:linedata})
provides
constrains on emission measure
for a given temperature
and a metal abundance,
as shown in
\Fig{fig:emissivities}.
If the line emission originate from
a single component CIE plasma
any two curves from a single element
cross at a single point
of the model temperature and emission measure.
Furthermore,
if the assumed metal abundances are correct,
curves from different elements
also cross at a single point.
Our measured profiles
from the entire core
intersect together at around
3--4~keV,
indicating that
the observed line fluxes and hence their ratios
can be approximated by a single component CIE plasma with the solar abundance ratios.
From these curves
we notice that
the Fe~\emissiontype{XXVI}~Ly$\alpha$
is the most sensitive to hotter ($>4$~keV) emission
and
He-like S and Ar lines
are the most sensitive to cooler  ($<3$~keV) emission.

\section{Effective area uncertainties}
\label{appendix:arf}

\begin{figure}
\begin{center}
\includegraphics[width=8cm]{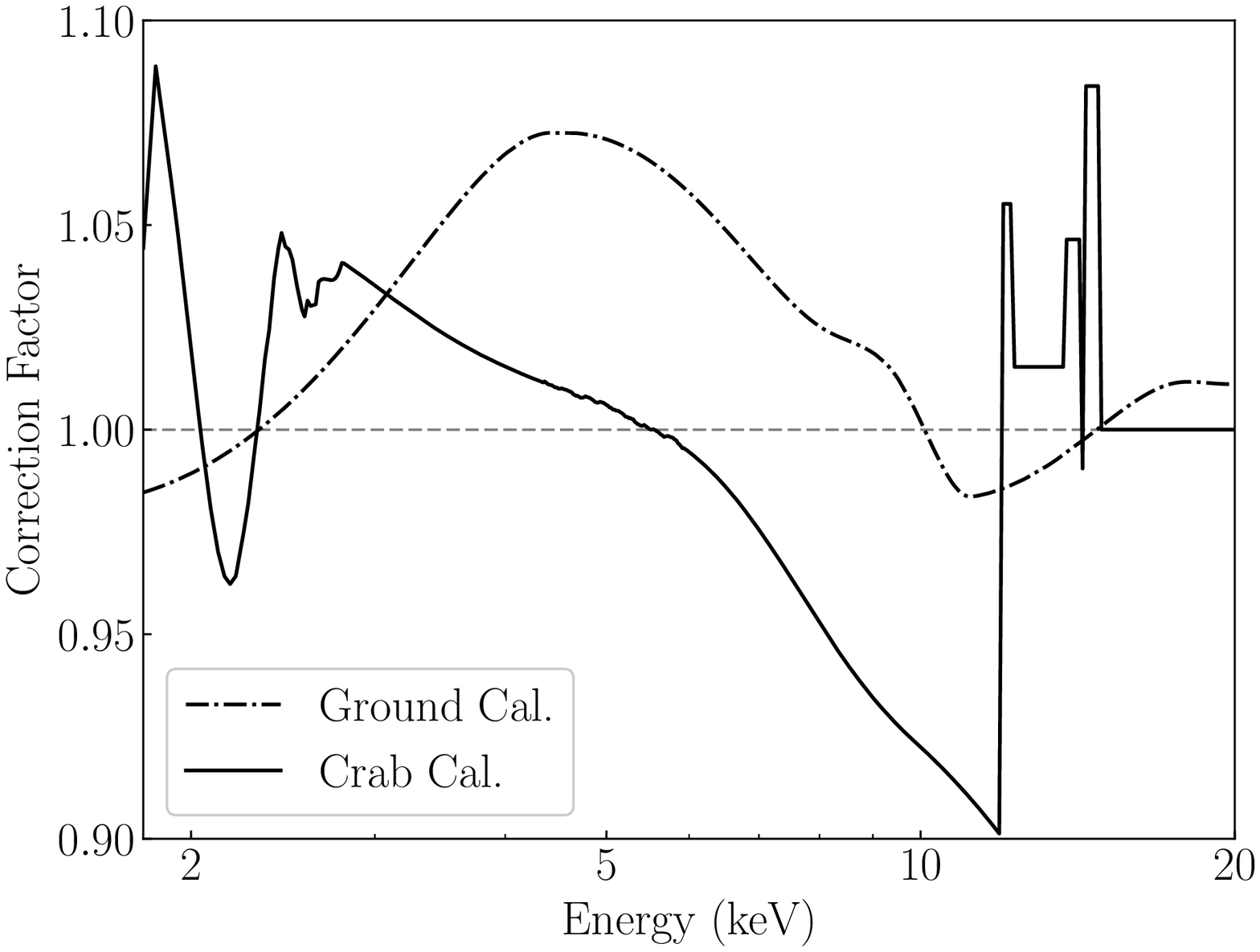}
\end{center}
\caption{Correction factor of the effective area with respect to the nominal value. The dot-dashed and solid lines indicate the calibration with the ground data and the actual Crab data, respectively.}
\label{fig:compare-arf}
\end{figure}

\begin{figure}
\begin{center}
\includegraphics[width=8cm]{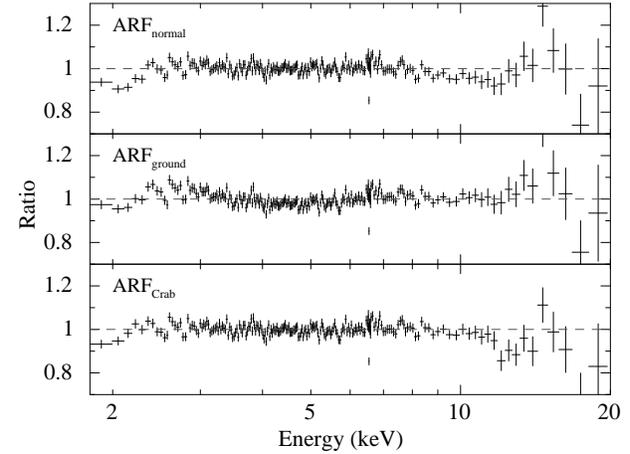}
\end{center}
\caption{Ratios of the Entire core spectrum to the best-fit modified 1CIE models with ARF$_\mathrm{normal}$ (top), ARF$_\mathrm{ground}$ (middle), and ARF$_\mathrm{Crab}$ (bottom).
}
\label{fig:compare-residuals}
\end{figure}

Because of the short life time of Hitomi, its in-flight calibration plan is not completed.
Data for the effective area calibration is especially limited.
In order to assess the uncertainty of the effective area, two kind of evaluations have been performed as follows. 
 
The instrument team compared the effective area derived from the ground calibration with that from ray-tracing simulations, and found residuals up to $\sim$7\% depending on the incident photon energy (\Fig{fig:compare-arf}).
According to this investigation, the correction factor for the ARF is provided as an \texttt{auxtransfile} in CALDB.
We corrected the ARF with this database (ARF$_\mathrm{ground}$).

\citet{tsujimoto17b} fitted the Crab spectrum with the canonical model ($\Gamma=2.1$; e.g., \cite{2015ApJ...801...66M}), and found residuals up to $\sim$10\% in the 1.8--20.0~keV band (\Fig{fig:compare-arf}).
The differences are probably due to uncertainties of not only the telescope reflectivity but also the transmission of the closed gate valve.  
This calibration method is not yet perfect because the spectral extraction region is smaller than that used for the canonical model due to the limited SXS FoV. 
Nevertheless, we made the local \texttt{auxtransfile} according to this result and corrected the ARF (ARF$_\mathrm{Crab}$).

We made the corrected ARF based on the above corrections factors (ARF$_\mathrm{ground}$ and ARF$_\mathrm{Crab}$). 
As described in \S\ref{ana:single-t}, we fitted the spectrum of the Entire core region with the modified 1CIE model using these corrected ARFs.
The residuals to the model with each ARF are shown in \Fig{fig:compare-residuals}.
The best-fit temperatures and normalizations are summarized in \Fig{fig:compare-arf-1cie}.
We also fitted the spectrum with the 2CIE model in the same manner, and show the results in \Fig{fig:compare-arf-2cie}.

\begin{figure*}
\begin{center}
\subfloat[Nebula]{\includegraphics[width=7.5cm]{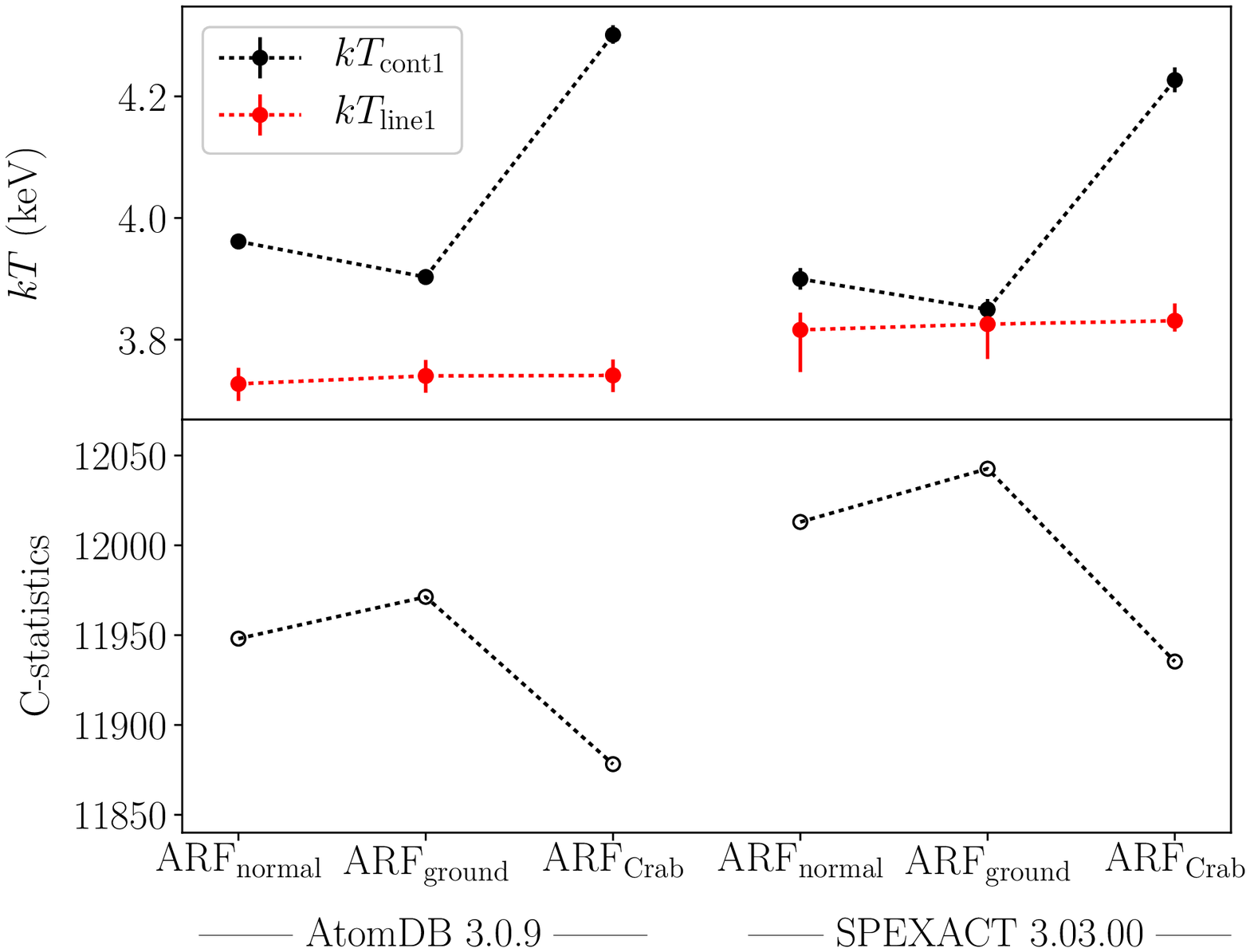}}
\hspace{1.0cm}
\subfloat[Rim]{\includegraphics[width=7.5cm]{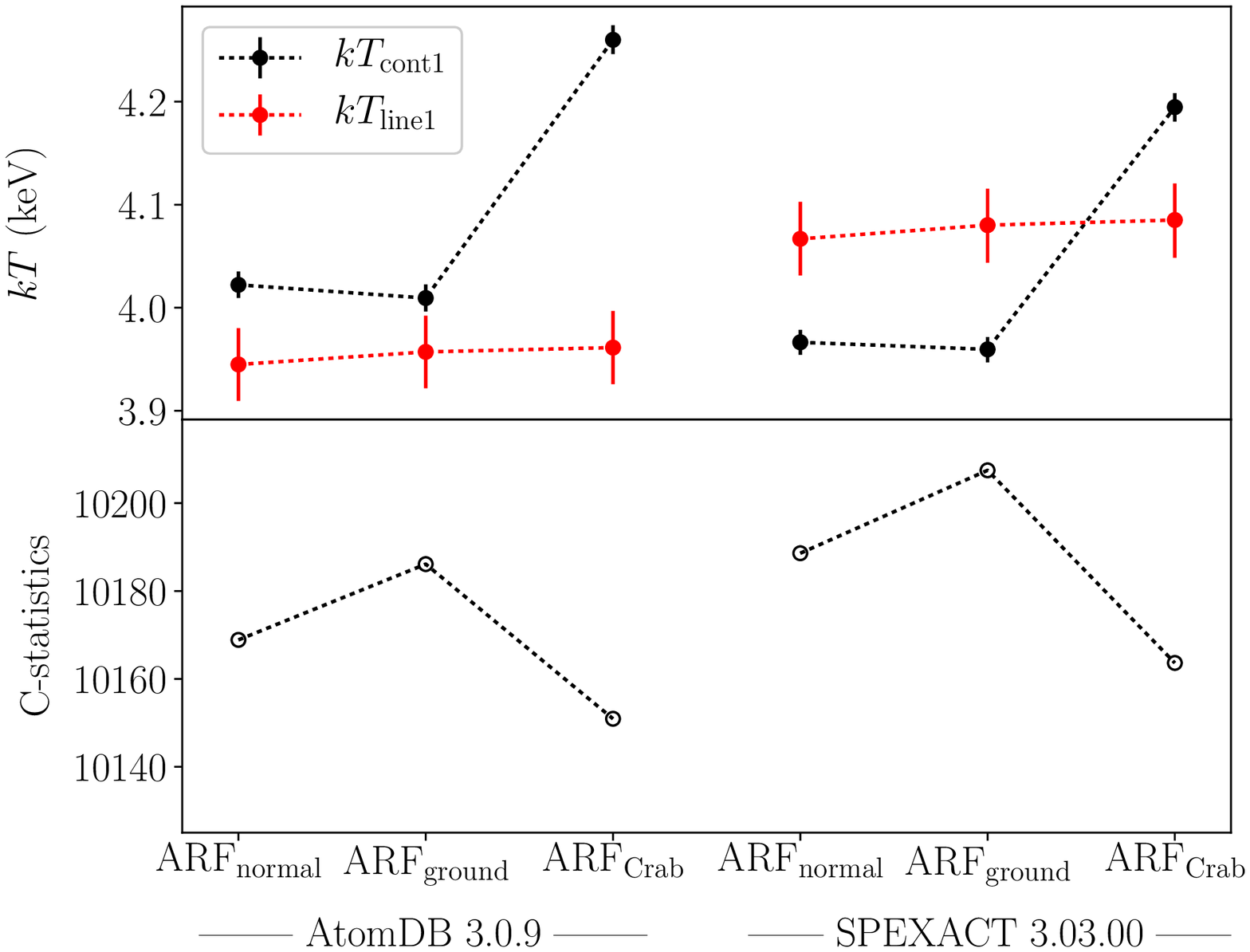}}
\vspace{0.5cm}
\subfloat[Outer]{\includegraphics[width=7.5cm]{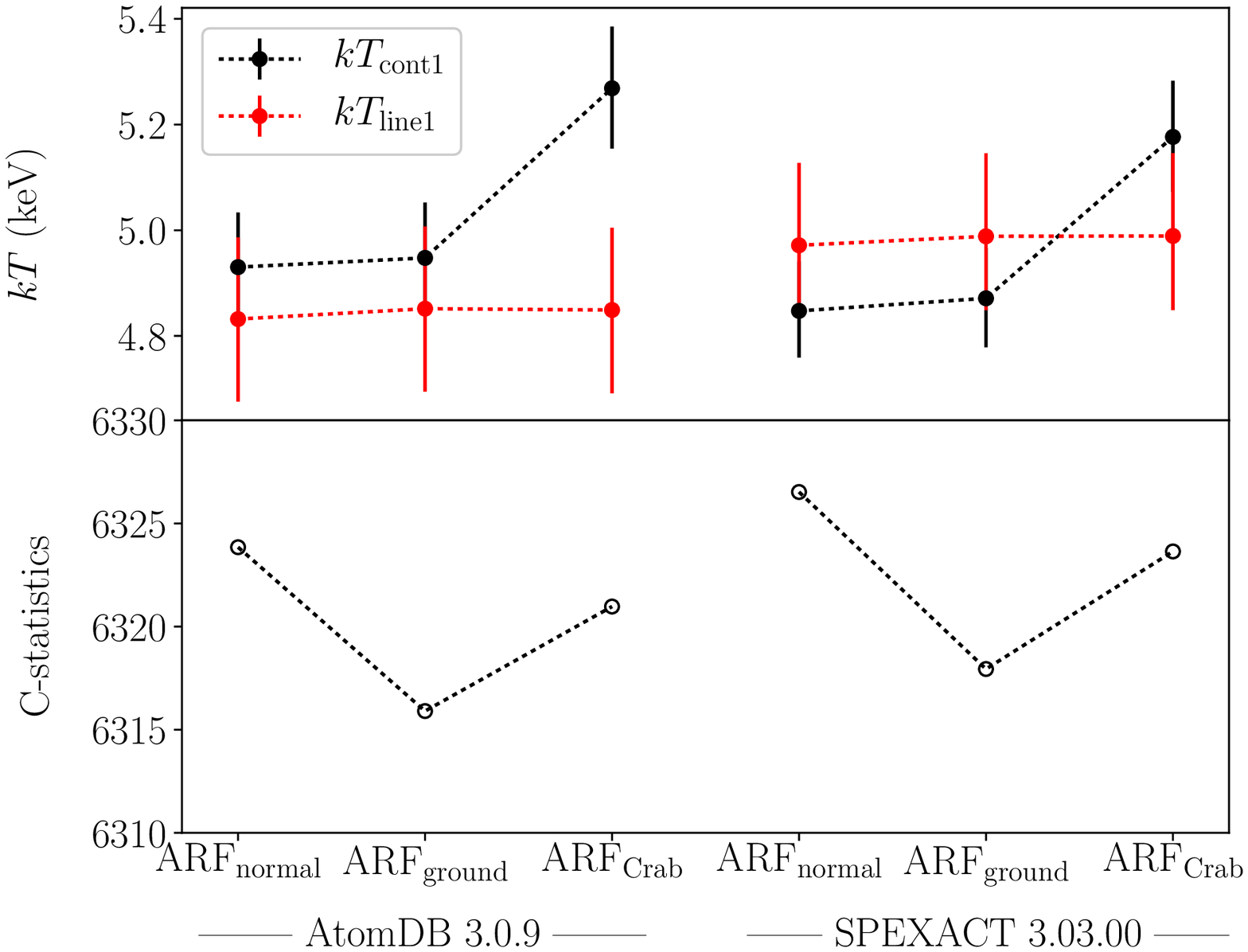}}
\end{center}
\caption{Fitting results of the modified 1CIE model with different ARFs in the Nebula, Rim, and Outer regions.}
\label{fig:compare-arf-divided-1cie}
\end{figure*}

\begin{figure*}
\begin{center}
\subfloat[Nebula]{\includegraphics[width=7.5cm]{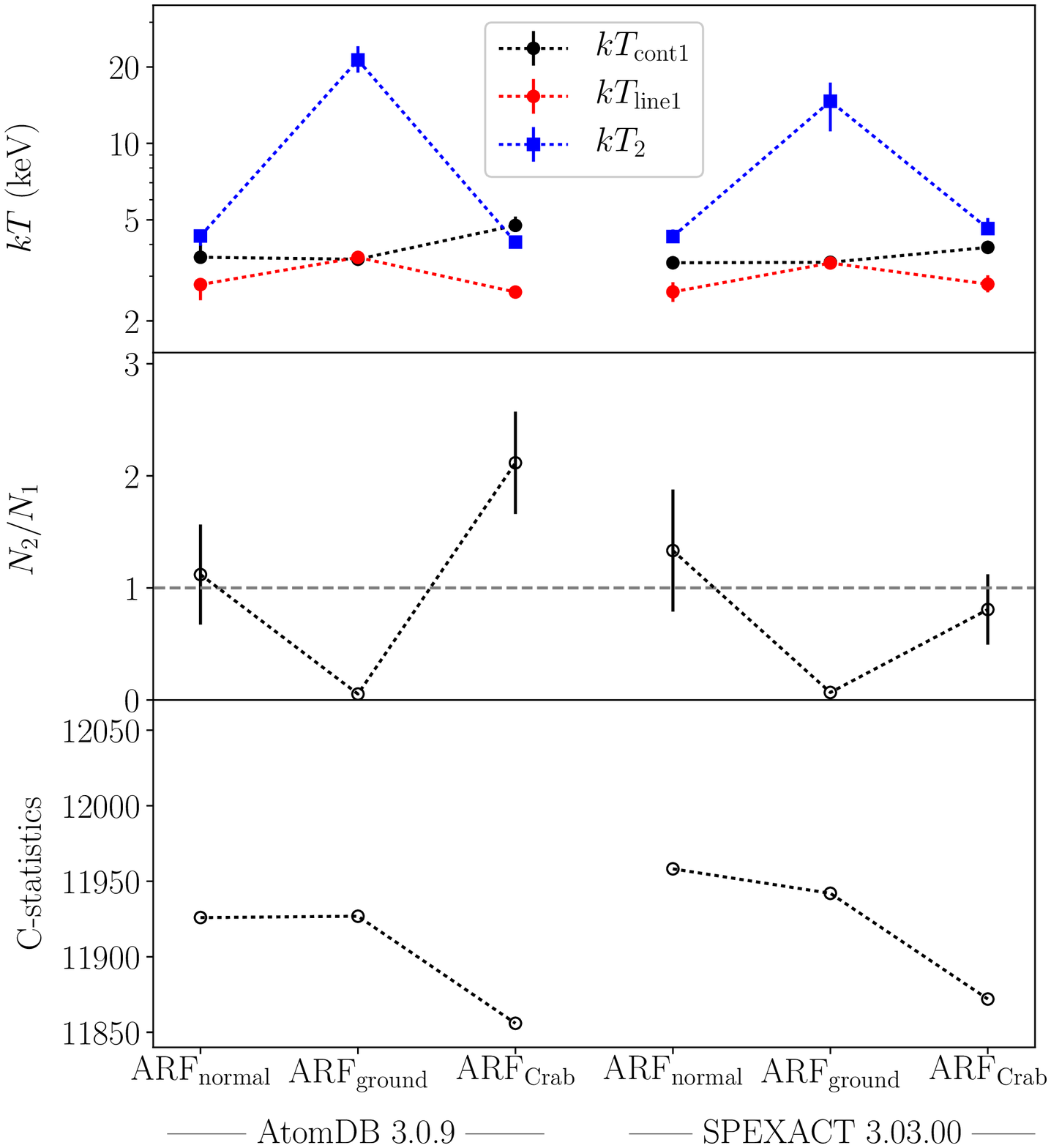}}
\hspace{1.0cm}
\subfloat[Rim]{\includegraphics[width=7.5cm]{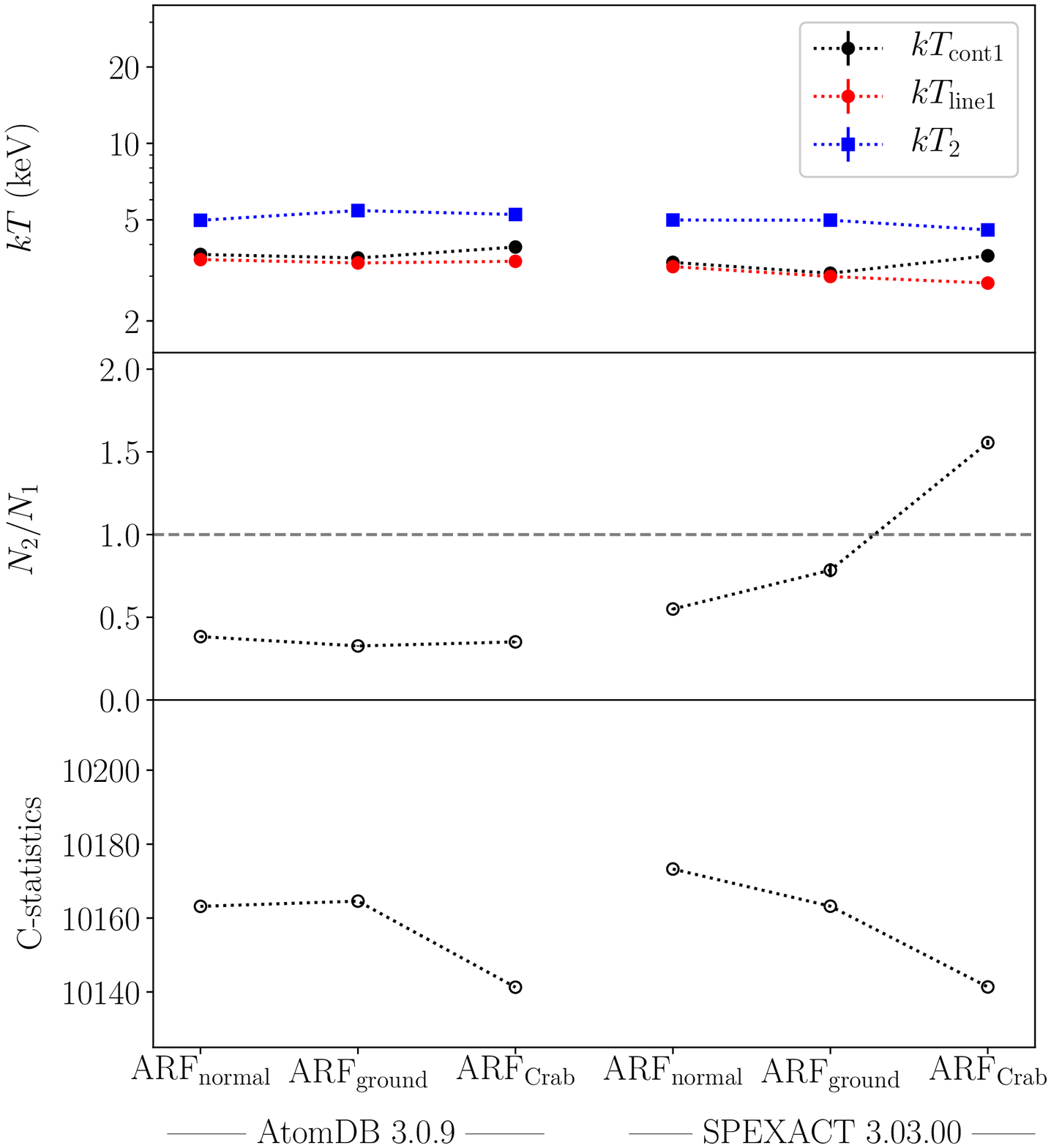}}
\end{center}
\caption{Fitting results of the 2CIE model with different ARFs in the Nebula and Rim regions.}
\label{fig:compare-arf-divided-2cie}
\end{figure*}

In \Fig{fig:compare-arf-divided-1cie} and \Fig{fig:compare-arf-divided-2cie}, we show the effect of the different ARFs on the fitting results for the Nebula, Rim, and (where applicable) Outer regions.
Similarly to the Entire core region, the measured values of $kT_\mathrm{line}$ are consistent with each other, while those of $kT_\mathrm{cont}$ significantly vary among the ARFs.  
The 2CIE model in the Nebula region also shows a similar tendency with the Entire core region, with ARF$_{\rm ground}$ giving an unphysically high temperature of the second component, while the thermal structures inferred for the Rim region are consistent for all the assumed ARFs.

\section{Line-on-sight velocity dispersion in the 2CIE model}
\label{appendix:sigmav}
In the 2CIE model described in \S\ref{ana:two-t}, the line-of-sight velocity dispersion is $145\pm3$~km~s$^{-1}$ and is consistent with the result of the single temperature model in the V~paper.
In order to investigate difference of the line-of-sight velocity dispersion between the lower and higher temperature components, we untied the link of the parameter between the two plasma models and refit the spectrum.
As a result, the velocity dispersion of the lower and higher temperature components in \texttt{SPEXACT} are $130\pm6$~km~s$^{-1}$ and $210^{+47}_{-24}$~km~s$^{-1}$, respectively, whereas those in \texttt{AtomDB} are both $145\pm7$~km~s$^{-1}$, suggesting that uncertainties of the atomic codes significantly affect the results.
Therefore, we found no significant difference of the line-of-sight velocity dispersion between the different temperature components.

\section{Comparison between AtomDB versions 3.0.8 and 3.0.9}
\label{appendix:atomdb}
\begin{figure}
\begin{center}
\includegraphics[width=8cm]{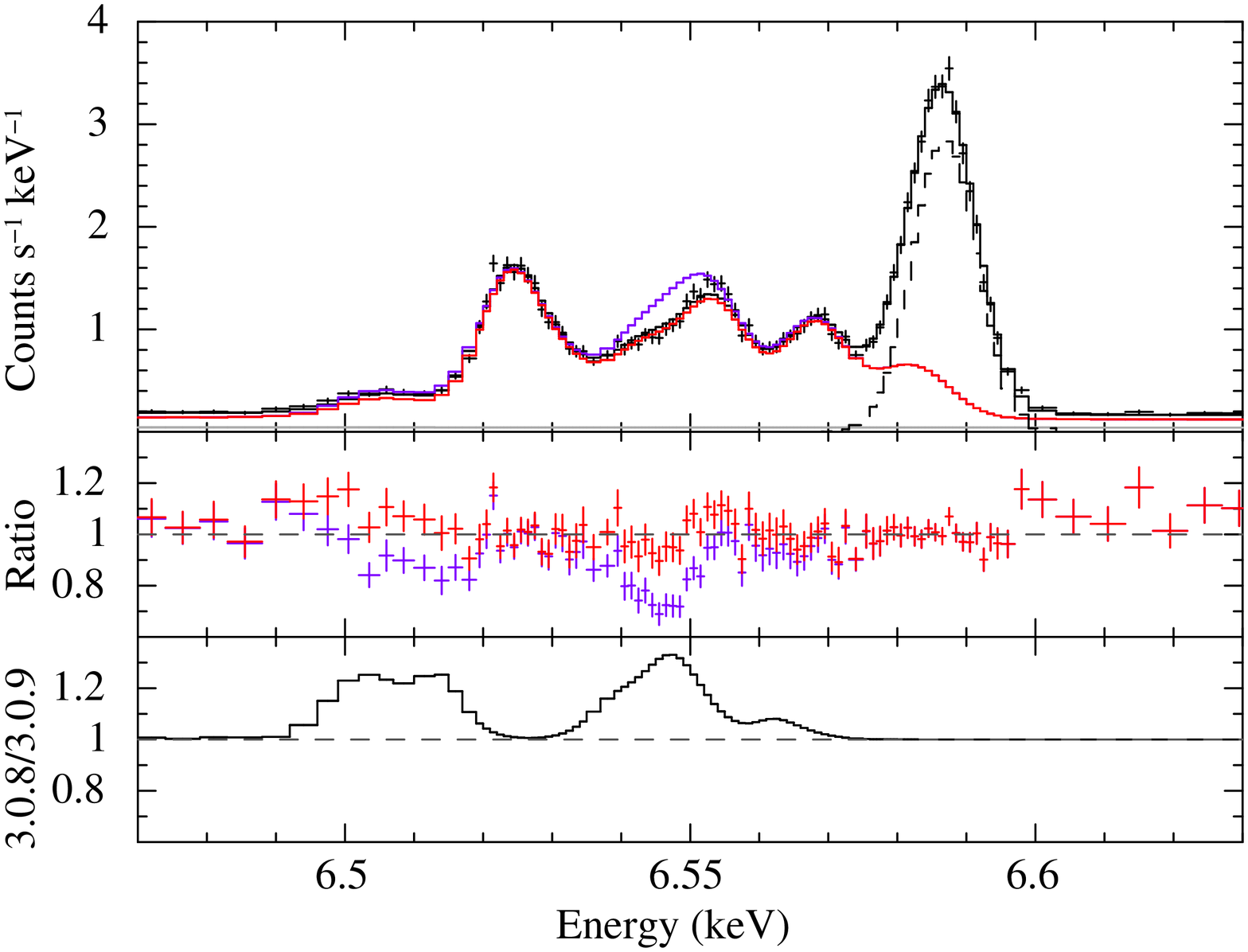}
\end{center}
\caption{Comparison between \texttt{AtomDB} version 3.0.9 (red) and 3.0.8 (purple) with the same parameters.
The top panel shows the SXS 6.47--6.63~keV spectrum in the Entire core region with the plasma models.
The middle panel is the ratio of the data to the model.
The bottom panel shows the ratio of the version 3.0.8 to the version 3.0.9.
}
\label{fig:compare-atomdb-308-309}
\end{figure}

In the latest release of \texttt{AtomDB} (version 3.0.9), the emissivities of the dielectric-recombination satellite lines for highly charged Fe are changed with respect to version 3.0.8.
The difference between versions 3.0.9 and 3.0.8 is shown in \Fig{fig:compare-atomdb-308-309}.
This difference does not appear to be large, but it significantly affects the result of the 2CIE model as shown in \Tab{tab:fit-compare-308-309}.
It demonstrates that uncertainties of the atomic code are very significant for the results obtained from high resolution spectroscopy measurements. 

\begin{table}[]
\centering

\tbl{Comparsion of the best-fit parameters between AtomDB 3.0.9 and 3.0.8}{%
\begin{tabular}{lcc}
\hline
Model/Parameter                                            & AtomDB v3.0.9           & AtomDB v3.0.8       \\
\hline \multicolumn{3}{l}{Modified 1CIE model}             \\
~~~$kT_{\mathrm{cont}}$~(keV)                              & 4.01$^{+0.01}_{-0.01}$  & 4.02$^{+0.01}_{-0.01}$  \\
~~~$kT_{\mathrm{line}}$~(keV)                              & 3.80$^{+0.02}_{-0.02}$  & 4.00$^{+0.02}_{-0.02}$  \\
~~~$N$~(10$^{12}$~cm$^{-5}$)                               & 22.77$^{+0.04}_{-0.04}$ & 23.07$^{+0.04}_{-0.04}$ \\
~~~C-statistics/dof                                        & 13085.9/12978           & 13183.5/12978           \\
\hline \multicolumn{3}{l}{2CIE model (modified CIE + CIE)} \\
~~~$kT_{\mathrm{cont}1}$~(keV)                             & 3.66$^{+0.01}_{-0.02}$  & 4.06$^{+0.01}_{-0.01}$  \\
~~~$kT_{\mathrm{line}1}$~(keV)                             & 3.06$^{+0.04}_{-0.03}$  & 4.03$^{+0.03}_{-0.08}$  \\
~~~$kT_2$~(keV)                                            & 4.51$^{+0.02}_{-0.03}$  & 1.59$^{+0.03}_{-0.17}$  \\
~~~$N_1$~(10$^{12}$~cm$^{-5}$)                             & 12.98$^{+0.05}_{-0.05}$ & 22.64$^{+0.05}_{-0.05}$ \\
~~~$N_2$~(10$^{12}$~cm$^ {-5}$)                            & 9.71$^{+0.06}_{-0.05}$  & 0.68$^{+0.11}_{-0.11}$  \\
~~~C-statistics/dof                                        & 13058.5/12976           & 13178.0/12976           \\
\hline
\end{tabular}
}\label{tab:fit-compare-308-309}
\end{table}


%
%
%

\end{document}